\documentclass{aa}  

\usepackage{graphicx}
\usepackage{txfonts}
\usepackage[allcolors=blue]{hyperref}
\usepackage{orcidlink}
\begin{document}

   \title{Long-term monitoring of large-scale magnetic fields across optical and near-infrared domains with ESPaDOnS, Narval and SPIRou}
\subtitle{The cases of EV~Lac, DS~Leo, and CN~Leo}
   \titlerunning{Monitoring the large-scale field of active M dwarfs}

   \author{S. Bellotti \inst{1,2}\orcidlink{0000-0002-2558-6920}
          \and
          J. Morin \inst{3}\orcidlink{0000-0002-4996-6901}
          \and
          L. T. Lehmann\inst{1}\orcidlink{0000-0001-5674-2116}
          \and
          P. Petit \inst{1}\orcidlink{0000-0001-7624-9222}
          \and
          G. A. J. Hussain \inst{2}\orcidlink{0000-0003-3547-3783}
          \and 
          J-F. Donati \inst{1}\orcidlink{0000-0001-5541-2887}
          \and
          C. P. Folsom \inst{4}\orcidlink{0000-0002-9023-7890}
          \and
          A. Carmona \inst{5}\orcidlink{0000-0003-2471-1299}
          \and
          E. Martioli \inst{6,7}\orcidlink{0000-0002-5084-168X}
          \and
          B. Klein \inst{8}\orcidlink{0000-0003-0637-5236}
          \and
          P. Fouqu\'e\inst{1}\orcidlink{0000-0002-1436-7351}
          \and
          C. Moutou \inst{1}\orcidlink{0000-0002-2842-3924}
          \and
          S. Alencar\inst{9}\
          \and
          E. Artigau\inst{10}\orcidlink{0000-0003-3506-5667}
          \and
          I. Boisse\inst{11}\orcidlink{0000-0002-1024-9841}
          \and
          F. Bouchy\inst{12}\orcidlink{0000-0002-7613-393X}
          \and
          J. Bouvier\inst{5}\orcidlink{0000-0002-7450-6712}
          \and
          N.~J. Cook\inst{9}\orcidlink{0000-0003-4166-4121}
          \and
          X. Delfosse \inst{5}\orcidlink{0000-0001-5099-7978}
          \and
          R. Doyon\inst{9}\orcidlink{0000-0001-5485-4675}
          \and
          G. H\'ebrard\inst{7,13}\orcidlink{0000-0001-5450-7067}          
          }
   \authorrunning{Bellotti et al.}

   \institute{
            Institut de Recherche en Astrophysique et Plan\'etologie,
            Universit\'e de Toulouse, CNRS, IRAP/UMR 5277,
            14 avenue Edouard Belin, F-31400, Toulouse, France\\
            \email{stefano.bellotti@irap.omp.eu}
        \and
             Science Division, Directorate of Science, 
             European Space Research and Technology Centre (ESA/ESTEC),
             Keplerlaan 1, 2201 AZ, Noordwijk, The Netherlands
        \and
             Laboratoire Univers et Particules de Montpellier,
             Universit\'e de Montpellier, CNRS,
             F-34095, Montpellier, France
        \and 
            Tartu Observatory, 
            University of Tartu, 
            Observatooriumi 1, Tõravere, 61602 Tartumaa, Estonia
        \and
            Univ. Grenoble Alpes, CNRS, IPAG, 38000 Grenoble, France
        \and     
           Laborat\'{o}rio Nacional de Astrof\'{i}sica, Rua Estados Unidos 154, 37504-364, Itajub\'{a} - MG, Brazil
        \and 
           Institut d'Astrophysique de Paris, CNRS, UMR 7095, Sorbonne Universit\'{e}, 98 bis bd Arago, 75014 Paris, France
        \and
            Sub-department of Astrophysics, Department of Physics, University of Oxford, Oxford, OX1 3RH, UK
        \and
            Universidade Federal de Minas Gerais, Belo Horizonte, MG, 31270-901, Brazil
        \and
            Universit\'e de Montr\'eal, D\'epartement de Physique, IREX,
            Montr\'eal, QC H3C 3J7, Canada
        \and
            Aix Marseille Univ, CNRS, CNES, LAM, Marseille, France
        \and
            Observatoire de Gen\`eve, Universit\'e de Gen\`eve, Chemin Pegasi, 51, 1290 Sauverny, Switzerland
        \and
            Observatoire de Haute Provence, St Michel l'Observatoire, France
}

   \date{Received ; accepted }

 
  \abstract
   {Dynamo models describing the generation of stellar magnetic fields for partly and fully convective stars are guided by observational constraints. Zeeman-Doppler imaging has revealed a variety of magnetic field geometries and, for fully convective stars in particular, a dichotomy: either strong, mostly axisymmetric, and dipole-dominated or weak, non-axisymmetric, and multipole-dominated. This dichotomy is explained either by dynamo bistability (i.e. two coexisting and stable dynamo branches) or by long-term magnetic cycles with polarity reversals, but there is no definite conclusion on the matter. }
   {Our aim is to monitor the evolution of the large-scale field for a sample of nearby M~dwarfs with masses between 0.1 and 0.6~$M_\odot$, which is of prime interest to inform distinct dynamo theories and explain the variety of magnetic field geometries studied in previous works. This also has the potential to put long-term cyclic variations of the Sun's magnetic field into a broader context.}
   {We analysed optical spectropolarimetric data sets collected with ESPaDOnS and Narval between 2005 and 2016, and near-infrared SPIRou data obtained between 2019 and 2022 for three well-studied, active M~dwarfs: EV~Lac, DS~Leo, and CN~Leo. We looked for secular changes in time series of longitudinal magnetic field, width of unpolarised mean-line profiles, and large-scale field topology as retrieved with principal component analysis and Zeeman-Doppler imaging.}
   {We retrieved pulsating (EV~Lac), stable (DS~Leo), and sine-like (CN~Leo) long-term trends in longitudinal field. The width of near-infrared mean-line profiles exhibits rotational modulation only for DS~Leo, whereas in the optical it is evident for both EV~Lac and DS~Leo. The line width variations are not necessarily correlated to those of the longitudinal field, suggesting complex relations between small- and large-scale field. We also recorded topological changes in the form of a reduced axisymmetry for EV~Lac and transition from a toroidal-dominated to poloidal-dominated regime for DS~Leo. For CN~Leo, the topology remained predominantly poloidal, dipolar, and axisymmetric, with only an oscillation in field strength.}
   {Our results show a peculiar evolution of the magnetic field for each M~dwarf individually, with DS~Leo and EV~Lac manifesting more evident variations than CN~Leo. These findings confirm that M~dwarfs with distinct masses and rotation periods can undergo magnetic long-term variations and suggest an underlying variety of cyclic behaviours of their magnetic fields.} 


   \maketitle

%

\section{Introduction}\label{sec:introduction}

The magnetic fields of low-mass stars ($M_*<1.2M_\odot$) are powered by dynamos \citep{Schrijver2000}, and their study is paramount to understanding stellar evolution and activity phenomena. The evolution of a star's rotation is linked to magnetic activity as the magnetic field couples with the stellar wind and results in angular momentum loss over the star's lifetime \citep{Skumanich1972,Vidotto2014,Finley2017,See2019}. Therefore, one can infer the stellar age and rotation period using magnetic activity as a proxy \citep{Noyes1984,LorenzoOliveira2018,Dungee2022}. Moreover, stellar magnetic activity is responsible for inhomogeneities in brightness (spots, faculae, plages) and local velocity fields (suppression of convection) that may prevent the unambiguous detection and characterisation of exoplanets, especially those similar to the Earth \citep{Queloz2001,Huerta2008,Meunier2021}. Finally, the stellar magnetism dictates the environment in which exoplanets orbit  \citep[e.g.][]{Folsom2020,Bellotti2023a}, influencing their potential habitability \citep[e.g.][]{Segura2010,Vidotto2013,Luger2015,Tilley2019}.

On the low-mass end of the main sequence, M~dwarfs are the most common spectral type in the solar neighbourhood \citep{Reid2004}, with masses ranging between 0.08 and 0.57\,$M_\odot$ \citep{Pecaut2013}. Above 0.35\,$M_\odot$ (approximately M3.5 type), the internal structure is solar-like, that is, with an inner radiative core and an outer convective envelope separated by the tachocline \citep{Chabrier1997}. The tachocline is an interface region of strong shear, where the magnetic field is thought to be amplified. Below 0.35\,$M_\odot$, M~dwarfs possess fully convective interiors, and the absence of a tachocline challenges dynamo theories relying on a deep-seated interface \citep{Durney1993,Chabrier2006,Browning2008,Yadav2015}. Overall, M~dwarfs represent exquisite laboratories to study dynamo-powered magnetic field generation, under similar and different physical interior conditions to those of  the Sun. 

Dynamo theories are informed by observations of stellar magnetic fields, which can be measured using two complementary approaches \citep{Reiners2012,Kochukhov2021}. The modulus of the magnetic field vector is estimated via radiative transfer modelling of Zeeman splitting and magnetic intensification of individual unpolarised spectral lines. Values of field strength between 0.2 and 7\,kG have been reported \citep{Shulyak2017,Shulyak2019,Reiners2022,Cristofari2023}, and they follow the activity-rotation relation exhibiting a (quasi-)saturated and non-saturated regime \citep{ReinersBasri2009,Shulyak2019,Reiners2022}. The geometry of the large-scale field can be inferred by means of tomographic inversion techniques, for example Zeeman-Doppler imaging (ZDI; \citealt{Semel1989,DonatiBrown1997}) applied to spectropolarimetric time series of linearly and circularly polarised spectra \citep{Morin2012}.

The application of ZDI to spectropolarimetric data of partly and fully convective low-mass stars have revealed a wide variety of large-scale field geometries \citep[e.g.][]{Donati2008,Morin2008,Morin2010,Fares2013,Hebrard2016}. Partly convective stars with masses higher than 0.5\,$M_\odot$ tend to have weak, predominantly toroidal, and non-axisymmetric fields, whereas stars with masses between 0.35 and 0.5\,$M_\odot$ harbour stronger, poloidal, and axisymmetric fields \citep{Donati2008,Phan-Bao2009}. For fully convective stars close to the 0.35\,$M_\odot$ boundary, the large-scale field is strong, mainly poloidal, and axisymmetric, while below $M<0.2M_\odot$ a dichotomy of topologies co-exist: weak, complex, and  non-axisymmetric or strong, simple, and  axisymmetric \citep{Morin2008,Morin2010}. This dichotomy can be explained by either two distinct and independent branches of dynamo known as bistability \citep{Morin2011,Gastine2013,Kochukhov2017} or by assuming that fully convective M~dwarfs undergo magnetic cycles, and previous ZDI reconstructions captured only a snapshot of a long-term topological variation \citep{Kitchatinov2014}. However, no definitive interpretation has been reached so far.

The Sun is an important benchmark for stellar cycles, because the large-scale dipolar component of the magnetic field undergoes a polarity reversal in a cyclical fashion every 11\,yr, and it is accompanied by an oscillation in the fraction of poloidal-to-toroidal magnetic energy \citep{Sanderson2003,Charbonneau2010}. More precisely, the poloidal component peaks at cycle minimum, and the toroidal components increase towards cycle maximum. However, our understanding of the cyclic nature of the solar magnetism is sill not complete \citep{Charbonneau2020}, and its contextualisation advocates for additional magnetic field observations of distinct stellar types. In this direction, informing dynamo theories requires a long-term spectropolarimetric monitoring of selected M~dwarfs, for which we can trace the secular evolution of the large-scale field geometry \citep{Klein2021,Klein2022,Bellotti2023b}. There is observational evidence of activity cycles for M~dwarfs, from photometric and chromospheric activity monitoring \citep[e.g.][]{SuarezMascareno2016,SuarezMascareno2018,Fuhrmeister2023,Mignon2023}, and  from radial velocity searches of exoplanets \citep{GomesDaSilva2012,Wargelin2017,LopezSantiago2020}, but we are only starting to capture long-term behaviour that may resemble solar-like magnetic cycles \citep{Bellotti2023b,Lehmann2024}. For instance, the recent work of \citet{Bellotti2023b} reports the long-term evolution of the magnetic field of AD~Leo, which has similarities with the evolution of the Sun's field.

With this work we investigate the long-term monitoring of three active M~dwarfs, EV~Lac, DS~Leo, and CN~Leo, by analysing spectropolarimetric data collected across the optical and near-infrared domain. To investigate the long-term evolution, we analysed the full time series of the longitudinal magnetic field (B$_l$) and the full width at half maximum (FWHM) of unpolarised (Stokes~$I$) mean-line profiles. We reconstructed the large-scale field topology via ZDI, and we inspected temporal changes in the morphology of circular polarisation profiles via principal component analysis (PCA; \citealt{Lehmann2022}).
The paper is structured as follows. In Sect.~\ref{sec:observations} we describe the near-infrared and optical spectropolarimetric observations. We  outline the analysis of the longitudinal magnetic field in Sect.~\ref{sec:Blon}, the analysis FWHM of Stokes $I$ in Sect.~\ref{sec:FWHM}, the PCA technique in Sect.~\ref{sec:pca}, and the ZDI reconstructions in Sect.~\ref{sec:magnetic_imaging}. Finally, we present our conclusions in Sect.~\ref{sec:conclusions}.

\section{Observations}\label{sec:observations}

We used spectropolarimetric data collected in near-infrared and optical domains, and performed a large-scale magnetic field monitoring for three M-type stars: EV~Lac (GJ~873), DS~Leo (GJ~410), and CN~Leo (GJ~406). A summary of their properties is given in Table~\ref{tab:star_properties}.

According to stellar models, the transition between partly and fully convective interiors spans from 0.20 to 0.35\,$M_\odot$ \citep{Dorman1989,Chabrier1997,Rabus2019}. Other factors like age, metallicity and magnetic field strength play a role \citep{Mullan2001,Maeder2000,VanSaders2012,Tanner2013}. With such a range, DS~Leo and CN~Leo are partly and fully convective, respectively. With a mass of 0.32\,$M_\odot$, EV~Lac is either partly convective or near the transition between the two regimes.

\begin{table*}[ht]
\caption{Properties of the M~dwarfs examined.} 
\label{tab:star_properties}     
\centering                       
\begin{tabular}{l c c c c c c c c c c}      
\hline     
Star & ID & Sp Type & M$_{*}$ & R$_{*}$ & P$_\mathrm{rot}$ & $v_\mathrm{eq}\sin i$ & $i$ & $\log(\mathrm{L}_\mathrm{X}/\mathrm{L}_\mathrm{bol})$ & $\mathrm{logR'}_\mathrm{HK}$\\
& & & [$M_\odot$] & [$R_\odot$] & [d] & [km s$^{-1}$] & [deg] & [dex] & [dex]\\
\hline
DS~Leo & GJ~410 & M1.0 & 0.58 & 0.53 & $13.91\pm0.01^\dagger$  & 2.0 & 60 & $-$3.80 & $-$4.16\\
EV~Lac & GJ~873 & M3.5 & 0.32 & 0.30 & $4.36\pm0.01^\dagger$   & 4.0 & 60 & $-$1.99 & $-$3.75\\
CN~Leo & GJ~406 & M5.5 & 0.10 & 0.12 & $2.70\pm0.01^\dagger$   & 2.0 & $45\pm20^\dagger$ & $\ldots$ & $-$4.01\\
\hline                                 
\end{tabular}
\tablefoot{The columns list the following quantities: 1) and 2) stellar name and alternative identifier, 3) spectral type, 4) mass, 5) radius, 6) rotation period at equator, 7) equatorial projected rotational velocity, 8) inclination, 9) X-ray-to-bolometric luminosity ratio \citep{Wright2011}, and 10) CaII H\&K index \citep{Noyes1984,BoroSaikia2018}. Columns 4) and 5) are  from \citet{Cristofari2023}, and columns 7) and 8) are from \citet{Morin2008}, \citet{Hebrard2016}, and \citet{Cristofari2023}. Columns 9) and 10) are  from \citet{Wright2011}, \citet{Stelzer2013} and \citet{BoroSaikia2018}, respectively. The dagger ($\dagger$) indicates that the parameter was estimated in this work.}
\end{table*}

\subsection{Near-infrared}

All near-infrared observations were performed in circular polarisation mode with the SpectroPolarim\`etre InfraRouge (SPIRou) as part of the large programme SPIRou Legacy Survey\footnote{\url{http://spirou.irap.omp.eu/Observations/The-SPIRou-Legacy-Survey}}
(SLS; id P42, PI: Jean-Fran\c{c}ois Donati). SPIRou is a stabilised high-resolution near-infrared spectropolarimeter \citep{Donati2020} mounted on the 3.6\,m Canada–France–Hawaii Telescope (CFHT) atop Maunakea, Hawaii. It provides a quasi-continuous coverage of the near-infrared spectrum from 0.96 to 2.5~$\mu$m ($YJHK$ bands) at a spectral resolving power of $R \sim 70\,000 $, with a 2-nm gap between 2.4371 and 2.4391~$\mu$m \citep{Donati2020}. Optimal extraction of SPIRou spectra was carried out with {\it A PipelinE to Reduce Observations} (\texttt{APERO} v0.6.132), a fully automatic reduction package installed at CFHT \citep{Cook2022}. The journal of the observations is given in Appendix~\ref{app:logs}.

Starting from the polarimetric products of \texttt{APERO} \citep[see][for more details]{Cook2022}, we computed Stokes~$I$ (unpolarised) and $V$ (circularly polarised) mean profiles using least-squares deconvolution \citep[LSD;][]{Donati1997,Kochukhov2010}.\footnote{We used the python LSD code available at \href{https://github.com/folsomcp/LSDpy}{https://github.com/folsomcp/LSDpy}} With this technique the observed spectrum is deconvolved with a line list, namely a series of Dirac delta functions located at each absorption line in the stellar spectrum and with the associated line features such as depth, and sensitivity to Zeeman effect (commonly known as Land\'e factor and indicated as $g_\mathrm{eff}$). The deconvolution results in an individual, high-signal-to-noise ratio (S/N) kernel summarising the properties of thousands of spectral lines, and allowing us to extract polarimetric information from the spectrum.

For our stars, we used two line lists corresponding to a local thermodynamic equilibrium model \citep{Gustafsson2008} characterised by $\log g=$ 5.0\,[cm s$^{-2}$], $v_{\mathrm{micro}}=$ 1\,km s$^{-1}$, and $T_{\mathrm{eff}}=3000$\,K (for CN~Leo) and $T_{\mathrm{eff}}=3500$\,K (for EV~Lac and DS~Leo). These two lists contain 1000 and 1400 atomic photospheric lines between 950--2600\,nm with depth larger than 3\,\% the continuum level. The depth threshold is chosen to remove shallow lines (with low effective S/N), while keeping a large number of lines with which to compute LSD profiles. The masks were synthesised using the Vienna Atomic Line Database\footnote{\url{http://vald.astro.uu.se/} using the Montpellier mirror to request locally MARCS model atmospheres.} \citep[VALD,][]{Ryabchikova2015}, and contain information on $g_\mathrm{eff}$ (ranging from 0 to 3), which is not accessible with an empirical or molecular mask. Given the spectral type of EV~Lac and DS~Leo, we respectively tested a mask associated with 3000\,K (590 lines) 4000\,K (1550 lines), to check the robustness of our results. { For both Stokes~$I$ and $V$, we adopted the weighting factor $d\mathrm{g}_\mathrm{eff}\lambda/(d_n\mathrm{g}_\mathrm{eff,n}\lambda_n)$, with $d$ is the line depth and $\lambda$ is the wavelength, whereas the quantities at denominator are the normalisation parameters \citep[see][for more details]{Kochukhov2010}.}

A forest of telluric absorption lines due to the Earth's atmosphere pollute the wavelength domain in which SPIRou operates. As described in \citet{Cook2022} and \citet{Artigau2022}, the removal of telluric contamination from science data is performed by \texttt{APERO} with a two-step algorithm. First, a pre-cleaning is carried out by means of a TAPAS \citep[Transmissions of the AtmosPhere for AStromomical data;][]{Bertaux2014} absorption model, which is also applied to a set of hot stars to obtain a library of telluric-correction residuals. From this, a telluric-residuals model is constructed and subtracted from the science frames. To account for potential residuals in the telluric correction, we ignored the following intervals of the spectrum when computing the LSD profiles: [950, 979], [1081, 1169], [1328, 1492], [1784, 2029], [2380, 2500]~nm. These correspond to spectral windows where the telluric absorption is highest. The final number of spectral lines used in LSD is 588 and 830 for the $T_{\mathrm{eff}}=3000$\,K  and $T_{\mathrm{eff}}=3500$\,K  line lists, respectively. 

\begin{figure}[!ht]
    \centering
    \includegraphics[scale=0.41]{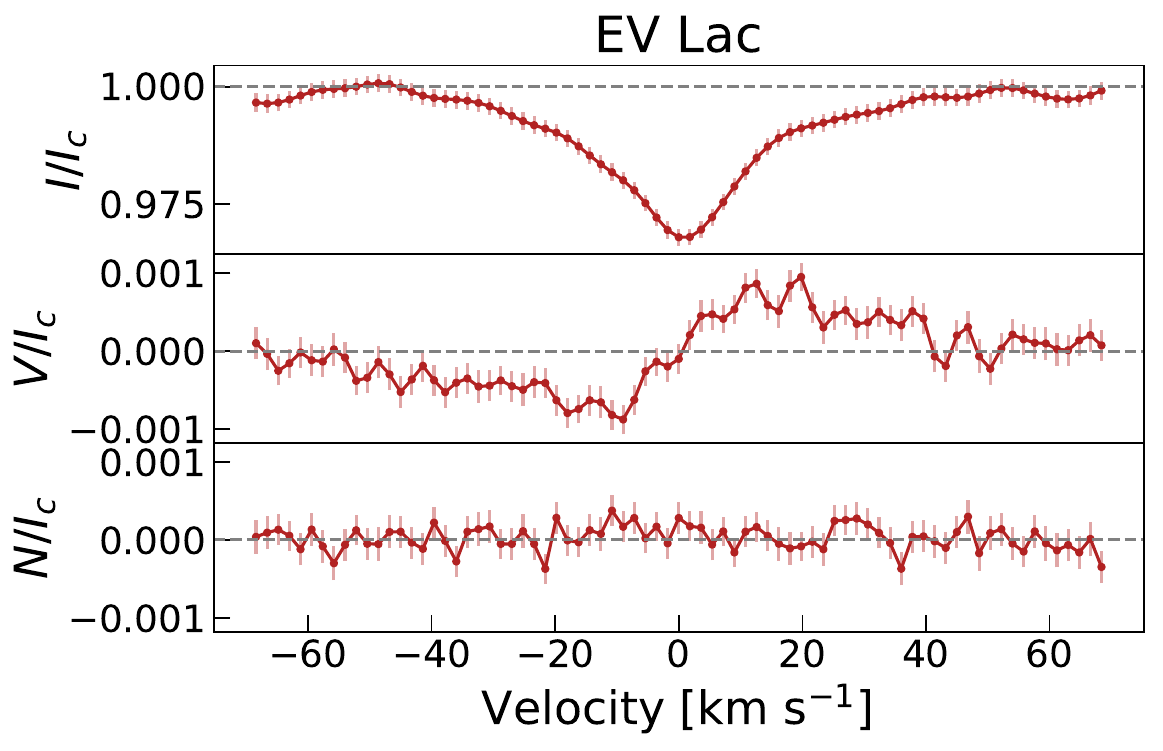}
    \includegraphics[scale=0.41]{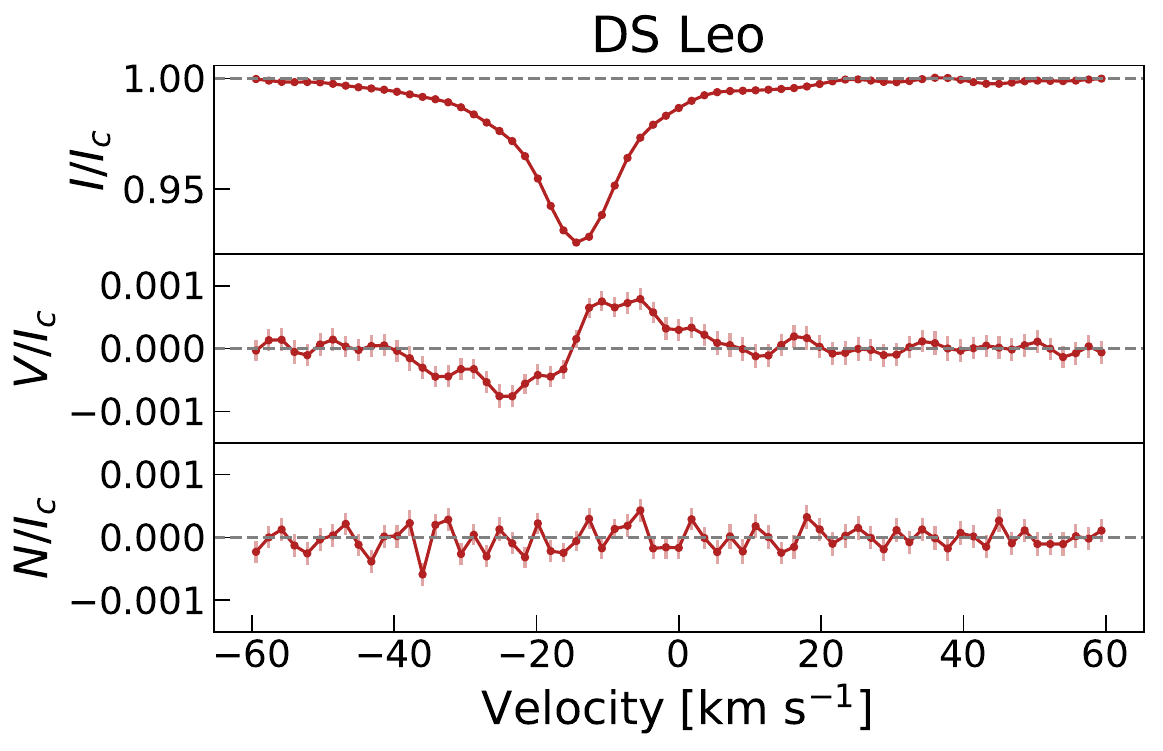}
    \includegraphics[scale=0.41]{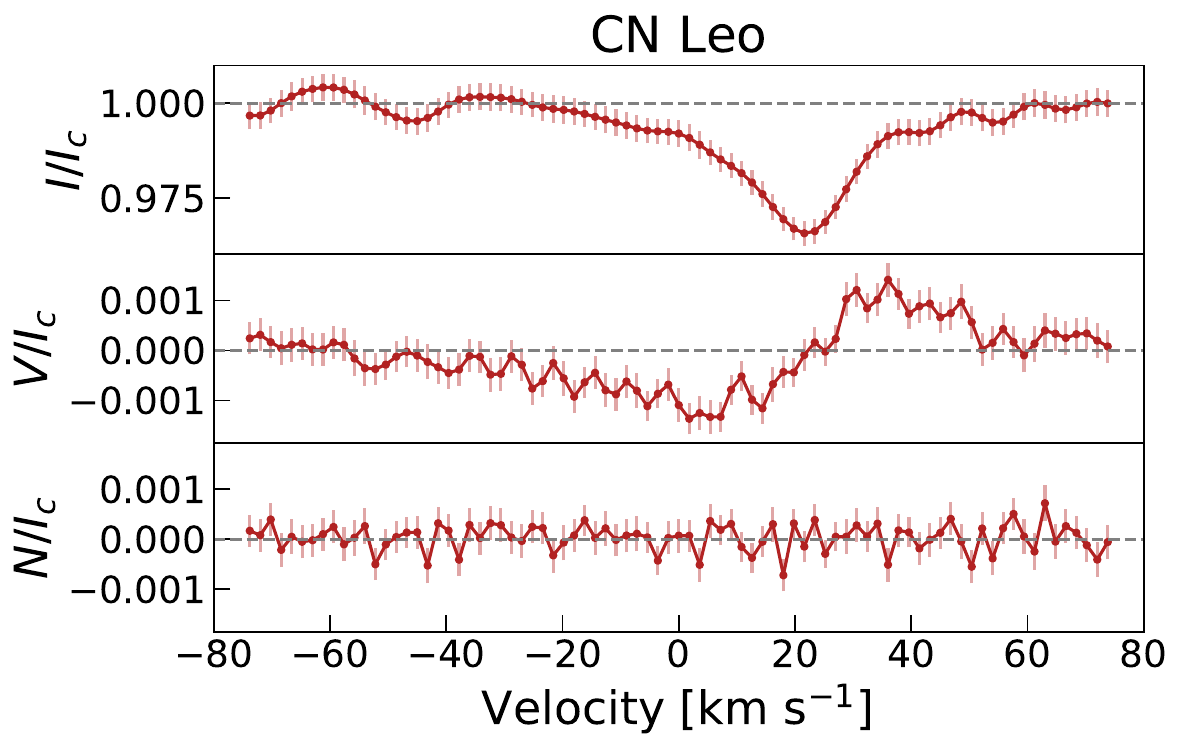}
    \caption{Examples of Stokes profiles obtained with SPIRou for EV~Lac, DS~Leo and CN~Leo. In each panel, Stokes $I$ (top), $V$ (middle), and null polarisation profile (bottom; see \citealt{Bagnulo2009}) are illustrated in units of unpolarised continuum ($I_c$). Clear circular polarisation signatures are present, with a S/N of 5\,000, 5\,600, and 3\,080 for EV~Lac, DS~Leo, and CN~Leo, respectively. The different magnetic activity level of the stars is shown by the different widths of the average line profiles in both unpolarised and polarised light.}
    \label{fig:Stokes}%
\end{figure}

We show an example of LSD Stokes profiles for the examined stars in Fig.~\ref{fig:Stokes}. As mentioned by \citet{Lavail2018}, the continuum of the Stokes~$I$ LSD profiles for M~dwarfs lies below unity owing to molecular spectral lines that are not accounted in the line lists. We thus re-normalised the profiles to unity by fitting a linear model to the region outside the line, to include residuals of continuum normalisation at the level of the spectra. The Stokes~$V$ and $N$ profiles were correspondingly re-scaled with the same fit. The operation of continuum normalisation does not alter the magnetic analyses presented here appreciably. For instance, the longitudinal field values computed with non-normalised LSD profiles are at most 5\% larger compared to the values obtained from normalised LSD profiles. In a similar manner, the retrieved field topology and evolution are robust against LSD profiles normalisation, as long as a multi-epoch time series for a particular target is analysed in a consistent way.

EV~Lac was observed for 163 nights between September 2019 and November 2021, spanning a period of $\sim$800 days. We split the entire time series in smaller epochs based on the instrumental scheduling and visibility gaps, following \citet{Bellotti2023b}. We obtained three epochs: 2019b (35 obs between 2019.86 and 2019.95), 2020b2021a (67 observations between 2020.57 and 2021.02) and 2021b (61 observations between 2021.47 and 2021.72). We recorded a S/N at 1650~nm per spectral element between 31 and 165, with a mean of 130, and an average airmass of 1.2.

DS~Leo was observed for 130 nights between November 2020 and June 2022, spanning a period of $\sim$580 days. The time series was split in two epochs: 2020b2021a (61 observations between 2020.27 and 2021.55) and 2021b2022a (69 observations between 2021.71 and 2022.46). The S/N ranged between 51 and 151, with a mean of 125, and the mean airmass is 1.3. 

CN~Leo was observed for 169 nights between April 2019 and June 2022, spanning a period of $\sim$1150 days. The time series was split in four epochs: 2019a (19 observations between 2019.29 and 2019.47), 2019b2020a (37 observations between 2019.83 and 2020.44) and 2020b2021a (63 observations between 2020.83 and 2021.50) 2021b2022a (45 observations between 2021.87 and 2022.44). The S/N ranged between 40 and 146, with a mean of 111, and the average airmass is 1.3. Six observations in February 2019 were excluded due to an optical component not working nominally at very low temperatures, in a similar way to \citet{Bellotti2023b}.

\subsection{Optical}

We used archival optical spectropolarimetric observations collected with ESPaDOnS and Narval. ESPaDOnS is the spectropolarimeter on the 3.6 m CFHT located atop Mauna Kea in Hawaii \citep{Donati2006esp}, and Narval is the twin instrument on the 2\,m T\'elescope Bernard Lyot (TBL) at the Pic du Midi Observatory in France \citep{Donati2003}. Data reduction was performed with \texttt{LIBRE-ESPRIT} \citep{Donati1997}, and the continuum normalised spectra were retrieved from PolarBase \citep{Petit2014}. 

The optical data sets were already examined in previous studies \citep{Donati2008,Morin2008,Morin2010,Hebrard2016}. For EV~Lac, there are 79 observations taken between 2005 and 2016. The 2010 observations were affected by a malfunction of the rhombs which affected the circular polarisation signal, hence only the Stokes~$I$ data are used in the following analyses. For DS~Leo, there are 94 observations between 2006 and 2014. For CN~Leo, there are four observations in 2008, which we included only for the FWHM analysis (see Sec.~\ref{sec:FWHM}).

Similarly to the near-infrared, we computed Stokes~$I$ and $V$ profiles with LSD. For EV~Lac and DS~Leo, we adopted an optical 3500\,K VALD mask containing 3240 lines in range 350-1080 nm and with depths larger than 40\% the continuum level, similarly to \citep{Morin2008,Bellotti2021}. For CN~Leo, we used a 3000\,K mask built in a similar manner and containing 3492 absorption lines. The number of lines in both masks already accounts for the removal of the wavelength intervals affected by telluric lines analogously to \citet{Bellotti2021}. These are: [627,632], [655.5,657], [686,697], [716,734], [759,770], [813,835], and [895,986] nm. In the next sections, the observations will be phased with the following ephemeris
\begin{align}
    \mathrm{HJD}_\mathrm{EVLac} = 2458738.0805 + \mathrm{P}_\mathrm{rot,EVLac}\cdot n_\mathrm{cyc}\\
    \mathrm{HJD}_\mathrm{DSLeo} = 2459158.1132 + \mathrm{P}_\mathrm{rot,DSLeo}\cdot n_\mathrm{cyc}\\
    \mathrm{HJD}_\mathrm{CNLeo} = 2458590.0095 + \mathrm{P}_\mathrm{rot,CNLeo}\cdot n_\mathrm{cyc}
    \label{eq:ephemeris}
\end{align}
where we separately used the first SPIRou observation for each star as heliocentric Julian Date reference, P$_\mathrm{rot,i}$ is the stellar rotation period of the $i$-th star (see Sec.~\ref{sec:Blon}), and $n_\mathrm{cyc}$ represents the rotation cycle (tabulated in Appendix~\ref{app:logs}).

\section{Longitudinal magnetic field}\label{sec:Blon}

The disk-integrated, line-of-sight component of the magnetic field (B$_l$) is computed from the first-order moment of a Stokes $V$ profile \citep{Rees1979,Folsom2016}. Formally, 
\begin{equation}
\mathrm{B}_l\;[G] = \frac{-2.14\cdot10^{11}}{\lambda_0 \mathrm{g}_{\mathrm{eff}}c}\frac{\int vV(v)dv}{\int(I_c-I)dv} \,,
\label{eq:Bl}
\end{equation}
where $\lambda_0$ and $\mathrm{g}_\mathrm{eff}$ are the normalisation wavelength (in nm) and Land\'e factor of the LSD profiles, $I_c$ is the continuum level, $v$ is the radial velocity associated to a point in the spectral line profile in the star's rest frame (in km\,s$^{-1}$) and $c$ the speed of light in vacuum (in km\,s$^{-1}$).

For all stars examined, the adopted normalisation wavelength is 1700\,nm and 700\,nm for the near-infrared and optical domains, respectively. The normalisation Land\'e factor is 1.191 in near-infrared and 1.154 in optical for EV~Lac, 1.181 and 1.086 for DS~Leo, and 1.223 for CN~Leo (only near-infrared). Considering the larger line width in near-infrared than optical, the integration is computed within $\pm$50\,km\ s$^{-1}$ and $\pm$ 30\,km\ s$^{-1}$ from line centre for EV\,Lac and within $\pm$60\,km\ s$^{-1}$ and $\pm$ 30\,km\ s$^{-1}$ for CN\,Leo. For DS\,Leo, the integration intervals are $\pm$25\,km\ s$^{-1}$ and $\pm$ 20\,km\ s$^{-1}$, since it is a slower rotator.

\begin{figure}[!t]
    \centering
    \includegraphics[width=\columnwidth]{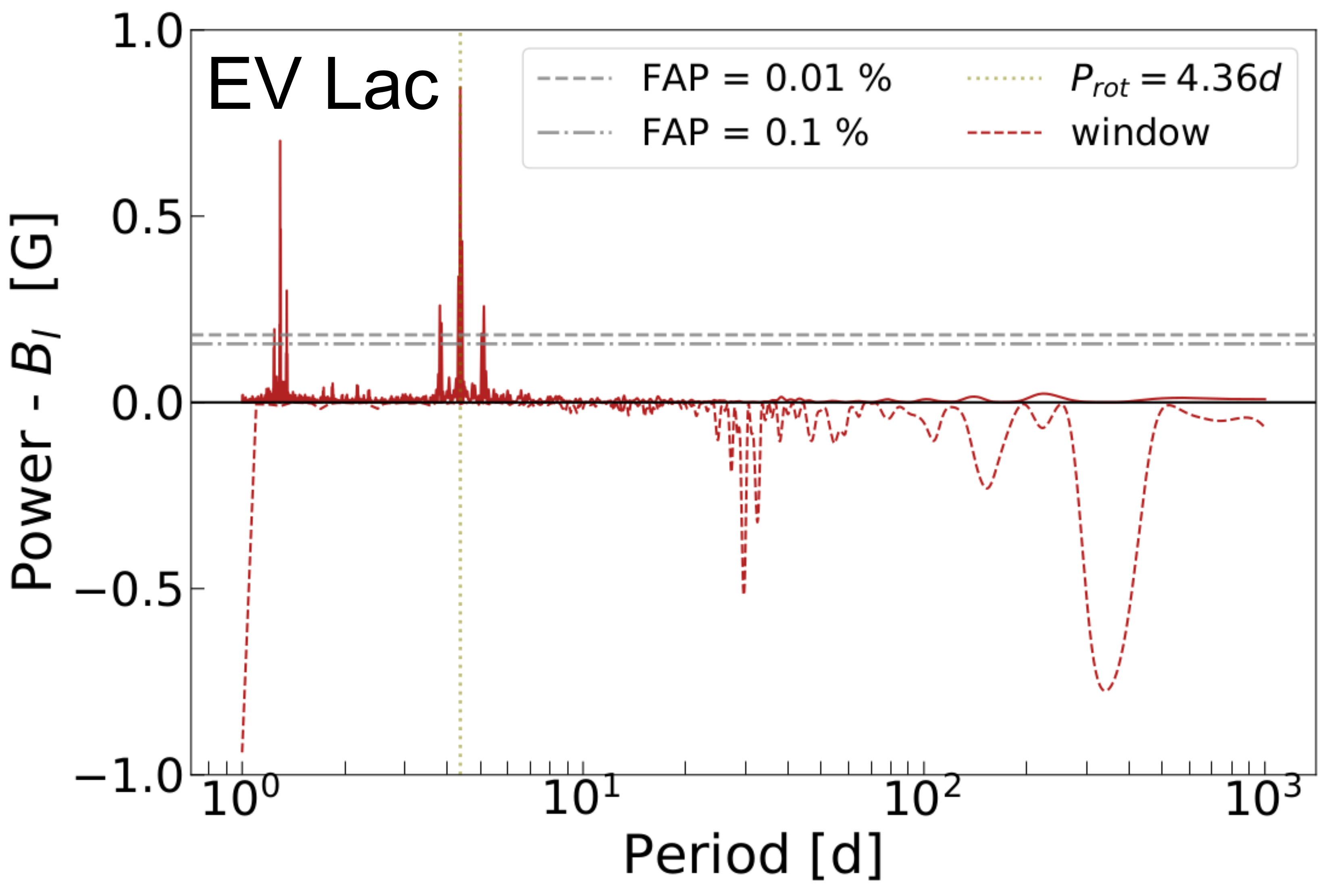}
    \includegraphics[width=\columnwidth]{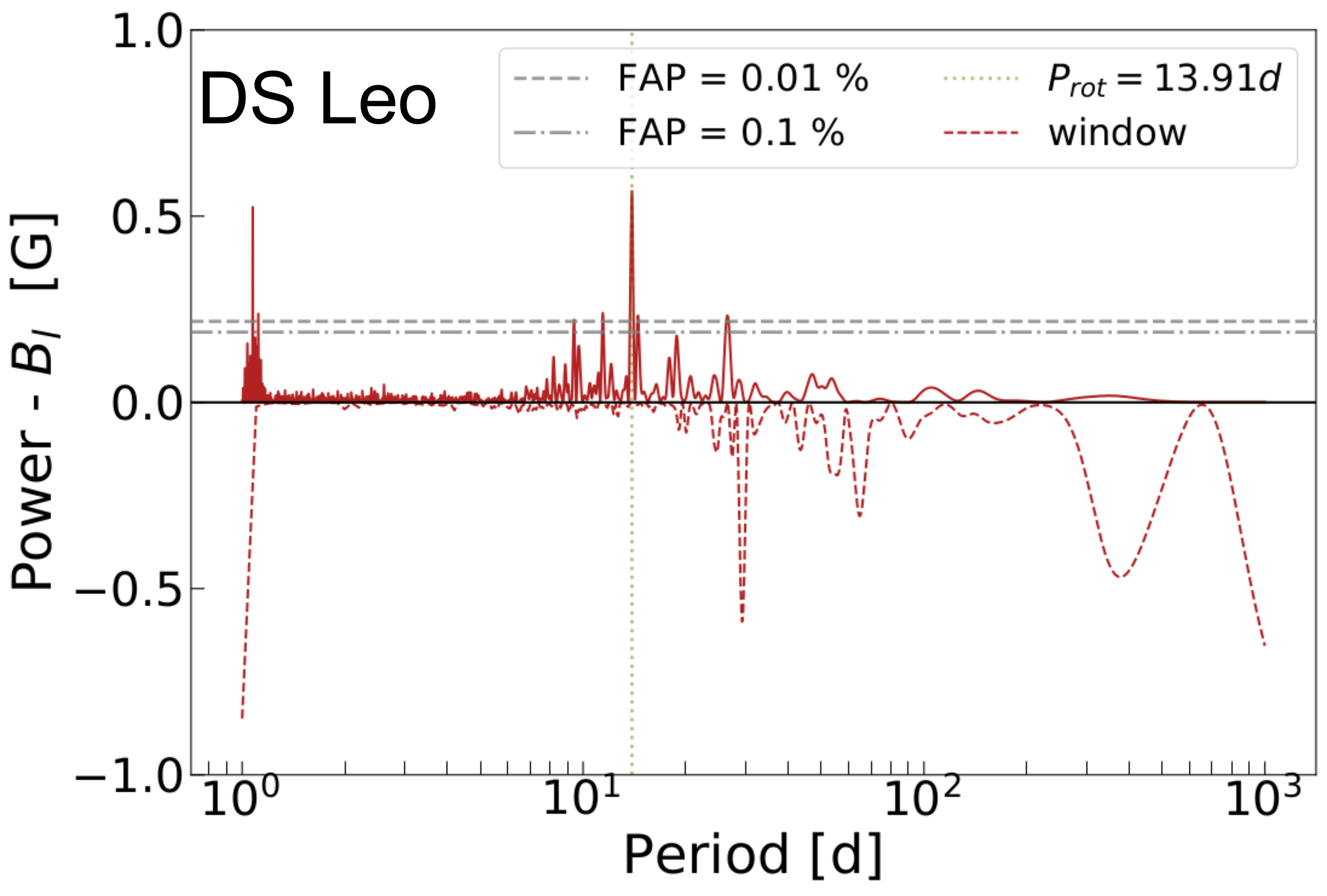}
    \includegraphics[width=\columnwidth]{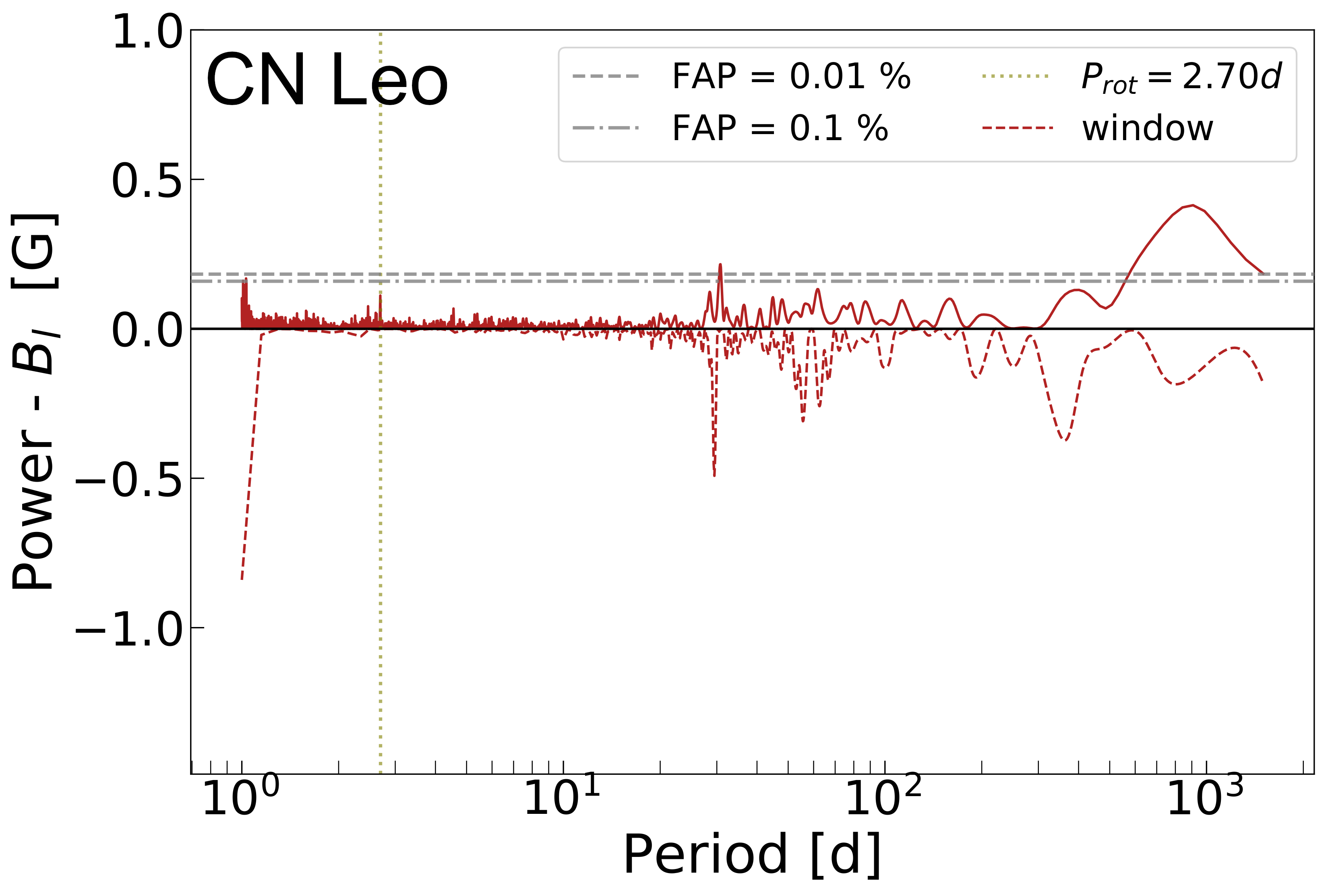}
    \caption{Generalised Lomb-Scargle periodogram of the longitudinal field near-infrared time series, from top to bottom EV\,Lac, DS\,Leo, and CN\,Leo. In each panel the rotation period corresponding to the highest peak is highlighted with a green dotted line, while two FAP levels (0.1\% and 0.01\%) are shown as grey horizontal lines. The window function of the entire time series is included and mirrored with respect to the $x$-axis \citep{VanderPlas2018} to highlight aliases due to the observation cadence and scheduling.}
    \label{fig:Bl_LSP}%
\end{figure}

The longitudinal field is a useful diagnostic of surface magnetic features, whose appearance on the visible disk is modulated at the stellar rotation period. For this reason, practical information can be extracted from the time series of B$_l$ values with standard techniques \citep{Folsom2018,Petit2021,Carmona2023,Fouque2023,Donati2023b}. We applied a Generalised Lomb-Scargle periodogram \citep{Zechmeister2009} to the entire time series as well as the individual subsets of data for all the examined stars. The results are illustrated in Fig.~\ref{fig:Bl_LSP}.

\begin{figure}[!t]
    \centering
    \includegraphics[width=\columnwidth]{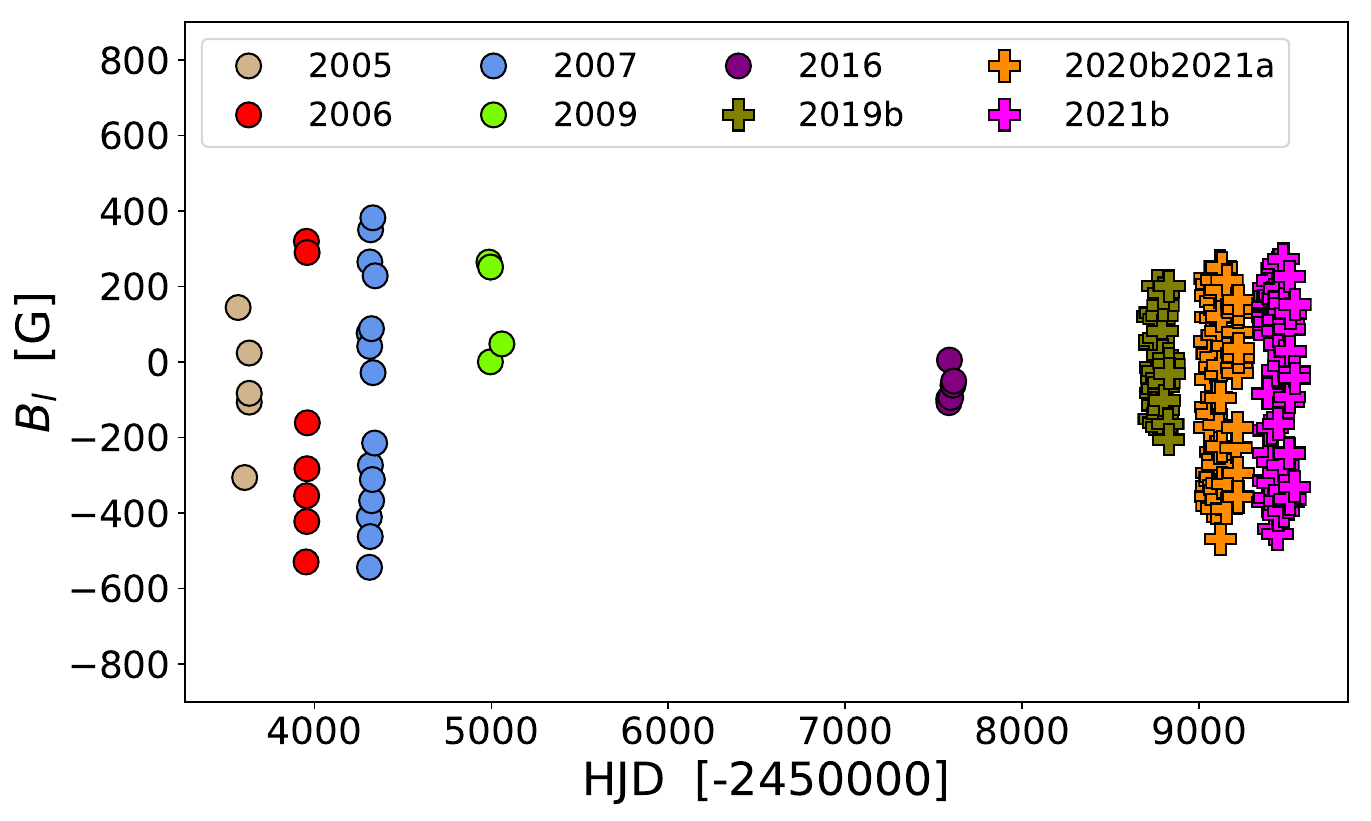}
    \includegraphics[width=\columnwidth]{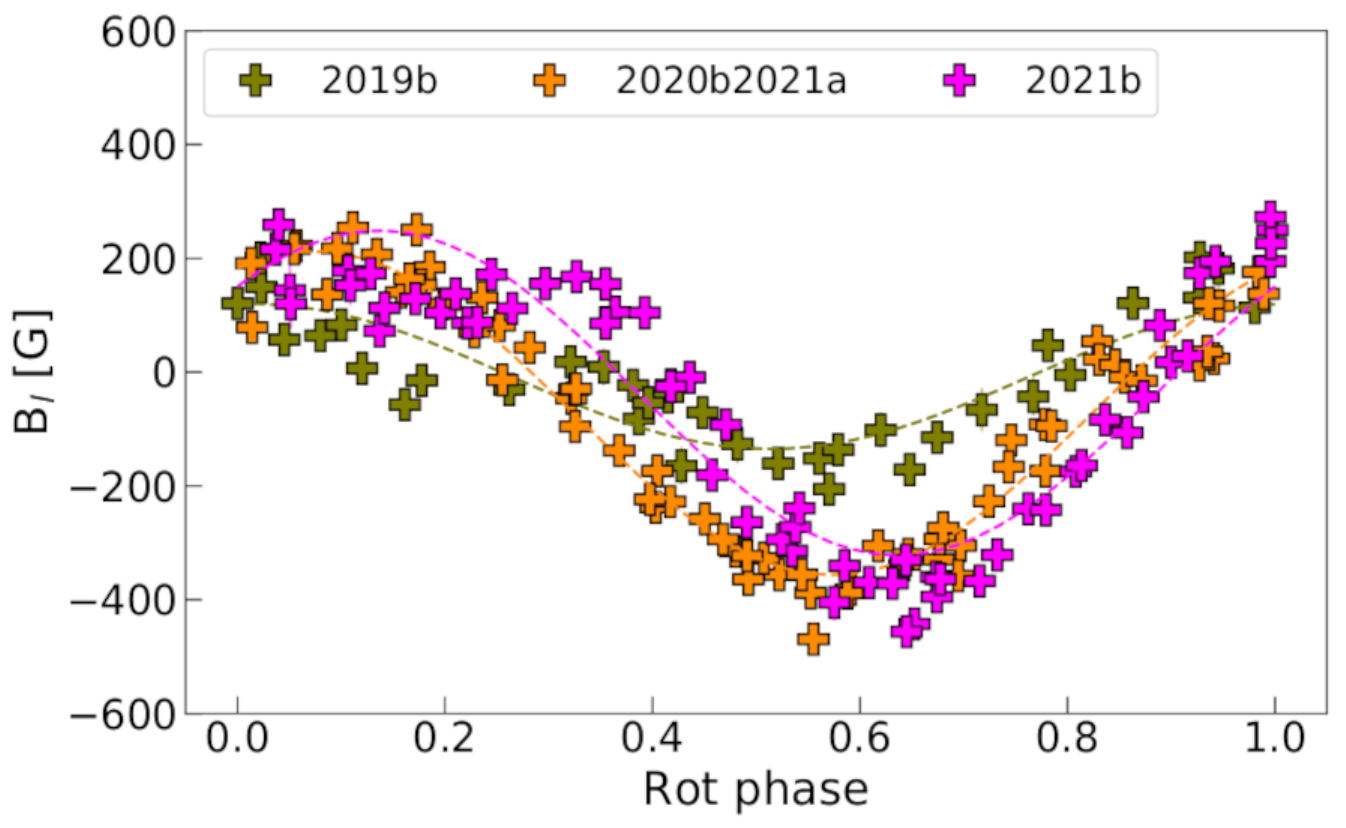}
    \caption{Temporal evolution of the longitudinal magnetic field for EV~Lac. The shape of the data points distinguishes optical (circles) from near-infrared (squares), and the colour represents the epoch in which the data were collected. Top: Full time series of measurements with ESPaDOnS, NARVAL, and SPIRou. Bottom: Phase-folded curves of SPIRou data points colour-coded by epoch;  the associated least-squares sine fits are shown as dashed lines. The rotation period used is listed in Table~\ref{tab:star_properties}.}
    \label{fig:Bl_evol_evlac}%
\end{figure}

For EV\,Lac and DS\,Leo, the peak at the expected rotation period is unambiguously higher than a false alarm probability (FAP) of 0.01\%, which emphasises the dominant stellar activity signal and confirms B$_l$ as suitable diagnostic. For CN\,Leo, the interpretation of the periodogram is more elaborate because of strong aliases of the observational cadence ($\sim$1 d), the gap between instrument runs ($\sim$30 d), and the time span of the time series ($\sim$1000 d). By ignoring these features, the periodogram shows a clear and unique peak at the expected stellar rotation period, for which there is no correspondence with the observing window function. If we restrict the periodogram analysis to each the four subsets of CN~Leo (see Sec~\ref{sec:observations}), only the last one 2021b2022a does not show a spike at the expected rotation period. For all stars, we retrieve P$_\mathrm{rot}$ in agreement to known values from the literature: P$_\mathrm{rot, EVLac}=$ 4.36$\pm$0.01\,d \citep{Morin2008}, P$_\mathrm{rot, DSLeo}=$ 13.91$\pm$0.01\,d \citep{Hebrard2016}, and P$_\mathrm{rot, CNLeo}$ 2.70$\pm$0.01\,d \citep{Reinhold2020,Lafarga2021,Irving2023}.

\begin{figure}[!t]
    \centering
    \includegraphics[width=\columnwidth]{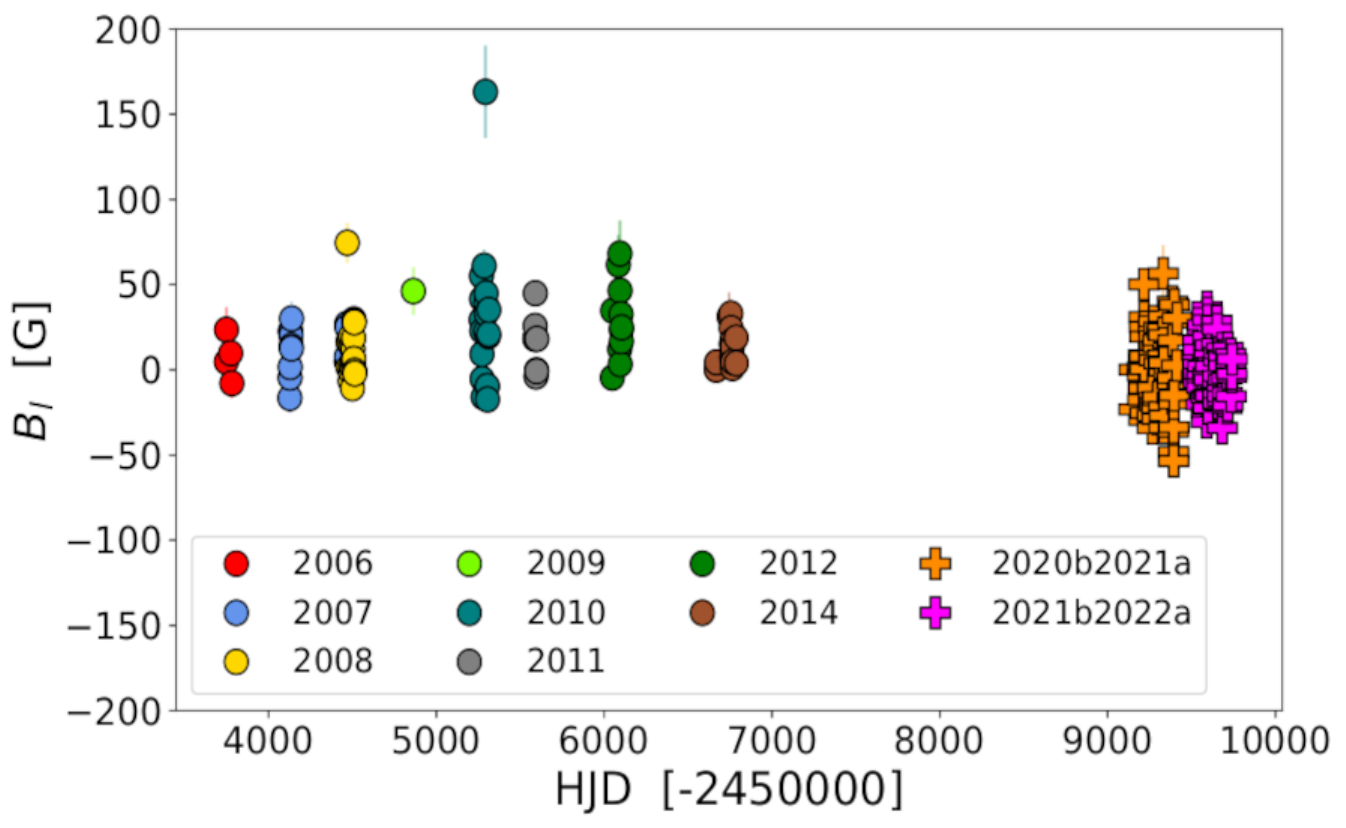}
    \includegraphics[width=\columnwidth]{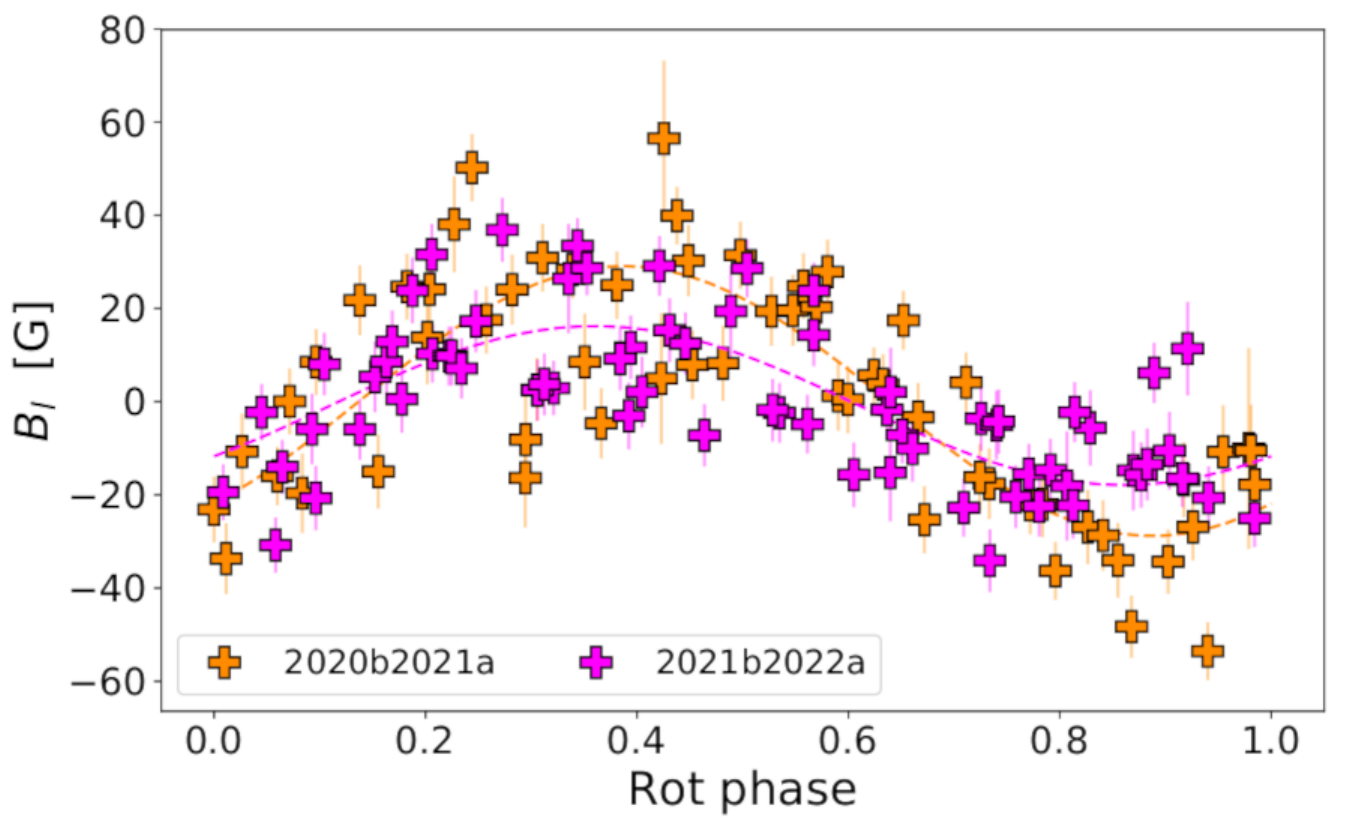}
    \caption{Temporal evolution of the longitudinal magnetic field for DS~Leo. The format is the same as in Fig.~\ref{fig:Bl_evol_evlac}.}
    \label{fig:Bl_evol_dsleo}%
\end{figure}

The longitudinal field computation was carried out for both optical and near-infrared domains for all stars, and is shown in Figs.~\ref{fig:Bl_evol_evlac}~to~\ref{fig:Bl_evol_cnleo}. The journal of B$_l$ values is reported in Appendix~\ref{app:logs}.

\subsection{EV~Lac}

EV~Lac's near-infrared measurements lie between 270 and $-$470\,G, with a mean of $-$50\,G and a median error bar of 19\,G. In comparison, the optical data have a larger spread, ranging between 380 and $-$540\,G with a mean of $-$90\,G and a median error bar of 16\,G. When a 3000\,K, near-infrared mask is used to compute the LSD profiles, the range of B$_l$ spans between $-$480 and 286\,G, with a mean of $-52$\,G and a median error bar of 21\,G. The Stokes~$I$ profiles derived with the colder mask are on average 20\% deeper than those associated with the 3500\,K mask, and the amplitude of Stokes~$V$ profiles is on average $\sim$20\% larger. These differences in the Stokes profiles cancel out according to Eq.~\ref{eq:Bl}, ultimately making the results consistent. Overall, the near-infrared B$_l$ values seem to progressively restore the amplitude with time, after a potential minimum in 2016, making the overall trend look like a beating signal. However, the dearth of data between 2010 and 2019 prevents us from constraining the phenomenon further.

When EV~Lac's near-infrared data are phase-folded at P$_\mathrm{rot, EVLac}=4.36$\,d, we note a clear rotational modulation for the three epochs (see Fig. \ref{fig:Bl_evol_evlac}). The phase variations are modelled with a Levenberg-Marquardt least squares sine fit, and we observe an increasing amplitude of the field variations, indicating a decrease in the field axisymmetry or a field enhancement over time. A detail investigation will be performed with Zeeman-Doppler imaging in Sec.~\ref{sec:magnetic_imaging}.

\subsection{DS~Leo}

DS~Leo is the least magnetically active star of our sample. The longitudinal field near-infrared measurements fall within $\pm$56\,G with a median error bar of 7\,G, whereas the optical ones range between $-$25 and 80\,G, with a median error of 7\,G (excluding a visible outlier at more than 150\,G). The dispersion of the near-infrared data is comparable to the optical (20\,G against 24\,G) and there is a shift of the near-infrared mean towards zero by 16\,G relative to the optical one. The phase-folded curves of the two epochs (2020b2021a and 2021b2022a) do not reveal particular changes, as the variations are both characterised by sinusoidal trends and similar amplitudes (see Fig.~\ref{fig:Bl_evol_dsleo}). The results do not change appreciably when a 4000\,K synthetic mask is used against the 3500\,K. There is a $\sim$40\% difference in Stokes~$I$ depth and amplitude of Stokes~$V$ between the two masks, which again cancels out in Eq.~\ref{eq:Bl}. The range of B$_l$ values with the hotter mask is between $-50$ and 60\,G with a median error bar of 8\,G.

\subsection{CN~Leo}

\begin{figure}[t]
    \centering
    \includegraphics[width=\columnwidth]{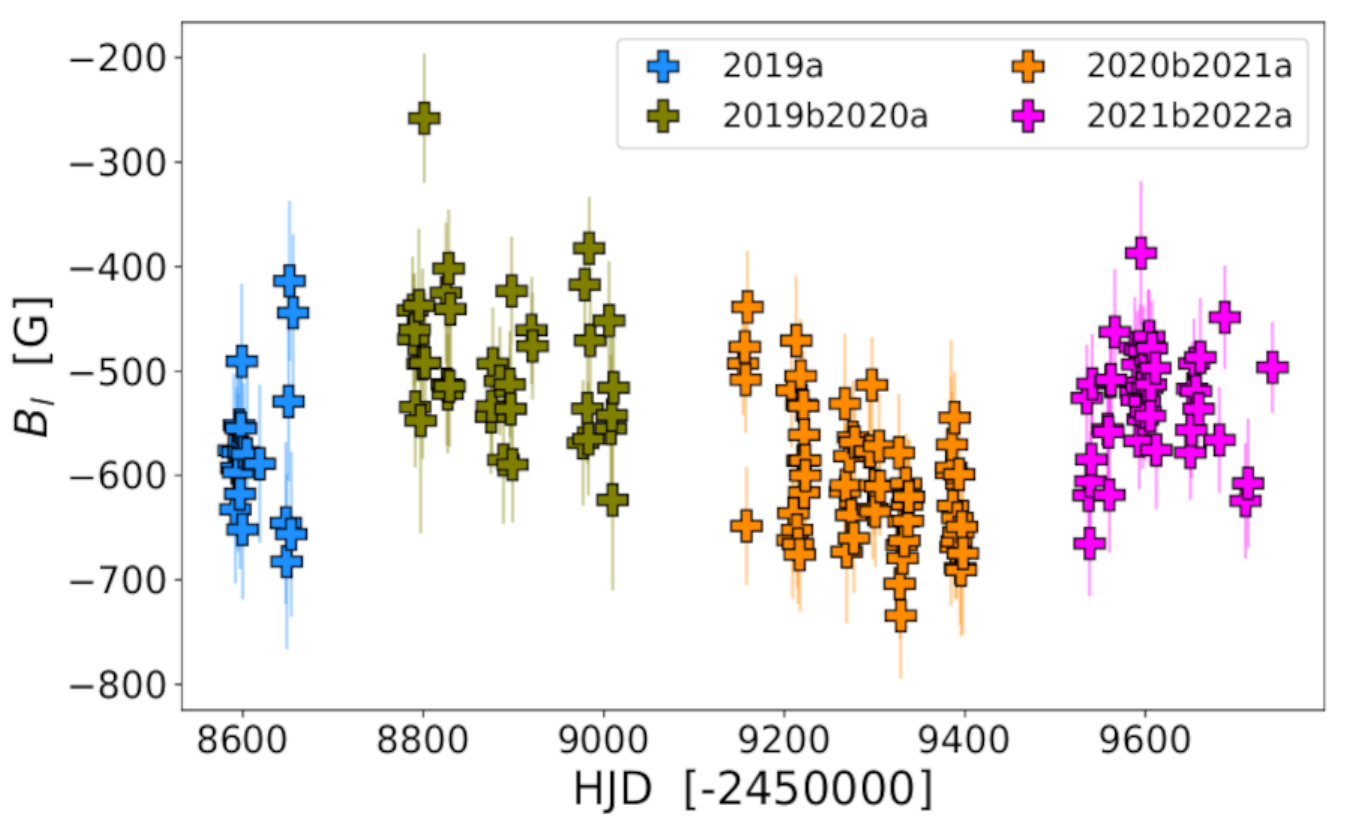}
    \includegraphics[width=\columnwidth]{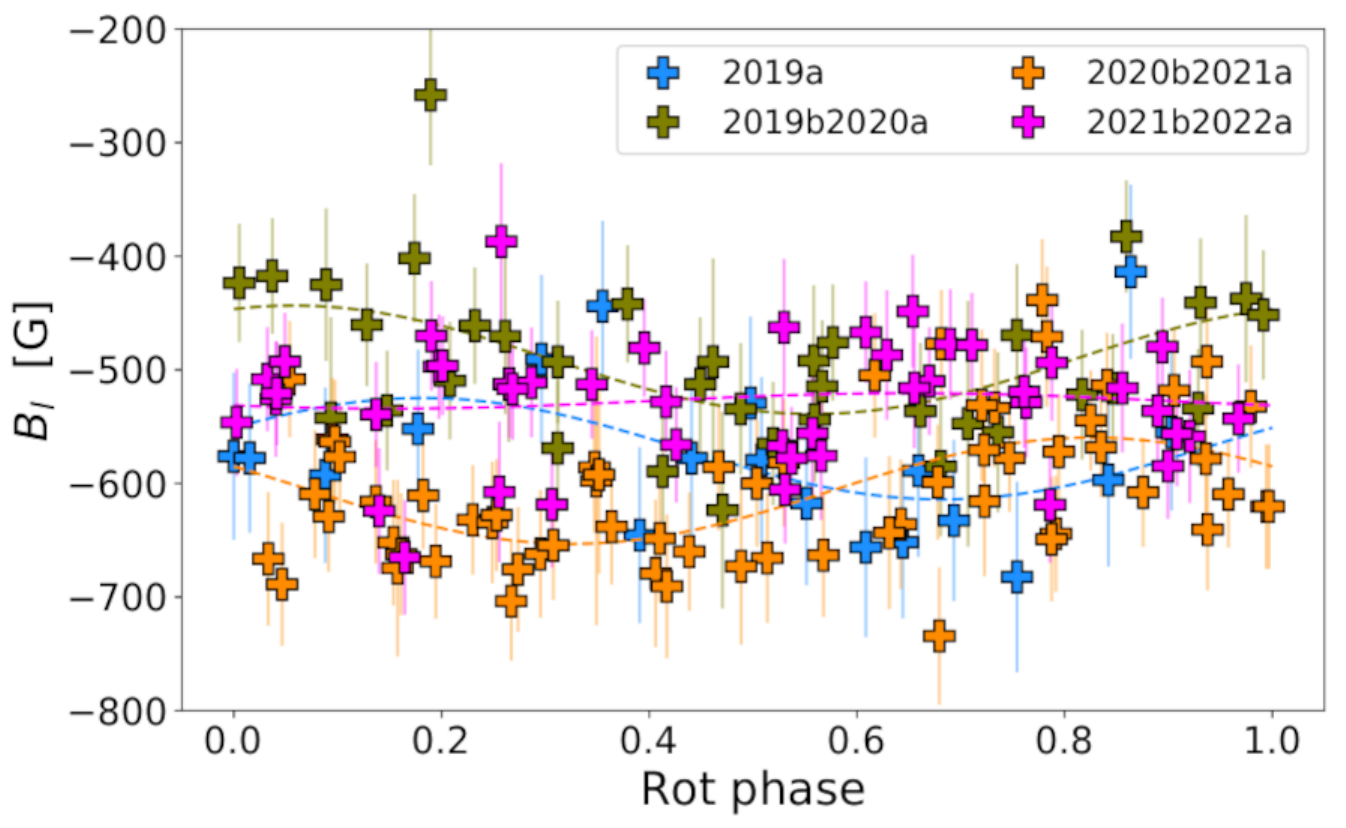}
    \caption{Temporal evolution of the longitudinal magnetic field for CN~Leo. The format is the same as in Fig.~\ref{fig:Bl_evol_evlac}.}
    \label{fig:Bl_evol_cnleo}%
\end{figure}

For CN~Leo, the B$_l$ measurements range from $-$630 to $-$240\,G with a median error bar of 40\,G, implying that we observe only the negative polarity of the large-scale field, similar to AD~Leo \citep{Bellotti2023b}. These are compatible with the average B$_l$ of $-$691$\pm$\,54\,G reported in \citet{Morin2010}. The field values exhibit a sine-like oscillation of about 1000\,d as well as fluctuations within each epoch. Although the scatter of the global and epoch-by-epoch time series is lower or compatible with the respective median error bar, which would make the oscillations dubious, the time series is dense, hence we are more sensitive to the epoch-to-epoch variability (see Fig.~\ref{fig:Bl_evol_cnleo}). If we bin the full time series with 50-d intervals, we are still able to capture a sinusoidal variation of 1000\,d. When we phase-fold the four epochs at the stellar rotation period, we do not note any evident rotational modulation (the amplitude of the variations is consistent with zero within uncertainties), so we expect the field topology to be a dipole with negative polarity, and with the magnetic axis almost aligned with the rotation axis of the star.

\section{FWHM of Stokes $I$}\label{sec:FWHM}

Since the Zeeman effect is proportional to Land\'e factor, modulus of the magnetic field ($B$), and the square of the spectral line wavelength, the width of Stokes~$I$ profiles computed in the near-infrared domain for stars with intense magnetic fields can be used as a proxy for Zeeman broadening measurements to first order \citep{Donati2023,Bellotti2023b}. The value $B$ encapsulates the magnetic field at both small and large scales, while the longitudinal field probes the large-scales only. Thus, the temporal analysis of the FWHM of Stokes~$I$ and its correlation to $B_l$ can shed more light on the link between the two distinct spatial scales. In addition, the FWHM may exhibit a long-term evolution that correlates with the large-scale field evolution, as shown for AD~Leo by \citet{Bellotti2023b}.

The FWHM analysis can inform the design of activity-filtering techniques for radial velocity searches of exoplanets.  \citet{Haywood2022} show that, for the Sun, the rotationally-modulated line broadening correlates with the azimuthal distribution of the small-scale magnetic flux, and \citet{Lienhard2023} report the efficient activity-filtering in solar radial velocity time series using a novel proxy based on the total unsigned magnetic flux. For AU\,Mic \citet{Klein2021} report the correlation between the radial velocity and FWHM variations, when the latter were computed with magnetically sensitive lines. In Appendix~\ref{app:fwhm_app}, we describe first-order computations of the unsigned magnetic field using the FWHM as a proxy.

Following \citet{Bellotti2023b}, we applied LSD using magnetically sensitive (g$_\mathrm{eff}>1.2$) and weakly-sensitive (g$_\mathrm{eff}<1.2$) line masks. In the near-infrared domain, the masks contain 420 and 400 lines for the 3500\,K case (EV~Lac and DS~Leo), respectively, and for the 3000\,K case (CN~Leo) they contain 280 and 300 lines. In optical, the masks contain 1650 and 1500 lines for the 3500\,K case. In the following, the reference data set is the one obtained using the full mask, for either stellar temperature or wavelength domain. To compute the FWHM, we modelled the Stokes~$I$ LSD profiles with a combination of a Voigt and a linear model, the latter to account for residuals of continuum normalisation.

\begin{table*}[!ht]
\caption{Comparison of a constant line against a sine fit for the FWHM phase variations of EV~Lac, DS~Leo, and CN~Leo} 
\label{tab:fwhm}     
\centering                       
\begin{tabular}{l l c c c | c c | c c}       
\hline     
Epoch & Mask & Mean & Mean Error & STD & RMS$_\mathrm{const}$ & $\chi^2_{r,\mathrm{const}}$ & RMS$_\mathrm{sine}$ & $\chi^2_{r,\mathrm{sine}}$\\
& & [km s$^{-1}$] & [km s$^{-1}$] & [km s$^{-1}$] & [km s$^{-1}$] & & [km s$^{-1}$] &\\
\hline
\multicolumn{9}{c}{EV~Lac}\\
\hline
2019b      & default         & 22 & 0.65 & 1.68 & 1.68 & 6.9 & 1.45 & 5.9 \\
           & g$_\mathrm{eff}>1.2$ & 29 & 1.12 & 4.39 & 4.39 & 19.7 & 4.17 & 19.7\\
2020b2021a      & default         & 21 & 0.44 & 1.00 & 1.00 & 6.2 & 0.82 & 4.4 \\
           & g$_\mathrm{eff}>1.2$ & 28 & 0.87 & 2.62 & 2.62 & 12.9 & 2.45 & 10.7\\
2021b      & default         & 21 & 0.46 & 0.98 & 0.98 & 4.8 & 0.90 & 4.5 \\
           & g$_\mathrm{eff}>1.2$ & 28 & 0.74 & 2.57 & 2.57 & 13.0 & 2.44 & 16.1\\
\hline
\multicolumn{9}{c}{DS~Leo}\\
\hline
2020b2021a      & default         & 13 & 0.18 & 0.29 & 0.29 & 2.8 & 0.16 & 0.9 \\
           & g$_\mathrm{eff}>1.2$ & 14 & 0.21 & 0.47 & 0.47 & 6.2 & 0.33 & 2.9\\
2021b2022a      & default         & 13 & 0.18 & 0.21 & 0.21 & 1.5 & 0.20 & 1.3 \\
           & g$_\mathrm{eff}>1.2$ & 14 & 0.20 & 0.32 & 0.32 & 3.1 & 0.30 & 2.7\\
\hline
\multicolumn{9}{c}{CN~Leo}\\
\hline
2019a      & default         & 15 & 0.34 & 1.27 & 1.27 & 16.7 & 1.22 & 19.7 \\
           & g$_\mathrm{eff}>1.2$ & 21 & 1.19 & 3.73 & 3.73 & 19.8 & 3.44 & 22.5\\
2019b2020a      & default         & 17 & 0.36 & 1.41 & 1.41 & 19.1 & 1.35 & 16.6 \\
           & g$_\mathrm{eff}>1.2$ & 25 & 1.57 & 4.44 & 4.44 & 15.3 & 4.27 & 13.6\\
2020b2021a      & default         & 16 & 0.37 & 1.69 & 1.69 & 45.8 & 1.61 & 39.8 \\
           & g$_\mathrm{eff}>1.2$ & 22 & 1.51 & 5.08 & 5.08 & 26.8 & 4.74 & 23.1\\
2021b2022a      & default         & 17 & 0.34 & 1.37 & 1.37 & 25.7 & 1.29 & 24.7 \\
           & g$_\mathrm{eff}>1.2$ & 22 & 1.41 & 4.94 & 4.94 & 16.3 & 4.65 & 15.8\\
\hline  
\end{tabular}
\tablefoot{The columns are: 1) subset of the time series, 2) line list used in LSD computation, 3) mean FWHM, 4) mean error bar, 5) standard deviation of the data sets, 6) RMS (root mean square) residual of a constant line fit equal to the average of the data set, 7) reduced $\chi^2$ of a constant model, 8) RMS residual of a sine fit at the stellar rotation period, and 9) reduced $\chi^2$ of a sine model.}
\end{table*}
 
\subsection{Rotational modulation and short-term variability}

We used Eq.\ref{eq:ephemeris} to phase-fold the FWHM time series at each epoch and for both the full and high-g$_\mathrm{eff}$ masks. With the latter mask, \citet{Klein2021} show enhanced rotationally-modulated variations for AU\,Mic, whereas \citet{Bellotti2023b} do not report a modulation for near-infrared SPIRou observations of AD~Leo, given the large dispersion of the data set. We proceeded in a similar manner to \citet{Bellotti2023b}, and we fitted the observed FWHM values with a constant and a sine model at the stellar rotation period. The analysis is summarised in Table~\ref{tab:fwhm} and illustrated in Fig.~\ref{fig:FWHM_rotmod_dsleo} for DS~Leo and in Appendix~\ref{app:fwhm_app} for EV~Lac and CN~Leo.

For EV Lac, the FWHM of Stokes~$I$ profiles from near-infrared observations does not exhibit evident rotational modulation (see Fig.~\ref{fig:FWHM_rotmod_app_evlac}), since the change in reduced $\chi^2$ between a constant and a sine model is only marginal. In some cases, like 2021b with high-g$_\mathrm{eff}$ lines, the $\chi^2_r$ increases when using a sine model, but it is not statistically significant \citep{Press1992}. Only in 2020b2021a we observe a $\chi^2_r$ improvement of about two when using the sine model for either of the chosen masks, and we correspondingly visualise a hint of rotational modulation, especially for the full mask. If we colour-code the high-g$_\mathrm{eff}$ data by rotational cycle, we do not observe any specific behaviour, as the dispersion over different cycles (and within an individual epoch) is comparable. 

For the optical observations, the FWHM of EV~Lac's Stokes~$I$ profiles manifests clear rotational modulation in 2006, 2007, and 2010 regardless of the mask employed (see Fig.~\ref{fig:FWHM_rotmod_app_evlac}). A plausible explanation for observing rotational modulation in optical but not in near-infrared could be given by the larger contribution of the Zeeman effect in the latter domain. { The distortions induced by the Zeeman effect on the line profile shape cannot be modelled effectively by a simple Voigt kernel, and act as a dispersive factor of the FWHM measurement. For this reason, the FWHM may lose its diagnostic power in the near-infrared for strong magnetic fields.}

For DS~Leo, in the SPIRou 2020b2021a epoch, the phase-folded data points vary with the stellar rotation, quantified by a decrease in $\chi^2$ from 2.8 to 0.9 (for the default mask) and from 6.2 to 2.9 (for the high-g$_\mathrm{eff}$ mask) when using a sine model rather than a constant (see Fig.~\ref{fig:FWHM_rotmod_dsleo}). The sine model for the default mask fits the data approximately down to noise level. In the SPIRou 2021b2022a epoch, there is no clear rotational modulation instead and the variations can be equivalently explained by a constant model. If present, the modulation is enhanced or quenched when high-g$_\mathrm{eff}$ or low-g$_\mathrm{eff}$ lines are adopted. Colour-coding the high-g$_\mathrm{eff}$ by cycle number reveals short-term variability, as distinct modulations for different rotational cycles can be extracted. This is more evident for 2020b2021a, as the variations of the data points appear `stacked', that is, FWHM values belonging to different rotational cycles exhibit a vertical offset. This feature could be explained by the presence of differential rotation, since the shear would have displaced magnetic regions on the surface during the time span of our observations, making the cycle-averaged FWHM oscillate. The quantification of differential rotation will be investigated in Sec.~\ref{sec:magnetic_imaging}. 

For optical observations of DS~Leo, we were able to capture the rotational modulation of the FWHM in most epochs (Fig.~\ref{fig:FWHM_rotmod_dsleo}), especially when using magnetically sensitive lines. This supports our weak-field considerations, because the contribution of Zeeman effect to the line shape is less important for DS~Leo both in near-infrared and in optical, compared to EV~Lac and CN~Leo. 

For CN~Leo, the FWHM of near-infrared profiles does not manifest rotational modulation, similar to EV~Lac (see Fig.~\ref{fig:FWHM_rotmod_app_cnleo}), as the $\chi^2_r$ associated with a constant and a sine model is marginally different. Colour-coding the data points by rotational cycle does not reveal any specific feature that could be associated to short-term variability. We observe an increased dispersion at phases larger than 0.5 for almost all epochs when using high-g$_\mathrm{eff}$ lines, which could bias the measurement of the average FWHM, and which we attributed to the magnetic field, since low-g$_\mathrm{eff}$ lines are approximately constant at all phases. For the optical, the limited number of observations (four) prevents us from performing a rotational modulation analysis.

\subsection{Correlation between FWHM and B$_l$}

\begin{figure}[!t]
    \centering
    \includegraphics[width=\columnwidth, trim={0 279.85 0 0}, clip]{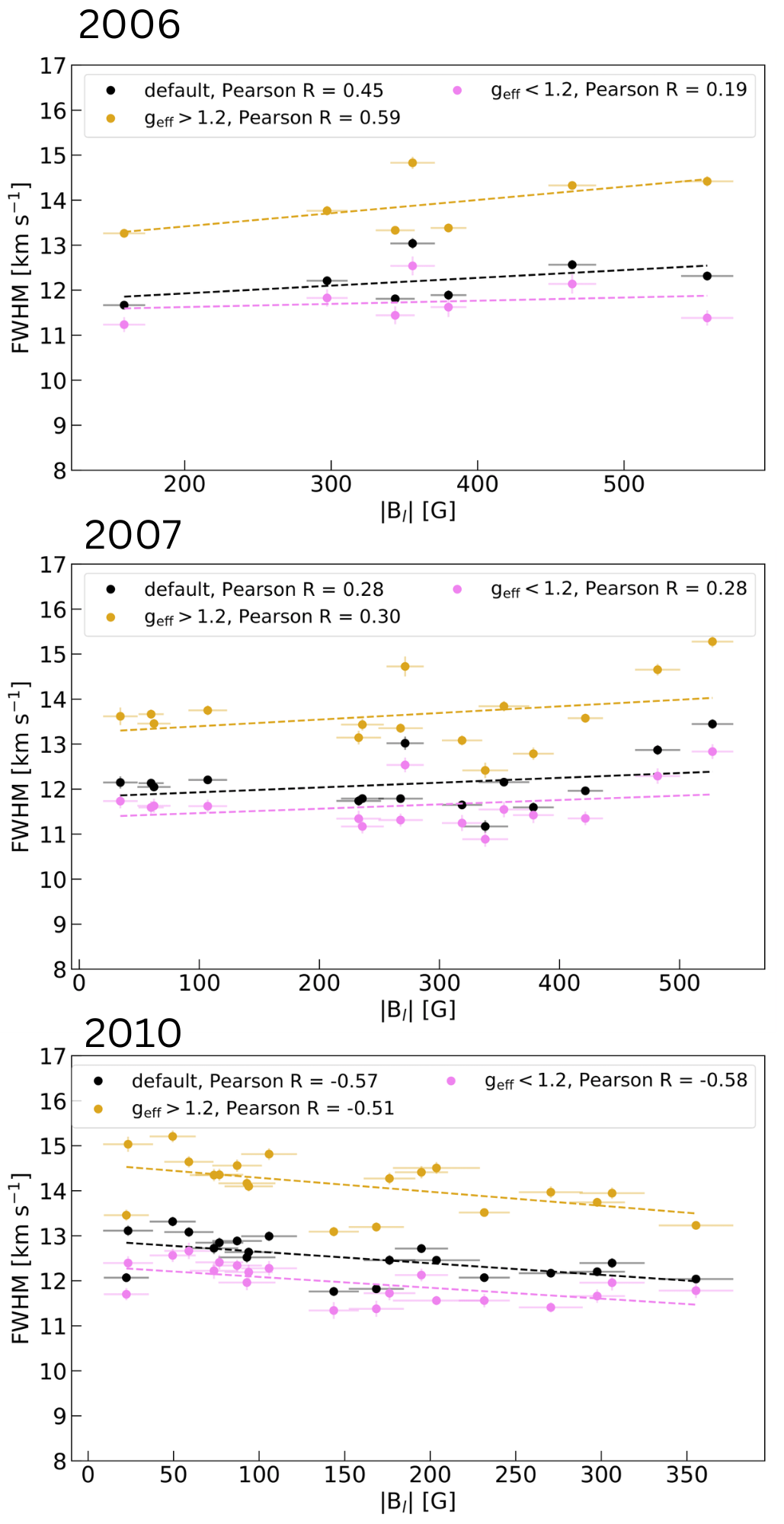}
    \caption{Correlation analysis between FWHM and $|$B$_l|$ for EV~Lac. The three panels correspond to the optical epochs in which rotational modulation is present: 2006, 2007, and 2010. In all panels the data points are colour-coded based on the line mask used for LSD: full (black), low-g$_\mathrm{eff}$ (purple), and high-g$_\mathrm{eff}$ (yellow).}
    \label{fig:FWHM_correl_EVLac}%
\end{figure}

The longitudinal magnetic field is a proxy for large-scale magnetic field, and the FWHM is a proxy for the small-scale field, so we inspected their correlations to yield more insights on the link between these two spatial scales. In particular, we computed Pearson correlation coefficients via 5,000 bootstrap iterations. The values summarised in Table~\ref{tab:fwhm_pearson} correspond to the mean and standard deviation of the coefficient distribution. The analysis was carried out for the FWHM computed with the three masks: full, low-g$_\mathrm{eff}$, and high-g$_\mathrm{eff}$. The analysis is illustrated in Fig.~\ref{fig:FWHM_correl_EVLac} for optical observations of EV~Lac and Fig.~\ref{fig:FWHM_correl_DSLeo} for optical and near-infrared observations of DS~Leo. CN~Leo was excluded because the FWHM from near-infrared lines is not modulated by rotation.

For EV~Lac, there is a positive correlation for 2006 data, which increases when going from low-g$_\mathrm{eff}$ lines (Pearson $R=0.19$) to high-g$_\mathrm{eff}$ lines (Pearson $R=0.59$). This is the expected behaviour from the principle of Zeeman effect: the line profile is broader when the field is stronger. A positive correlation is also seen in 2007, but its strength does not vary according to the line mask adopted. 

For DS~Leo, we observe a variety of relations: there is a positive correlation that increases going from low-g$_\mathrm{eff}$ to high-g$_\mathrm{eff}$ lines in 2007 and 2011, whereas in 2012 and 2014, there is an anti-correlation that becomes stronger for low-g$_\mathrm{eff}$ lines. For 2008 and 2010, there is one outlying data point at large B$_l$ values that likely biases the correlation coefficient, as the bulk of the data points does not manifest particular correlations. When looking at the SPIRou 2020b2021a epoch, we do not observe significant correlation, the Pearson $R$ coefficient being smaller than 0.2 for the three masks. 

While not straightforward to interpret, these results altogether suggest an underlying complexity in the link between small- and large-scale field, which may not be encapsulated in a simple scaling. The variety of correlations for different epochs for DS~Leo may stem from the fact that, for early M~dwarfs, the ratio of magnetic field recovered with polarised light relative to unpolarised light is only a few percent (and it increases to dozens of percents for mid-M, \citealt{ReinersBasri2009,Morin2010,Kochukhov2021}), hence there is an even less obvious scaling between small- and large-scale fields. 

\subsection{Long-term variations of FWHM}\label{sec:longfwhm}

\begin{figure}[!t]
    \centering
    \includegraphics[width=\columnwidth]{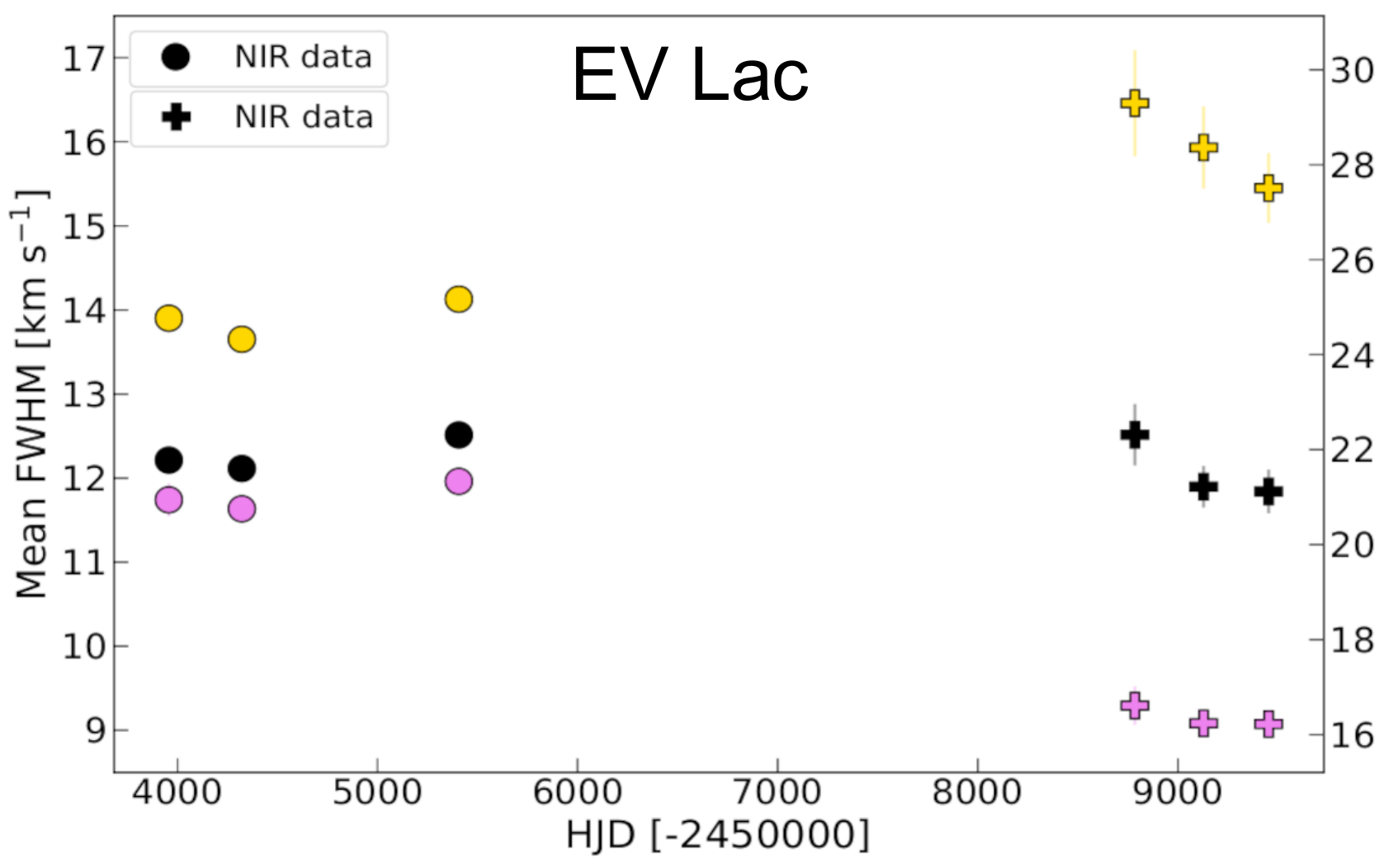}
    \includegraphics[width=\columnwidth]{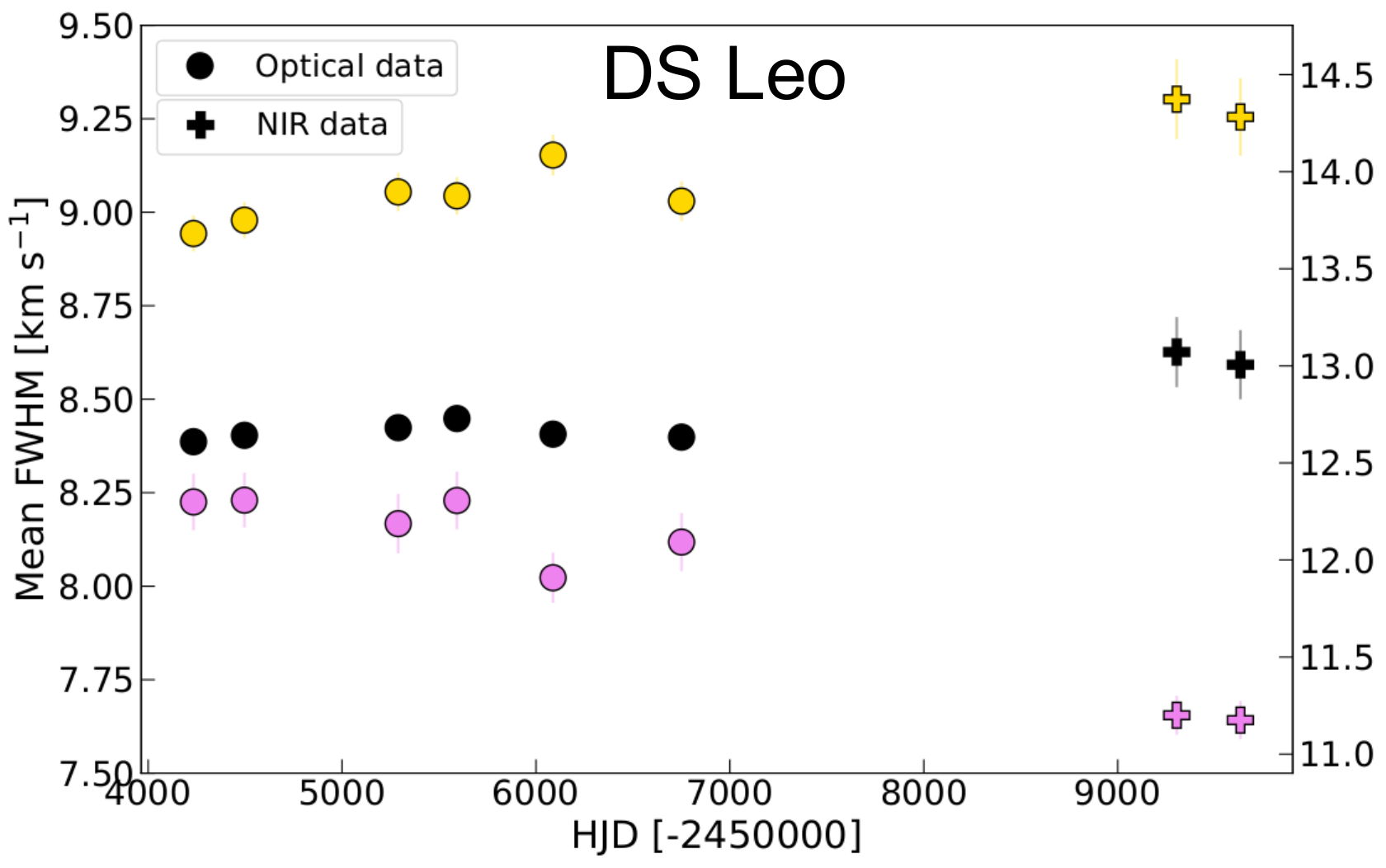}
    \includegraphics[width=0.92\columnwidth]{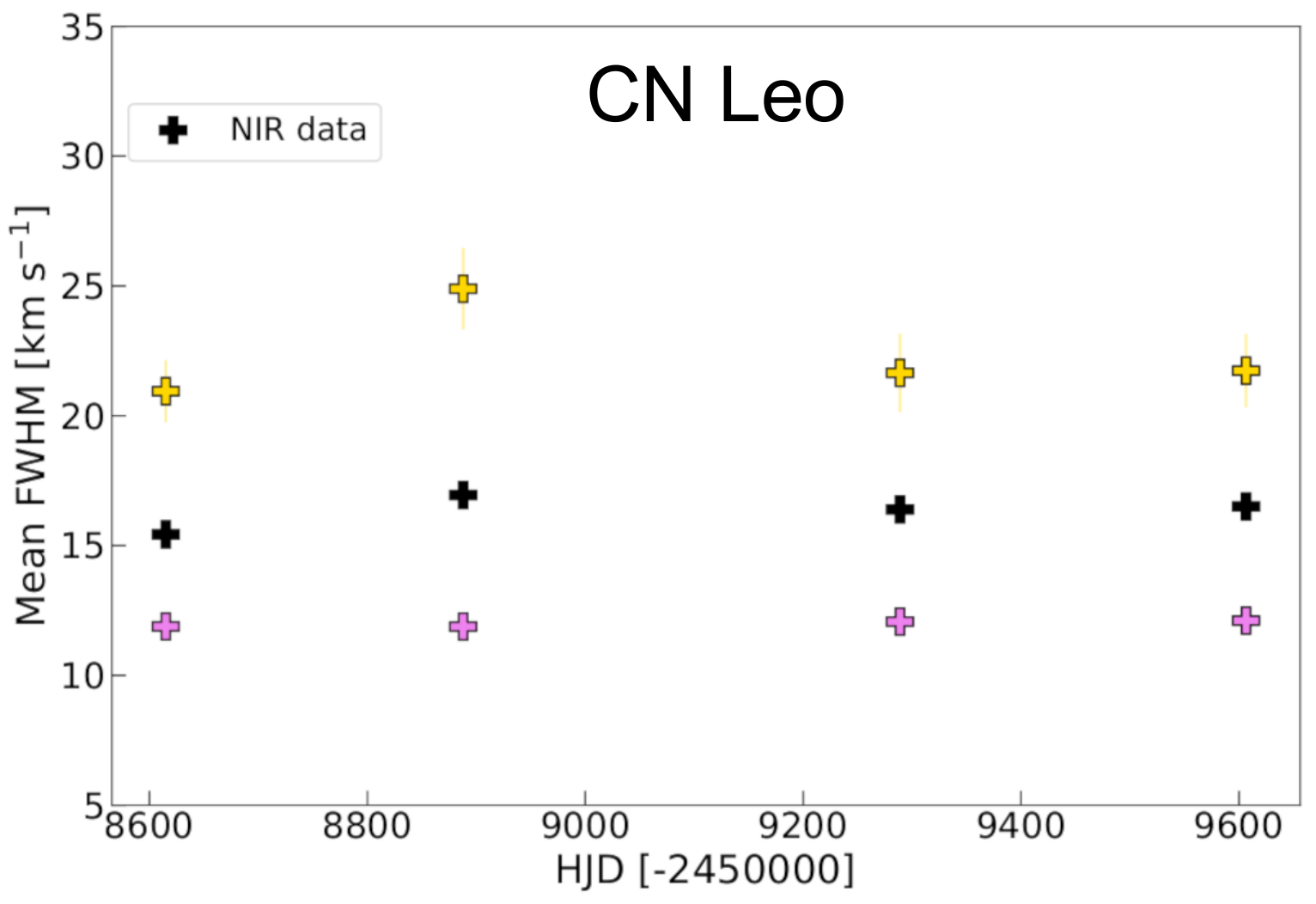}
    \caption{Long-term evolution of the epoch-averaged FWHM. The panels show the results from optical (circles) and near-infrared (pluses) observations of EV~Lac (top), DS~Leo (middle), and CN~Leo (bottom). In all panels, the data points are colour-coded based on the line mask used for LSD: full (black), low-g$_\mathrm{eff}$ (purple), and high-g$_\mathrm{eff}$ (yellow). The left $y$-axis refers to the optical observations, and the right $y$-axis to the near-infrared observations.}
    \label{fig:FWHM_evol}%
\end{figure}

We inspected the long-term behaviour of the epoch-averaged FWHM of Stokes~$I$, when computed with the full, high-g$_\mathrm{eff}$, and low-g$_\mathrm{eff}$ line masks. For AD~Leo, \citet{Bellotti2023b} show that the evolution of the epoch-averaged FWHM correlates with the secular trend of B$_l$ and the magnetic flux measured by modelling the Zeeman broadening and intensification. The results for the stars examined are shown in Fig.~\ref{fig:FWHM_evol}.

For EV~Lac, no clear trend is observed. In near-infrared, the average FWHM decreased from 22 to 21\,km\,s$^{-1}$ for the full mask, while it was stable around 16\,km\,s$^{-1}$ for the low-g$_\mathrm{eff}$ mask. For high-g$_\mathrm{eff}$, there is a slight temporal decrease from 29\,km\,s$^{-1}$ to just below 28\,km\,s$^{-1}$. In optical, we observe the same slight oscillation when adopting the three line masks, around 14\,km\,s$^{-1}$, above 12\,km\,s$^{-1}$ and below 12\,km\,s$^{-1}$ for the high-g$_\mathrm{eff}$, full, and low-g$_\mathrm{eff}$ masks, respectively.

For DS~Leo, the mean FWHM from near-infrared observations is 13\,km\,s$^{-1}$, 14\,km\,s$^{-1}$ and 11\,km\,s$^{-1}$ for the full, high-g$_\mathrm{eff}$ and low-g$_\mathrm{eff}$ mask and for both epochs. In optical, the mean FWHM is stable around 8.0\,km\,s$^{-1}$, 9.0\,km\,s$^{-1}$ and 8.4\,km\,s$^{-1}$ for the full, high-g$_\mathrm{eff}$ and low-g$_\mathrm{eff}$ masks. Therefore, we do not observe a significant evolution. The same conclusion is reached when computing near-infrared LSD profiles with a 4000\,K mask.

For CN~Leo, the low-g$_\mathrm{eff}$ lines are stable around 12\,km\,s$^{-1}$, the full mask values evolve from 15\,km\,s$^{-1}$ to 17\,km\,s$^{-1}$ and then back to 16\,km\,s$^{-1}$, while the high-g$_\mathrm{eff}$ mask values from 21\,km\,s$^{-1}$ to 25\,km\,s$^{-1}$ and back to 22\,km\,s$^{-1}$. Such temporal variation is anti-correlated with the long-term sine oscillation of B$_l$ (in absolute value), the average FWHM being largest when the $B_l$ is smallest. 

\begin{table}[!t]
\caption{Correlation coefficients between $|$B$_l|$ and FWHM of EV~Lac and DS~Leo, for the epochs in which we observed rotational modulation at the stellar rotation period.} 
\label{tab:fwhm_pearson}     
\centering                       
\begin{tabular}{l r r r}       
\hline     
Epoch & full & high-g$_\mathrm{eff}$ & low-g$_\mathrm{eff}$\\ 
\hline
\multicolumn{4}{c}{EV~Lac}\\
\hline
2006 & $0.33\pm0.39$ & $0.49\pm0.36$ & $0.08\pm0.48$\\
2007 & $0.27\pm0.29$ & $0.30\pm0.28$ & $0.29\pm0.30$\\
\hline
\multicolumn{4}{c}{DS~Leo}\\
\hline
2007 & $0.24\pm0.26$ & $0.10\pm0.29$ & $0.34\pm0.23$\\
2008 & $-0.35\pm0.15$ & $-0.34\pm0.15$ & $-0.51\pm0.12$\\
2010 & $-0.07\pm0.33$ & $0.19\pm0.24$ & $-0.49\pm0.42$\\
2011 & $0.84\pm0.12$ & $0.86\pm0.09$ & $0.20\pm0.28$\\
2012 & $-0.51\pm0.25$ & $0.27\pm0.29$ & $-0.66\pm0.18$\\
2014 & $-0.38\pm0.20$ & $-0.15\pm0.23$ & $-0.55\pm0.19$\\
2020b2021a & $0.09\pm0.12$ & $0.20\pm0.12$ & $-0.11\pm0.11$\\
\hline
\end{tabular}
\tablefoot{The columns are: 1) subset of the time series, 2) average Pearson correlation coefficient from bootstrap analysis using the full LSD mask, 3) the high-g$_\mathrm{eff}$ mask, and 4) the low-g$_\mathrm{eff}$ mask.}
\end{table}

\begin{figure*}[!t]
    \centering
    \includegraphics[width=0.85\textwidth]{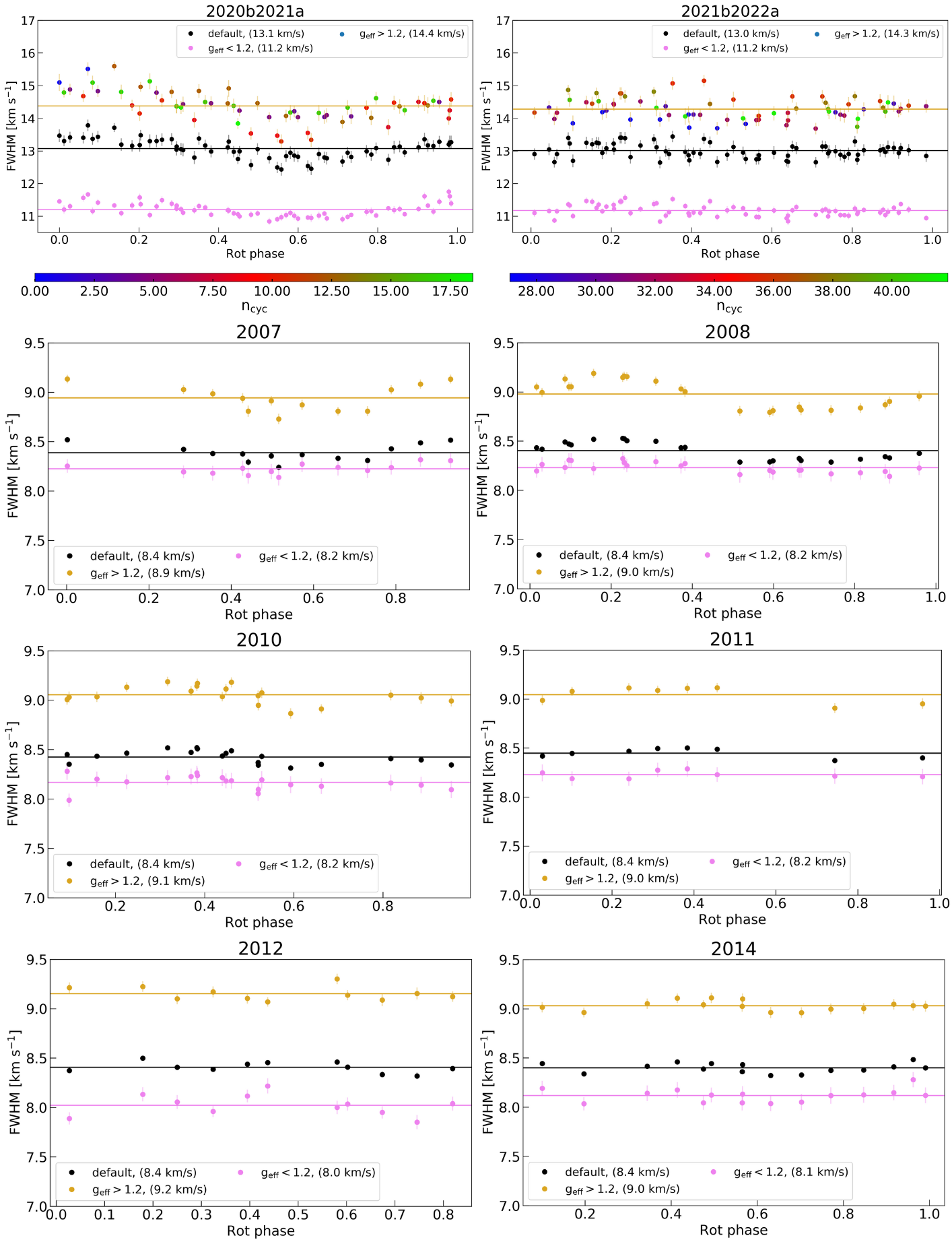}
    \caption{Rotational modulation analysis of FWHM for DS~Leo. The  top two panels correspond to the near-infrared epochs (2020b2021a and 2021b2022a), whereas the remaining panels correspond to the optical epochs (2007, 2008, 2010, 2011, 2012, and 2014). In each panel are shown the phase-folded time series of FWHM computed with the full, low-g$_\mathrm{eff}$, and high-g$_\mathrm{eff}$ mask with a horizontal line representing the mean FWHM value (also reported in each legend). For the near-infrared epochs the high-g$_\mathrm{eff}$ time series is colour-coded by rotational cycle to inspect signs of short-term variability. Observed stacked data points: observations within the first five cycles tend to fall at high FWHM values, then progressively at lower FWHM between cycle number 5 and 11, and at higher FWHM values again in the latest cycles.}
    \label{fig:FWHM_rotmod_dsleo}%
\end{figure*}

\section{Principal component analysis}\label{sec:pca}

\cite{Lehmann2022} presented a method based on principal component analysis (PCA) that reveals key properties of the stellar magnetic topology and its temporal evolution directly from the LSD Stokes~$V$ profiles, without prior assumptions about stellar parameters such as $v_\mathrm{eq}\sin i$ and inclination. The PCA method provides information about the degree of axisymmetry, the poloidal-to-toroidal fraction of the axisymmetric field, the field complexity (i.e. dipolar or more complex topology), and the temporal evolution of these three parameters. Information about the axisymmetric field are captured by the mean Stokes~$V$ profile of the observed time series and information about the non-axisymmetric field in the mean-subtracted Stokes~$V$ profiles. The PCA method analyses the mean Stokes~$V$ profile to infer the degree of axisymmetry and whether the axisymmetric field is more poloidal or toroidal. Furthermore, by determining the eigenvectors and coefficients of the mean-subtracted Stokes~$V$ profiles using PCA, information about the field complexity and its evolution with time can be obtained. For more information about the PCA method see \cite{Lehmann2022} and \cite{Lehmann2024}.

We studied the near-infrared LSD Stokes~$V$ time series collected by SPIRou for EV~Lac, DS~Leo and CN~Leo similar to our analysis for AD~Leo \citep{Bellotti2023b}. As in \cite{Lehmann2024}, we used the S/N-weighted mean profiles and the weighted PCA analysis \citep{Delchambre2015}, as they provide better results for long time series with varying S/N.

\begin{figure*}
        \raggedright \textbf{a.} \hspace{4.5cm} \textbf{b.} \\
        \centering
        \includegraphics[width=0.257\textwidth, trim={0 0 0 0}, clip]{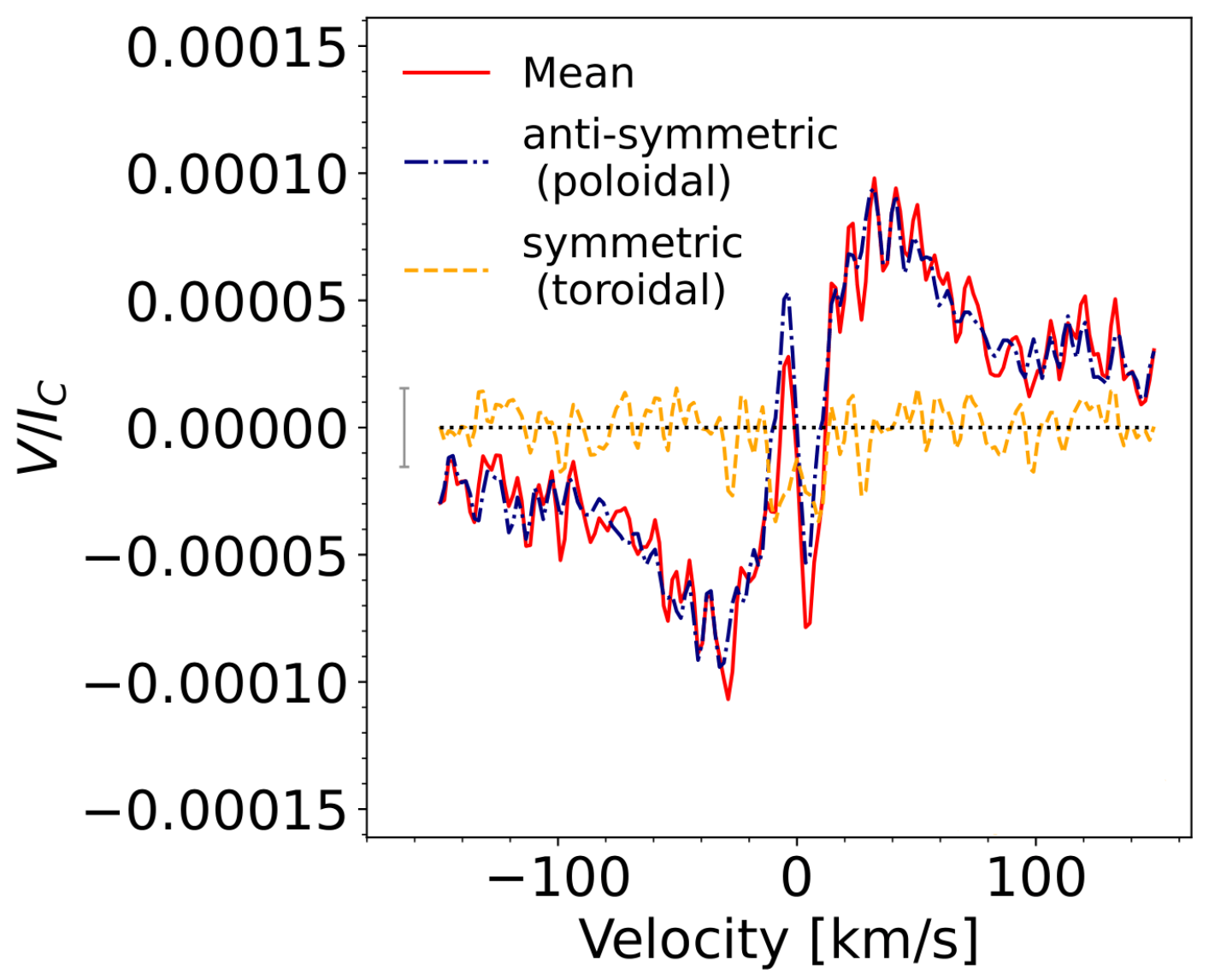}
        \includegraphics[width=0.73\textwidth, trim={0 400 0 0}, clip]{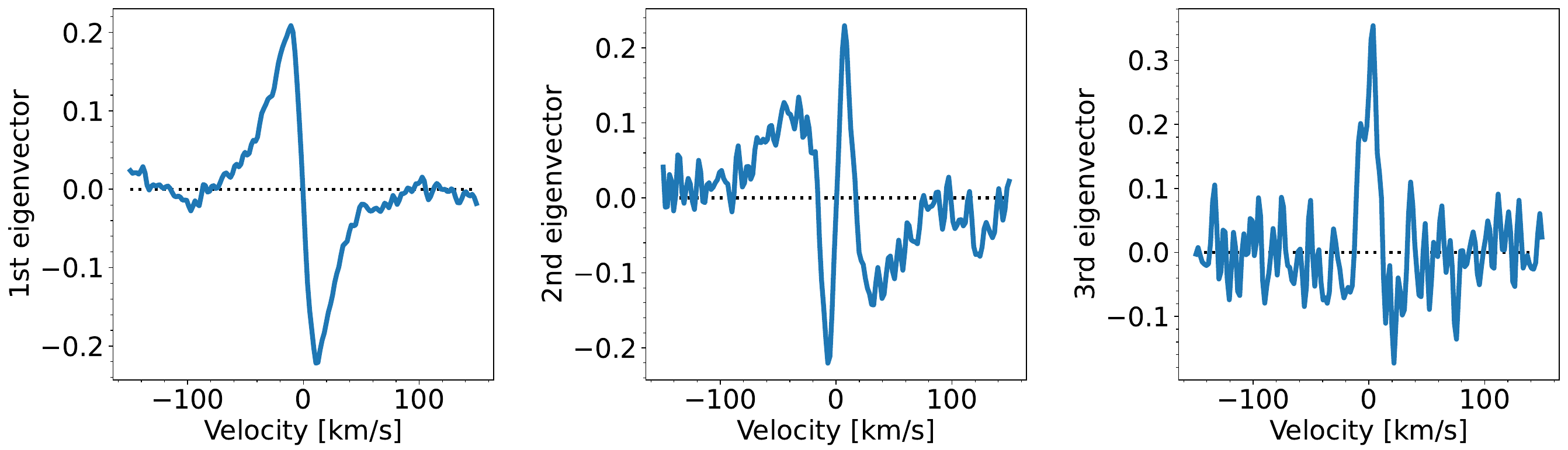}
	 
        \rule{14cm}{0.3mm}\\
        \raggedright \textbf{c.} \\
        \centering
 \textbf{2019b}\\
        \includegraphics[width=0.258\textwidth, trim={0 0 0 0}, clip]{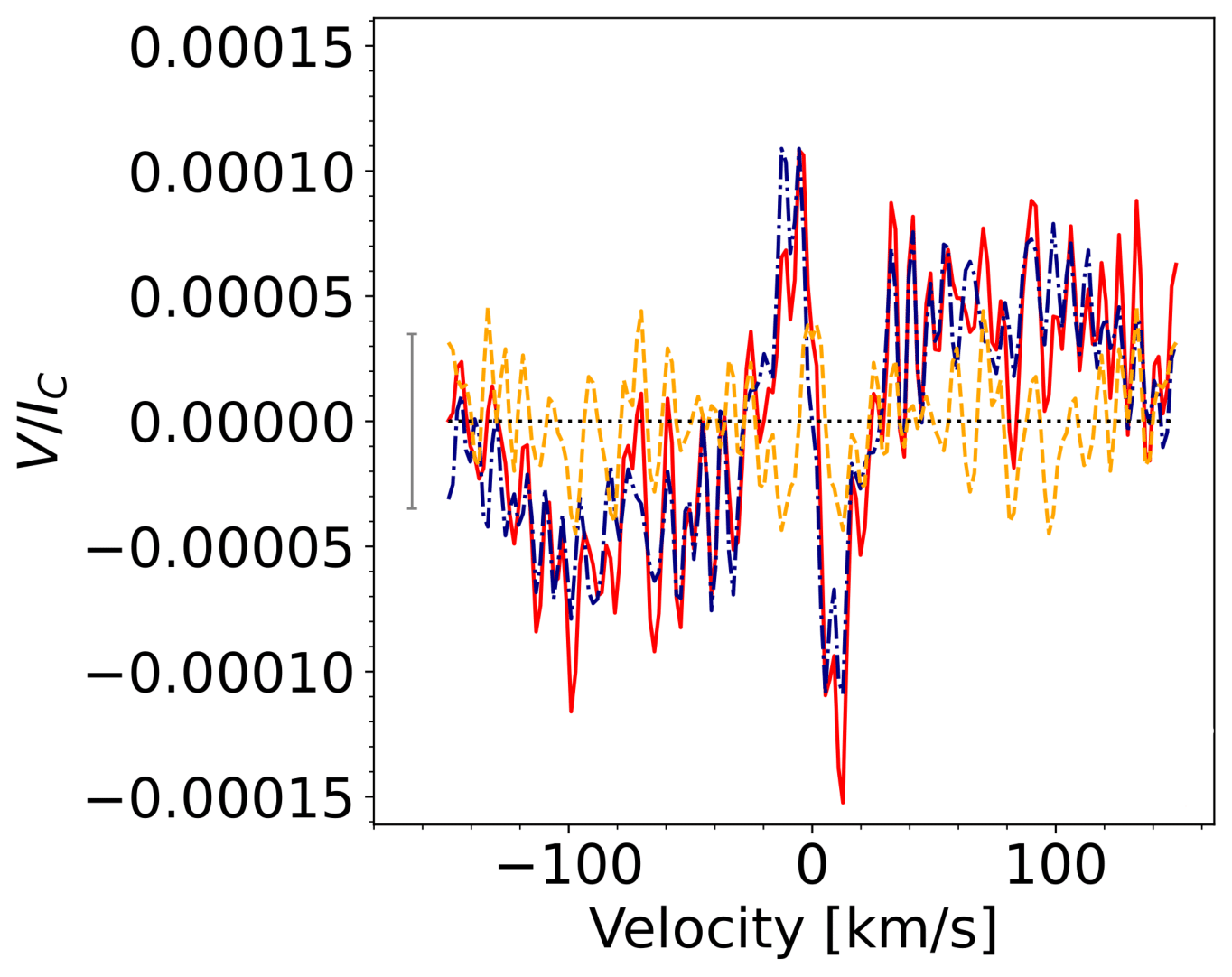}
        \includegraphics[width=0.73\textwidth, trim={0 400 0 0}, clip]{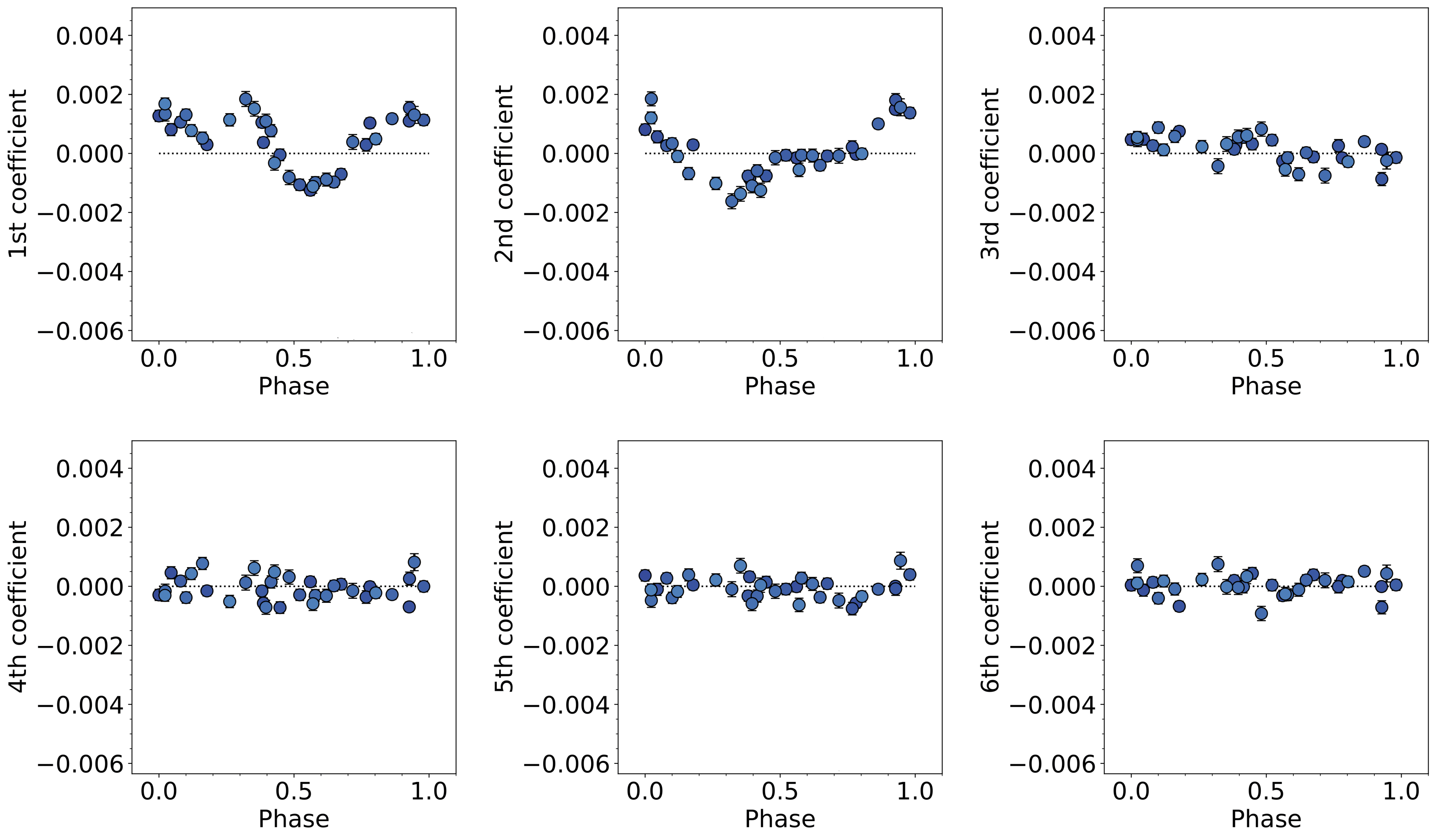}
	
 \textbf{2020b2021a}\\
        \includegraphics[width=0.258\textwidth, trim={0 0 0 0}, clip]{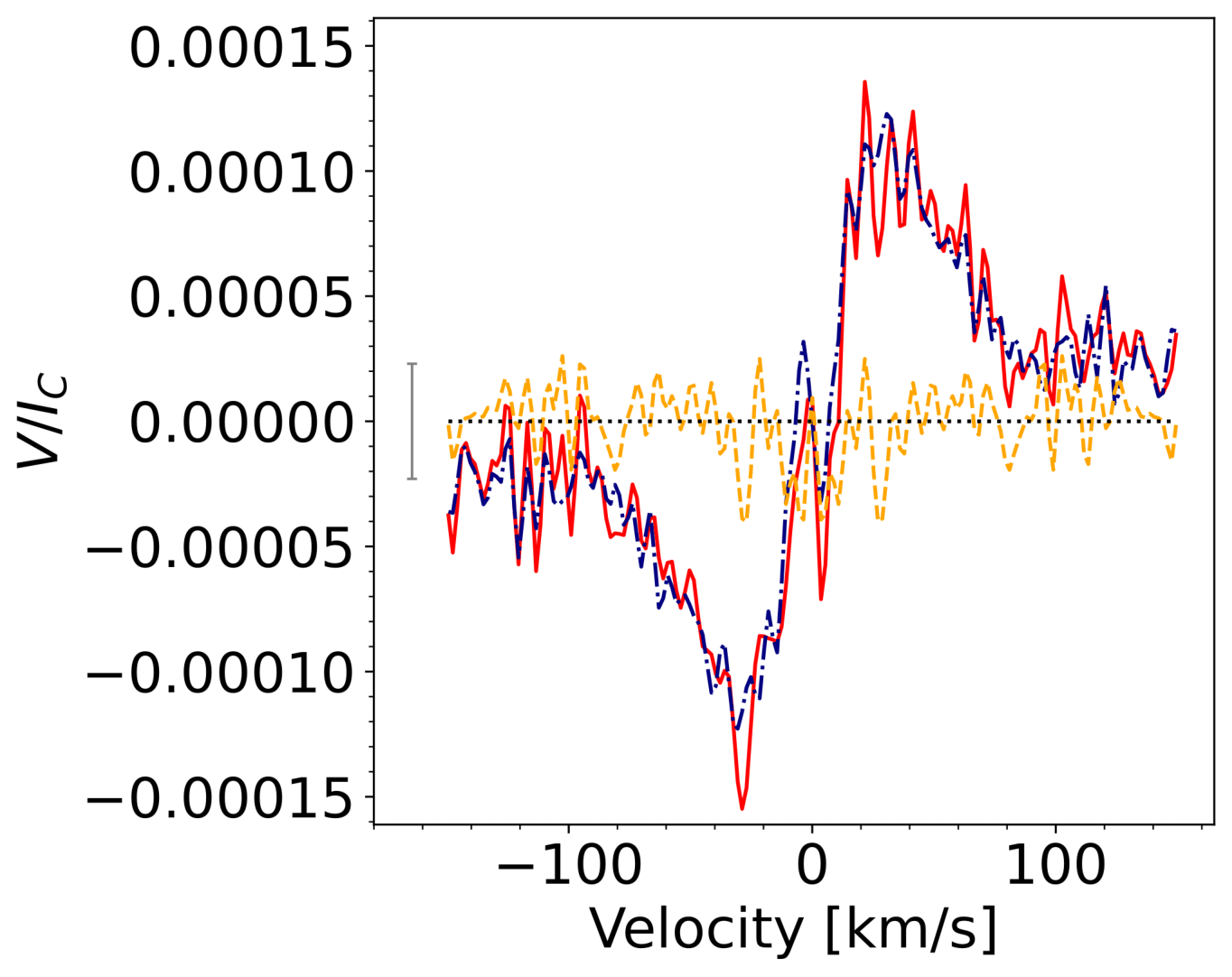}
        \includegraphics[width=0.73\textwidth, trim={0 400 0 0}, clip]{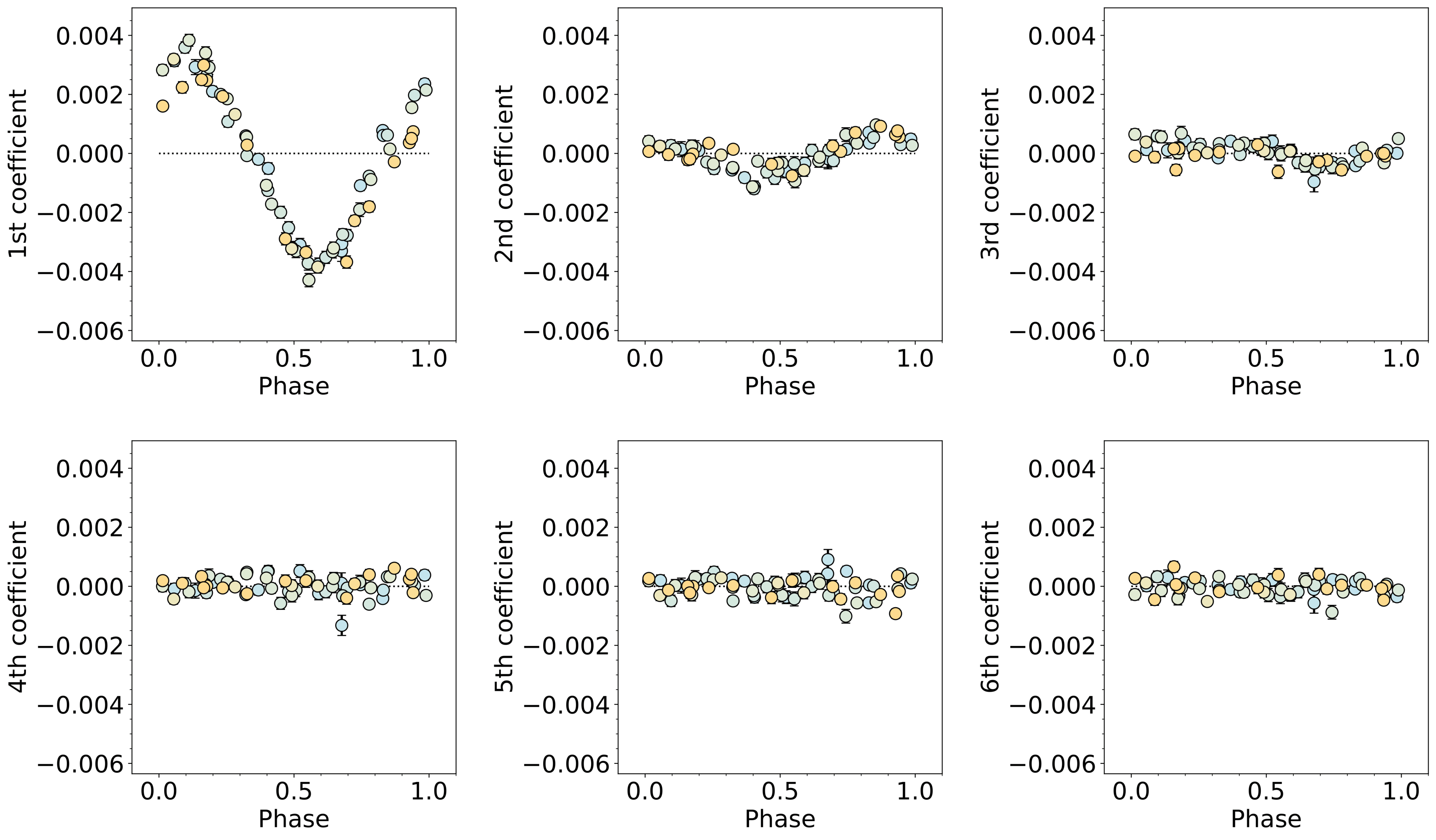}
	
 \textbf{2021b}\\
        \includegraphics[width=0.258\textwidth, trim={0 0 0 0}, clip]{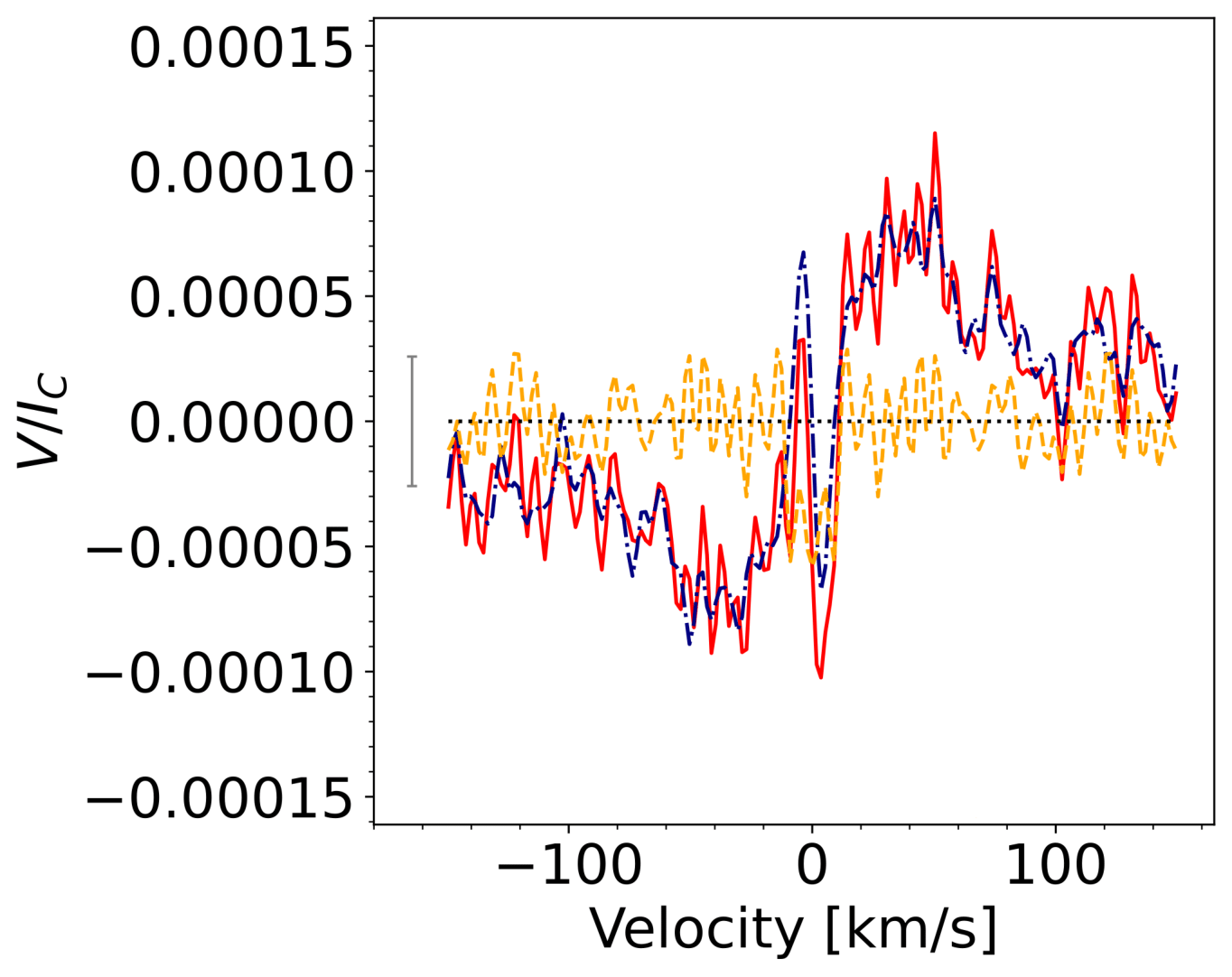}
        \includegraphics[width=0.73\textwidth, trim={0 400 0 0}, clip]{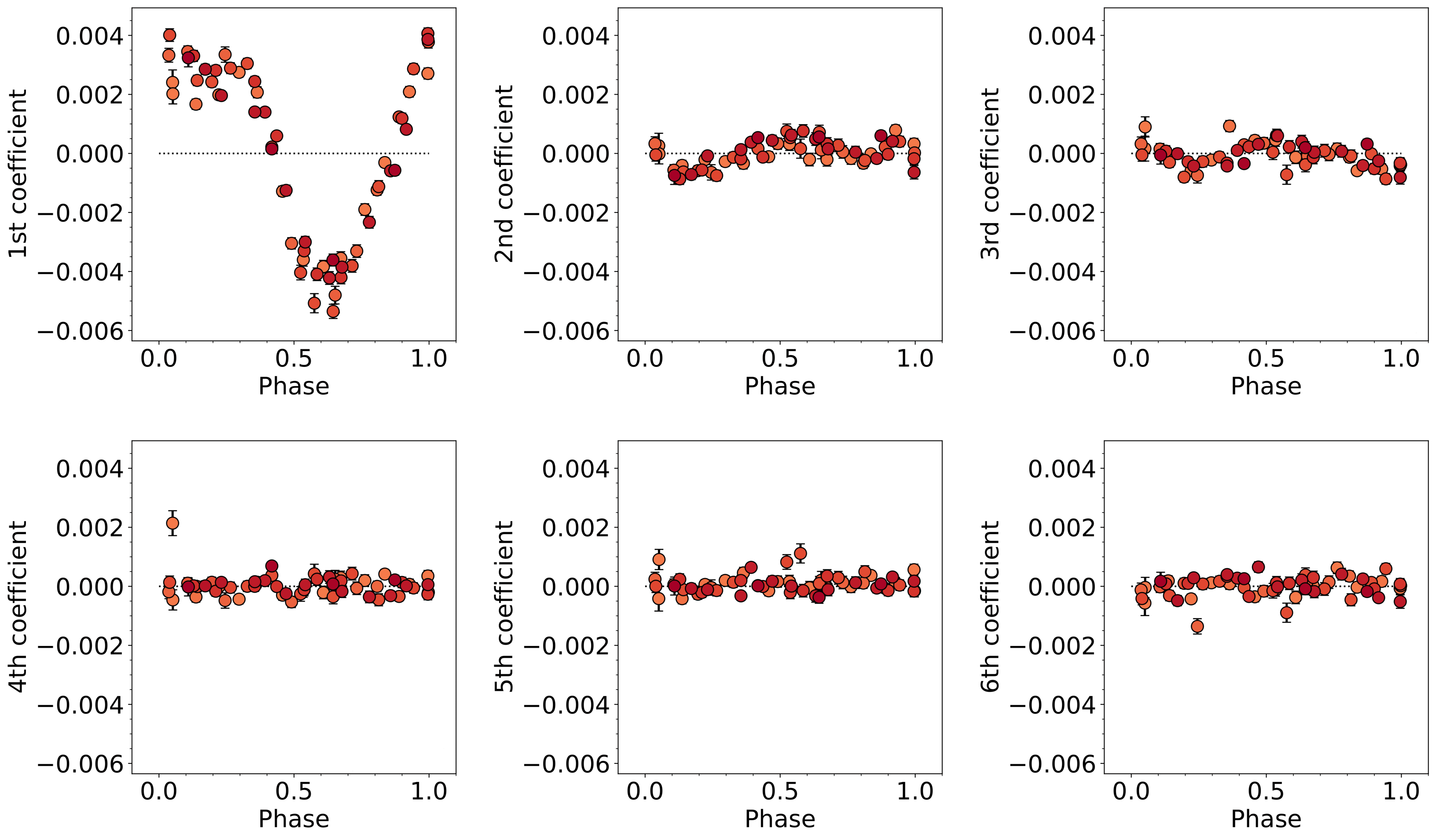}
    \caption{PCA analysis for EV~Lac. \textbf{a.} Mean profile (red) for all observations and its decomposition in the antisymmetric (blue dashed) and symmetric (yellow dotted) component (with respect to the line centre) related to the poloidal and toroidal axisymmetric field, respectively. 
    \textbf{b.} First three eigenvectors of the mean-subtracted Stokes~$V$ profiles. \textbf{c.} Mean profile (left column), and the coefficients of the first three eigenvectors (three columns to the right) for each season (one season per row). The mean profiles of the individual seasons are plotted in the same format as in panel a. The coefficients are colour-coded by rotation cycle.}
    \label{Fig:EVLac_PCA}
\end{figure*}

\subsection{EV~Lac}

The mean Stokes~$V$ profile of EV~Lac is antisymmetric to the line centre, suggesting that the axisymmetric field is predominantly poloidal (see Fig.~\ref{Fig:EVLac_PCA}a), and begins to split at the line centre owing to the strong magnetic field of EV~Lac. The PCA method is based on the weak-field approximation valid for $B << 1\mathrm{kG}$. { For EV Lac the total unsigned magnetic field values are above this limit with 4.3~kG \citep{Shulyak2017}, which causes the visible splitting at line centre of the mean Stokes~$V$ profile. Therefore, we need to be cautious about our conclusions derived with the PCA method. This also applies for CN~Leo which exhibits a total field value of $\sim2.3$~kG \citep{Shulyak2019}, although we do not record a clear Stokes~$V$ splitting.} As a double-check, we simulated the Stokes~$V$ profile using synthetic and observed magnetic field maps while artificially increasing or decreasing the magnetic field strength. { Our simulations for the three stars of our sample show that the conclusions derived with the PCA method remained valid for $B>1$~kG in the case of EV~Lac (and CN~Leo), even if it is based on the weak-field approximation.}

A signal is present in the first three eigenvectors of the PCA analysis of the mean-subtracted Stokes~$V$ profile capturing the information about the non-axisymmetric component of EV~Lac's topology, with the first two displaying an antisymmetric shape and the third a symmetric shape relative to the line centre (see Fig.~\ref{Fig:EVLac_PCA}b). The large number of eigenvectors containing a signal symbolises a non-axisymmetric field topology.

Figure~\ref{Fig:EVLac_PCA}c shows the per-epoch analysis of EV~Lac for the Stokes~$V$ mean profile capturing the information of the axisymmetric field and the PCA coefficients of the first three eigenvectors allowing the analysis of the non-axisymmetric field. The coefficients are derived from the weighted PCA of the mean-subtracted Stokes~$V$ time series using the weighted mean profile computed over all epochs (e.g.\ Fig.~\ref{Fig:EVLac_PCA}a), and not the weighted mean profile of each individual epoch (e.g.\ Fig.~\ref{Fig:EVLac_PCA}c left column). This because, in the latter case, the mean value of the coefficients would be centred for each epoch, and the amplitudes of the variations could not be compared between epochs.

We observe that the amplitude of the mean Stokes~$V$ profile increases from 2019b to 2020b2021a. It remains antisymmetric with respect to the line centre for all epochs, which indicates a poloidal-dominated axisymmetric component for 2019b to 2021b. Since the amplitude of the coefficients also increases from 2019b to 2021a, we conclude that the field of EV~Lac strengthens between 2019b and 2021a. The most complex phase variations of the coefficients occur in the first epoch 2019b, suggesting that the topology of EV~Lac has the lowest dipole fraction and therefore is most complex in 2019b. Somehow opposite is the second epoch 2020b2021a: the coefficients show sinusoidal curves for all three eigenvectors, reflecting a tilted, dipole-dominated topology. In the last epoch 2021b, the coefficients still show a sinusoidal trend, but become slightly more flat between rotational phase $0.0-0.3$.

\subsection{DS~Leo}

Figure~\ref{Fig:DSLeo_PCA} shows the PCA analysis for DS~Leo. The mean profile reflecting the axisymmetric component of the topology is mainly symmetric to the line centre, indicating a significant toroidal field fraction ($>5\%$) of the axisymmetric field, see Fig.~\ref{Fig:DSLeo_PCA}a. The PCA analysis of the mean-subtracted profiles allows the analysis of the non-axisymmetric component. Two eigenvectors emerge from the noise, one antisymmetric and one symmetric, indicating a non-axisymmetric topology. The degree of axisymmetry is however larger than EV~Lac's, as fewer eigenvectors show a signal (see Fig.~\ref{Fig:DSLeo_PCA}b). 

The per-epoch analysis shows a decrease in amplitude of the mean Stokes~$V$ profile, as well as of the phase variations of the coefficients, indicating a weakening of the magnetic field from the first to the second epoch. For both epochs, the coefficient of the first eigenvector exhibits the typical sinusoidal trend of a tilted dipole, and the pointing phase of the dipole appears to be stable. The coefficient of the second eigenvector manifests a more complex trend that evolves from 2020b2021a to 2021b2022a. This reflects the evolution of other magnetic field features in addition to the tilted dipole. We clearly see that the coefficients also vary within the epochs, implying that the magnetic field of DS~Leo evolves on the timescale of its rotation period, as already suggested by \cite{Lehmann2022} in their analysis of the 2008 optical observations of DS~Leo.

\subsection{CN~Leo}

CN~Leo exhibits a strong and antisymmetric Stokes~$V$ mean profile, indicating a predominantly poloidal axisymmetric field, as illustrated in Fig.~\ref{Fig:CNLeo_PCA}a. The first eigenvector is antisymmetric and there is evidence for a second, symmetric eigenvector, albeit the larger noise (Fig.~\ref{Fig:CNLeo_PCA}b). This suggests that CN~Leo is the most axisymmetric M~dwarf of our sample.

The per-epoch Stokes~$V$ mean profiles show a varying amplitude, together with the amplitude of the rotational modulation of the coefficients, which could be due to variability in either field strength or axisymmetry (see Fig.~\ref{Fig:CNLeo_PCA}c). The first epoch 2019a appears to be highly axisymmetric, as the coefficients cluster around zero. The second epoch 2019b2020a has a clear sinusoidal variation in the coefficients of the first eigenvector, implying a larger tilt angle of the dipole relative to the other epochs, and the mean of the coefficients is offset towards positive values. The third epoch 2020b2021a is characterised by a less clear sinusoidal variation in the coefficients, with their mean offset towards negative values. The fourth epoch 2021b2022a resemble the first one, with the mean of the coefficients around zero.

\section{Magnetic imaging}\label{sec:magnetic_imaging}

We applied Zeeman-Doppler imaging to reconstruct the large-scale magnetic field at the surface of EV~Lac, DS~Leo and CN~Leo. The magnetic geometry is described as the sum of a poloidal and a toroidal component, which are both expressed through spherical harmonics decomposition. In particular, we employed the formalism outlined in \citep{Lehmann2022}. The algorithm proceeds iteratively, by synthesising Stokes~$V$ profiles and comparing them with the observations, in order to fit the spherical harmonics coefficients $\alpha_{\ell,m}$, $\beta_{\ell,m}$, and $\gamma_{\ell,m}$ (with $\ell$ and $m$ the degree and order of the mode, respectively), until a maximum-entropy solution at a fixed reduced $\chi^2$ is reached \citep{Skilling1984,Semel1989,DonatiBrown1997}. We employed the \texttt{zdipy} code described in \citet{Folsom2018}, and which was implemented by \citet{Bellotti2023b} to incorporate the Unno-Rachkovsky's solutions to polarised radiative transfer equations in a Milne-Eddington atmosphere \citep{Unno1956,Rachkovsky1967,Landi2004} and the filling factor formalism adopted by \citet{Morin2008}. 

The filling factors f$_I$ and f$_V$ represent the fraction of the cell of the stellar surface grid covered by magnetic regions and magnetic regions producing net circular polarisation, respectively \citep{Morin2008,Kochukhov2021}. With the inclusion of f$_V$, we assume that the polarisation signal comes from a multitude of magnetic spots whose local field strength is $B/\mathrm{f}_V$ distributed following a certain large-scale structure such that the magnetic field modulus averaged over a grid cell is equal to $B$. In practice, using f$_V$ enables us to reproduce the amplitude (scaling with the magnetic field $B$) and the Zeeman splitting (scaling with $B/\mathrm{f}_V$) observed for Stokes~$V$ LSD profiles. The values of filling factors are assumed constant throughout the stellar surface grid.

The Unno-Rachkovsky models of the local line profiles (for both Stokes~$I$ and $V$) are described by the following parameters \citep{delToroIniesta2003,Landi2004}: the Gaussian width ($w_G$, related to thermal broadening), the Lorentzian width ($w_L$, related to pressure broadening), the ratio of the line to continuum absorption coefficients ($\eta_0$), and the slope of the source function in the Milne-Eddington atmosphere ($\beta$). Before applying ZDI, and for each epoch of our three stars, we perform a parameter optimisation based on a $\chi^2$ minimisation approach. In practice, we generate a series of synthetic Stokes~$I$ profiles for a grid of parameters, and compare them with the median observed Stokes~$I$ for a specific epoch, until a minimum $\chi^2$ is found. The parameters corresponding to the $\chi^2$ minimum are then used for the ZDI reconstruction. For EV~Lac, we used $w_G=0.1$~km\,s$^{-1}$, $w_L=12.0$~km\,s$^{-1}$, and $\eta_0=9.8$, for DS~Leo, we used $w_G=0.5$~km\,s$^{-1}$, $w_L=4.4$~km\,s$^{-1}$, and $\eta_0=17.0$, and for CN~Leo, we used $w_G=0.5$~km\,s$^{-1}$, $w_L=10.0$~km\,s$^{-1}$, and $\eta_0=9.5$. { The value of $\beta$ is derived considering that, for a Milne-Eddington atmosphere, the local continuum flux relative to the flux at disc centre is
\begin{equation}\label{eq:limb_beta}
    I_c/I^0_c = (1+\beta\cos\theta)/(1+\beta),
\end{equation}
where $\theta$ is the angle between the line of sight and stellar surface normal \citep{Landi2004}. A standard linear limb darkening law is given, for instance, by
\begin{equation}\label{eq:limb_eta}
    I_c/I_c^0 = 1-\eta+\eta\cos\theta,
\end{equation}
where $\eta$ is the limb darkening coefficient \citep{Gray2005}. From Eq.~\ref{eq:limb_beta} and \ref{eq:limb_eta}, we derive
\begin{equation}
    \beta = \eta / (1-\eta).
\end{equation}
In our case, we adopted a linear limb darkening coefficient in H band of 0.2 \citep{Claret2011}, hence $\beta$ is consistently fixed to $0.25$ in all ZDI reconstructions \citep[for more details, see][]{Erba2024}.} 

Once the line parameters were fixed, we searched for the optimised value of filling factor f$_V$ for each epoch of our stars. The procedure is similar to an optimisation of stellar parameters or differential rotation search \citep[see][]{Petit2002}. We run ZDI for a grid of f$_V$ values between 1.0\% and 100\%, and we recorded the $\chi^2$ reached by ZDI each time. The $\chi^2$ distribution is fit with a parabola around the minimum, and the value of f$_V$ corresponding to the minimum $\chi^2$ represents the optimal value. The $1\sigma$ error bars on f$_V$ are determined as the $\Delta\chi^2=1$ variation away from the $\chi^2$ minimum \citep{Press1992}.

Finally, the maximum degree of the harmonic expansion was set to $\ell_\mathrm{max}=$\,8 for all stars, consistently with the spatial resolution dictated by $v_\mathrm{eq}\sin i$ { and previous ZDI reconstructions \citep{Morin2008,Hebrard2016}}. { We note that only the modes with $\ell$ up to three have the most significant contribution in the reconstructed field maps}. We did not observe appreciable differences in the ZDI reconstructions for filling factor f$_I$ between 0.0 and up to 0.5 in some cases, hence we fixed it to f$_I=0.0$. 

We summarise the properties of all the magnetic field reconstructions in Table~\ref{tab:zdi_output}. { The obliquity refers to the colatitude of the maximum of the dipolar component. It is obtained by first computing the poloidal-dipolar component of the field using the $\alpha_{lm}$ coefficient, and then by locating the colatitude associated with the maximum.}

\subsection{EV~Lac}

The optical maps of EV~Lac where reconstructed by \citet{Morin2008} using the 2006 and 2007 data collected with ESPaDOnS and Narval. They found a strong (B$_\mathrm{mean}=$\,500\,G), non-axisymmetric, mostly dipolar field, composed mainly by two magnetic spots of distinct polarity at opposite longitudes. They were able to constrain a differential rotation rate of 1.7\,mrad\,d$^{-1}$ which was consistent with solid body rotation within 3$\sigma$.

To carry out tomographic inversion for EV~Lac near-infrared SPIRou data, we assumed $i=60^{\circ}$, $v_\mathrm{eq}\sin i=4.0$\,km\,s$^{-1}$, P$_\mathrm{rot}=4.36$\,d, and solid body rotation \citep{Morin2008}. The Stokes~$V$ time series is shown in Fig.~\ref{fig:zdi_stokesV_evlac}. The profiles are fitted down to a $\chi^2_r$ level of 0.9, 1.2, and 1.4 from an initial value of 1.2, 5.0, and 6.5 for 2019b, 2020b2021a, and 2021b, respectively.

\begin{figure*}[t]
    \centering
    \includegraphics[width=\textwidth]{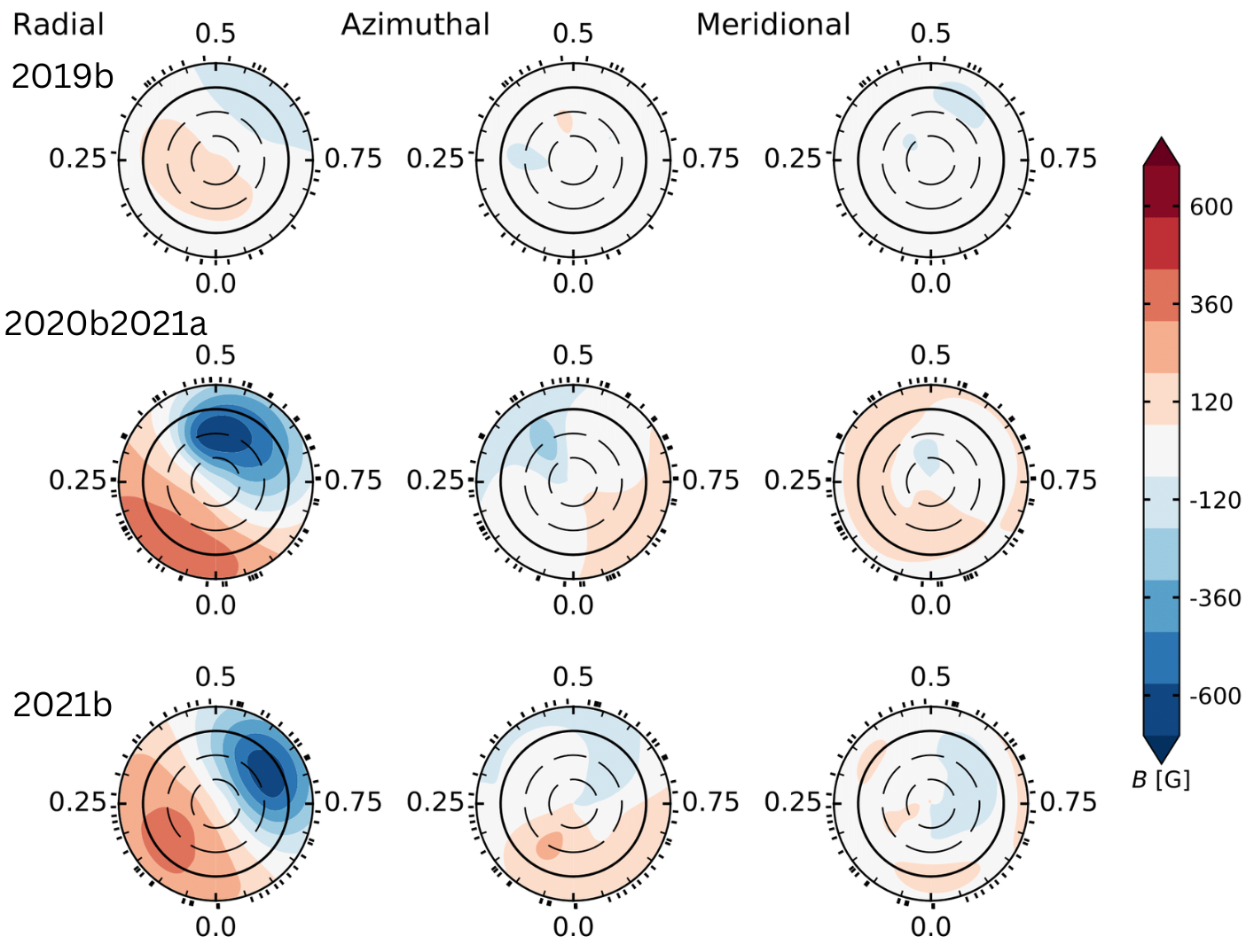}
    \caption{Reconstructed ZDI maps in flattened polar view of EV~Lac for (from left to right) 2019b, 2020b2021a, and 2021b. In each column the radial (top), azimuthal (middle), and meridional (bottom) components of the magnetic field vector are displayed. The radial ticks are located at the rotational phases when the observations were collected, while the concentric circles represent different stellar latitudes: +30\,$^{\circ}$ and +60\,$^{\circ}$ (dashed lines), and equator (solid line). The colour bar range is set by the maximum (in absolute value) of the magnetic field and illustrates the positive (red) and negative (blue) magnetic polarity for each epoch.}
    \label{fig:zdi_evlac}%
\end{figure*}

We constrained the filling factor $f_V$ to 9\% and 19\% for 2020b2021a and 2021b, while we could not for 2019b, hence we set it to 100\%. The typical error bar on $f_V$ is at most 1\%. Using either 9\% or 19\% on 2019b has only marginal effects on the map and the magnetic energy repartition, and the same level of $\chi^2$ is achieved. This could result from a lower effective S/N at this epoch, which prevents us from reliably constraining $f_V$. The fact that we can fix $f_V=100\%$ for this epoch means that modelling the horizontal splitting of the Stokes~$V$ lobes is not necessary, and weak-field approximation could be equivalently assumed.

For the two remaining epochs, it is not straightforward to claim an evolution of the small-scale features represented by a change in $f_V$ from 9\% to 19\%. If we set $f_V$ to the median value of the two epochs (i.e. 14\%), we observe only slight differences in the reconstructed maps: the field is more axisymmetric (an additional 4\%) and in 2020b2021a the field is also more octupolar (additional 4\%). While still limited, a variation of $f_V$ has more impact on 2020b2021a than 2021b because of the stronger constraint on $f_V$ in the former epoch. 

The maps of the magnetic field and their characteristics are shown in Fig.~\ref{fig:zdi_evlac} and Table~\ref{tab:zdi_output}. In all epochs, most of the magnetic energy is stored in the poloidal ($>99\%$) and dipolar ($>65\%$) components, with a substantial contribution from the quadrupolar modes ($>$10\%). The field is non-axisymmetric, and the axisymmetric energy fraction decreases over time from 25\% to 2\%. Likewise, the obliquity of the field increases from 50$^{\circ}$ to 87$^{\circ}$. The average magnetic field strength for the 2019b epoch is lower than the two other epochs. Considering that the Stokes~$V$ models of the 2019b epoch does not fully reproduce the amplitude of the LSD profiles, the magnetic field strength is likely underestimated. A differential rotation search on the SPIRou time series was inconclusive.

\subsection{DS Leo}

The large-scale magnetic field map for DS~Leo was reconstructed initially by \citet{Donati2008} using Narval data collected in 2007 and 2008. They obtained a predominantly toroidal field geometry encircling the star, whereas the poloidal component was mainly dipolar. The average field strength was 100\,G, and the axisymmetry of the poloidal component decreased between 2007 and 2008. Later, \citet{Hebrard2016} applied ZDI to HARPS-Pol and Narval observations collected in 2014, and consistently found a mostly-toroidal, axisymmetric geometry, with the poloidal component accounting for less than half the magnetic energy and mostly dipolar. Compared to the previous reconstructions, the average field strength decreased down to 80\,G and the dipolar component increased from 50\% to 88\%, already suggesting a rapidly evolving magnetic field. There are two Narval epochs with unpublished data, namely 2010 and 2012, for which we recovered the magnetic field map in this work, as outlined in Appendix~\ref{app:dsleo_app}.

We reconstructed the large-scale magnetic field for the 2020b2021a and 2021b2022a SPIRou epochs. The input parameters are $i=$\,60$^{\circ}$, $v_\mathrm{eq}\sin i=$\,2\,km\,s$^{-1}$ \citep{Hebrard2016}, P$_\mathrm{rot}=$\,13.91\,d, and we initially postulated solid body rotation. The time series of Stokes~$V$ profiles for the two epochs is shown in Appendix~\ref{app:zdi_app}. 
The profiles were fitted to a $\chi^2_r$ level of 1.4 and 1.2 from an initial value of 2.7 and 1.6 for 2020b2021a and 2021b2022a, respectively. We also constrained $f_V$ values of 10\% and 6\% for the two epochs (with error bars of 1\%).

\begin{figure}[!t]
    \centering
    \includegraphics[width=\columnwidth]{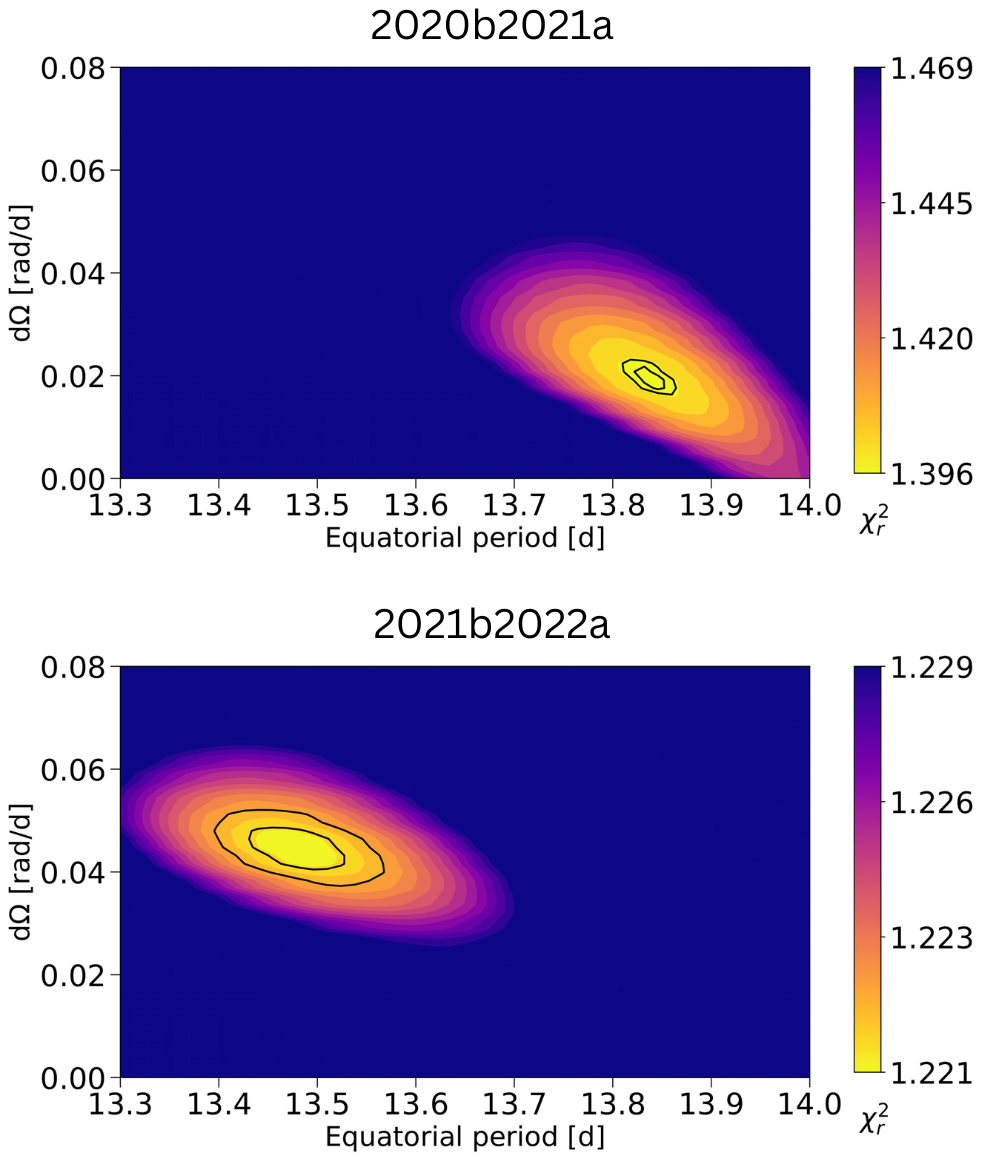}
    \caption{Differential rotation search for DS~Leo. Top: 2020b2021a. Bottom: 2021b2022a. The plots show the $\chi^2_r$ landscape over a grid of (P$_\mathrm{rot,eq}$,$d\Omega$) pairs, with the $1\sigma$ and $3\sigma$ contours. The best values are obtained by fitting a 2D paraboloid around the minimum, while their error bars are estimated from the projection of the $1\sigma$ contour on the respective axis \citep{Press1992}.}
    \label{fig:diff_rot_dsleo}%
\end{figure}

We searched for differential rotation using the method of \citet{Donati2000} and \citet{Petit2002}. Basically, we fixed the entropy at a certain value and inspected a dense grid of (P$_\mathrm{rot,eq}$, $d\Omega$) pairs to find the combination that minimises the $\chi^2_r$ between observations and synthetic models, as illustrated in Fig.~\ref{fig:diff_rot_dsleo}. The best parameters are measured by fitting a 2D paraboloid to the $\chi^2$ distribution, and the error bars are obtained from a variation of $\Delta\chi^2=1$ away from the minimum \citep{Press1992,Petit2002}. Differential rotation is implemented in \texttt{zdipy} with the following equation
\begin{equation}
    \Omega(\theta)=\Omega_\mathrm{eq}-d\Omega\sin^2(\theta)
    \label{eq:diff_rot}
\end{equation}
with $\Omega(\theta)$ the rotation frequency at colatitude $\theta$, $\Omega_\mathrm{eq}$ the rotation frequency at equator, and $d\Omega$ the differential rotation rate \citep{Folsom2018}.

We found P$_\mathrm{rot,eq}=13.836\pm0.016$\,d and $d\Omega=0.0192\pm0.0020$\,rad\,d$^{-1}$ for 2020b2021a, implying a rotation period at the pole of $14.447\pm0.069$\,d (see Eq.\ref{eq:diff_rot}). For 2021b2022a, we obtained P$_\mathrm{rot,eq}=13.469\pm0.054$\,d and $d\Omega=0.0448\pm0.0045$\,rad\,d$^{-1}$, and the rotation period at the pole was $14.899\pm0.172$\,d. Assuming the (P$_\mathrm{rot,eq}$, $d\Omega$) pairs as input parameters for ZDI, the Stokes~$V$ fit is improved down to a $\chi^2_r$ level of 1.25 and 1.10 for 2020b2021a and 2021b2022a, respectively. Finally, the inverse of the differential rotation rate represents the pole-equator lap time ($t_\mathrm{lap}=2\pi/d\Omega$). For 2020b2021a (time span of 256\,d) and 2021b2022a (time span of 205\,d) respectively, we obtained a $t_\mathrm{lap}$ of $327\pm34$\,d and $140\pm14$\,d. During this time, magnetic regions on the surface are presumably distorted by the shear of differential rotation. The short-term variability emerging from the SPIRou time series of FWHM values (see Fig.~\ref{fig:FWHM_rotmod_dsleo}) can be attributed to differential rotation since the equator-pole lap time is comparable to the time span of the epochs considered, meaning that differential rotation could have shifted the magnetic regions on the surface affecting the mean level of FWHM for that specific cycle.

\begin{figure*}[!t]
    \centering
    \includegraphics[width=\textwidth]{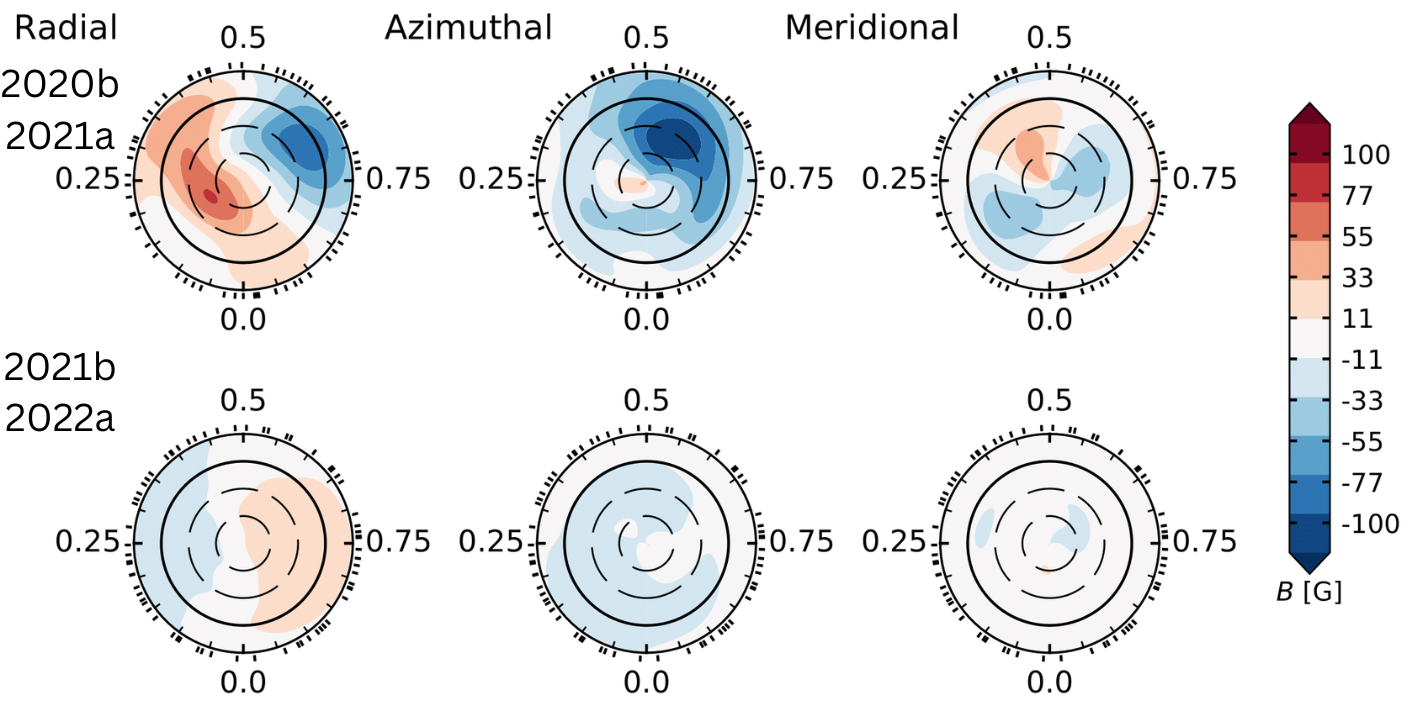}
    \caption{Reconstructed ZDI maps in flattened polar view of DS~Leo. Left: 2020b2021a. Right: 2021b2022a. The maps account for the constrained differential rotation. The format is the same as in Fig.~\ref{fig:zdi_evlac}.}
    \label{fig:zdi_dsleo}%
\end{figure*}

We obtained tighter constraints on $d\Omega$ for 2020b2021a relative to 2021b2022a because the field is less complex in the second epoch, that is to say it is characterised by a lower number of magnetic features tracking differential rotation. Besides, the error bars are smaller than the literature measurements because our time series contained a larger number of observations \citep{Donati2008,Hebrard2016}. In general, the error bars on the differential rotation parameters are only statistical and do not account for systematics, hence they are likely underestimated.
The values of equator/pole period encompass our measurement obtained from the periodogram analysis of B$_l$, and fall within the range of literature values \citep{Donati2008,Hebrard2016}. Our d$\Omega$ values are generally lower than estimated in the literature, but compatible within $3\sigma$ from \citet{Donati2008}, and $1\sigma$ from \citet{Hebrard2016}. Such difference could be due to an evolution of the shear at the surface of the star, as studied also by \citet{Donati2023} for AU\,Mic, but deciphering the exact mechanism is not a straightforward task. 
Fig.~\ref{fig:diff_rot_dsleo} also shows that there is an anti-correlation between P$_\mathrm{rot,eq}$ and $d\Omega$ (or equivalently an $\Omega_\mathrm{eq}-d\Omega$ correlation), likely due to the fact that we mainly trace one latitude when searching for differential rotation.

The maps are shown in Figure~\ref{fig:zdi_dsleo} and their properties are reported in Table~\ref{tab:zdi_output}. In 2020b2021a, the magnetic energy is distributed almost equally in poloidal and toroidal components, with the dipolar mode accounting for most of the energy (64\%). While the toroidal component is mostly axisymmetric (90\%), the poloidal component is largely non-axisymmetric (4\%). In 2021b2022a, the poloidal component takes over the toroidal one, counting 73\% of the magnetic energy against 26\%. The dipolar mode remains the dominant one (83\%), featuring a moderate increase similarly to the one reported in \citet{Hebrard2016}. In terms of axisymmetry, the toroidal component is stable, while the poloidal one increases from 4\% to 9\%. Between the two epochs, the average field strength decreased from 44 to 18\,G.

\subsection{CN Leo}

There is no reconstruction of the magnetic field in the literature for this star. From four spectropolarimetric observations, \citet{Morin2010} detected strong Zeeman signatures corresponding to longitudinal fields of 600\,G (as large as the maximum of EV~Lac). The shape of Stokes~$V$ profiles and the absence of intermittency in the amplitude of the profiles suggested an axisymmetric, poloidal dipole.

\begin{figure*}[!t]
    \centering
    \includegraphics[width=0.88\textwidth]{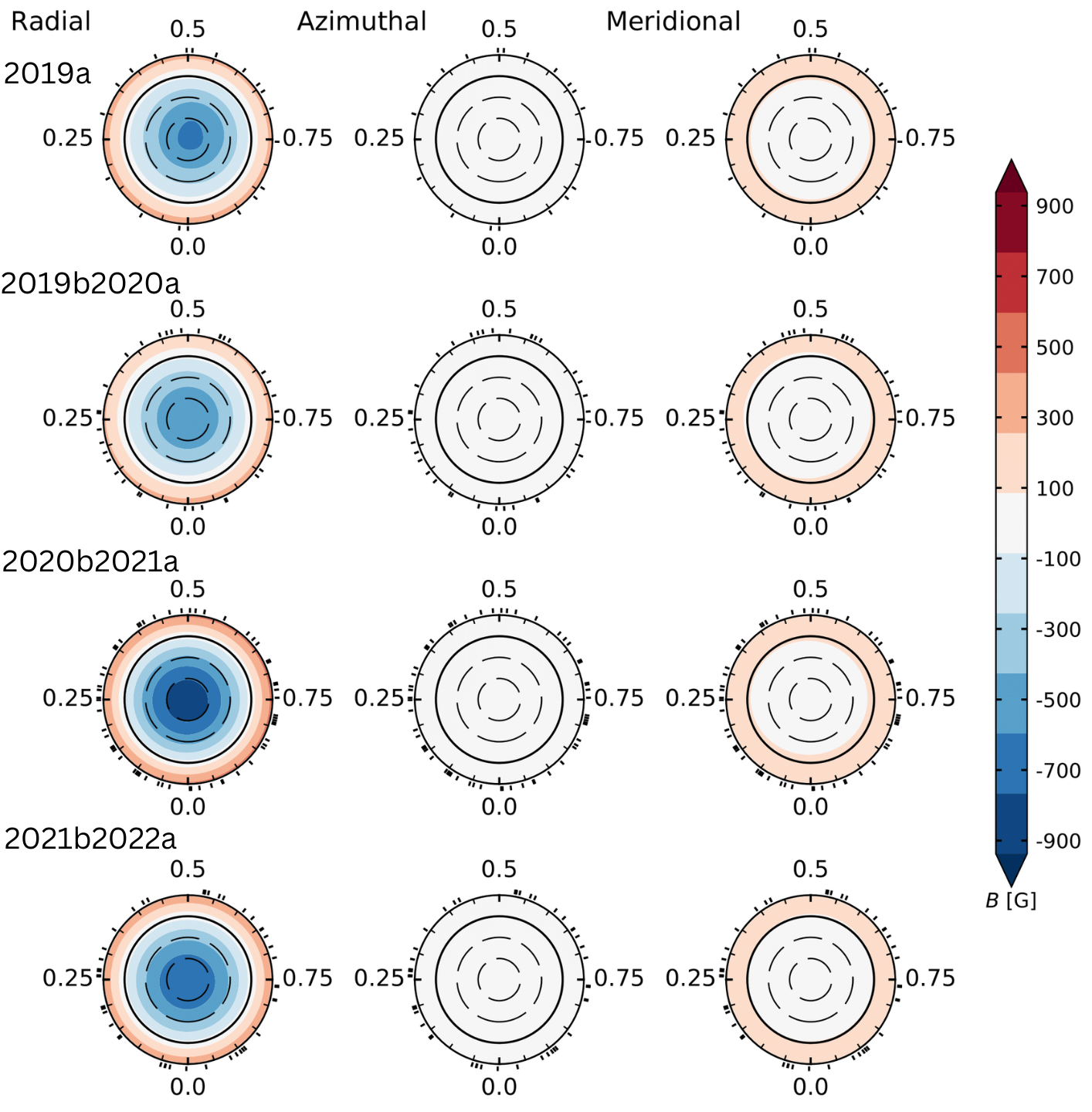}
    \caption{Reconstructed ZDI maps in flattened polar view of CN~Leo. From left to right: 2019a, 2019b2020a, 2020b2021a, and 2021b2022a. The format is the same as in Fig.~\ref{fig:zdi_evlac}.}
    \label{fig:zdi_cnleo}%
\end{figure*}

From the rotation period P$_\mathrm{rot}=2.70\pm0.01$~d we measured (see Sec.~\ref{sec:Blon}) and the equatorial projected velocity $v_\mathrm{eq}\sin i=2\pm1$~km\,s$^{-1}$ estimated by \citet{Cristofari2023}, we obtained $R\sin i=\mathrm{P}_\mathrm{rot}v_\mathrm{eq}\sin i/50.59=0.107\pm0.053R_\odot$, where the denominator accounts for the unit conversion of the variables involved. Considering the radius measurement for CN~Leo by \citet{Rabus2019} of $R=0.151\pm0.005$~$R_\odot$, we infer an inclination of $i=45\pm20^{\circ}$.

The reconstruction of the large-scale magnetic field at the surface of CN~Leo was performed for the 2019a, 2019b2020a, 2020b2021a and 2021b2022a epochs. The input parameters are $i=$\,45$^{\circ}$, $v_\mathrm{eq}\sin i=$\,2.0\,km\,s$^{-1}$, P$_\mathrm{rot}=$\,2.7\,d, and solid body rotation.

The profiles were fitted to a $\chi^2_r$ level of 0.9, 1.0, 1.2, and 1.2 from an initial value of 2.9, 3.5, 6.8, and 6.7 for 2019a, 2019b2020a, 2020b2021a, and 2021b2022a, respectively. The $f_V$ values were constrained to be 16\%, 13\%, 19\% and 16\% for the same epochs, with an error bar of 1\%. 
The Stokes~$V$ time series is shown in Appendix~\ref{app:zdi_app}. 

The maps of the magnetic field and their characteristics are shown in Fig.~\ref{fig:zdi_cnleo} and Table~\ref{tab:zdi_output}. In all epochs, most of the magnetic energy is stored in the poloidal ($>99\%$), dipolar ($>99\%$) and axisymmetric ($>99\%$) components, without any significant variation over time.
The field strength follows the sine-like long-term oscillation exhibited by B$_l$, meaning that it starts from an average field of 350\,G, it decreases to 320\,G, then increases to 490\,G and finally decreases to 420\,G. 

\setlength{\tabcolsep}{5pt}
\begin{table*}[!ht]
\caption{Properties of the magnetic map for EV~Lac, DS~Leo, and CN~Leo, reconstructed from different epochs of SPIRou near-infrared time series.} 
\label{tab:zdi_output}  
\centering    
\centering                       
\begin{tabular}{l | r r r | r r | r r r r}
\hline
& \multicolumn{3}{c}{EV~Lac} & \multicolumn{2}{|c}{DS~Leo} & \multicolumn{4}{|c}{CN~Leo}\\
\hline     
& 2019b & 2020b2021a & 2021b & 2020b2021a & 2021b2022a & 2019a & 2019b2020a & 2020b2021a & 2021b2022a\\ 
\hline
$f_V$ [\%] & $\ldots$ & 9 & 19 & 10 & 6  & 15 & 12 & 18 & 15\\
B$_\mathrm{mean}$ [G]   & 72.5   & 297.7 & 265.1 & 44.3 & 17.9 & 350.2  & 321.8 & 486.9  &420.1\\
B$_\mathrm{max}$ [G]    & 175.8  & 700.0 & 690.5 & 115.2 & 34.9 & 650.4  & 607.8 & 937.6  &820.9 \\
B$_\mathrm{pol}$ [\%]   & 99.1   & 99.7  & 99.2  & 55.9 & 72.9 & 99.9   & 99.9  & 99.9   &99.9\\
B$_\mathrm{tor}$ [\%]   & 0.9    & 0.3   & 0.8   & 44.1 & 27.0 & 0.1    & 0.1   & 0.1    &0.1\\
B$_\mathrm{dip}$ [\%]   & 63.1   & 88.7  & 82.8  & 64.0 & 83.2 & 99.4   & 99.4  & 99.3   &99.2\\
B$_\mathrm{quad}$ [\%]  & 25.4   & 9.4  & 14.6  & 28.5 & 13.1 & 0.5    & 0.6   & 0.6    &0.7\\
B$_\mathrm{oct}$ [\%]   & 10.2    & 1.3   & 2.4   & 3.9  & 2.6  & 0.0    & 0.0   & 0.1    &0.1\\
B$_\mathrm{axisym}$ [\%] & 25.1  & 12.1  & 1.8   & 39.4 & 30.6 & 99.5  & 99.6  & 99.9   &99.9\\
B$_\mathrm{axisym,pol}$ [\%] & 24.6 & 12.0 & 1.2 & 3.9 & 9.2 & 99.5 & 99.6 & 99.9 &99.9 \\
Obliquity [$^{\circ}$] & 50.5 & 68.5 & 86.5 & 75.5 & 69.5 & 4.5 & 3.5 & 1.5 & 1.5 \\
\hline                                 
\end{tabular}
\tablefoot{The following quantities are listed: filling factor on Stokes~$V$; mean magnetic strength; maximum magnetic strength; poloidal and toroidal magnetic energy as a fraction of the total energy; dipolar, quadrupolar and octupolar magnetic energy as a fraction of the poloidal energy; axisymmetric magnetic energy as a fraction of the total energy; and obliquity of the dipolar component relative to the rotation axis. To compute the local magnetic field (i.e. within a grid cell of the ZDI stellar model) the field strength should be divided by the associated filling factor $f_V$.} 
\end{table*}
\setlength{\tabcolsep}{6pt}

\begin{figure*}[!t]
    \centering
    \includegraphics[scale=0.7]{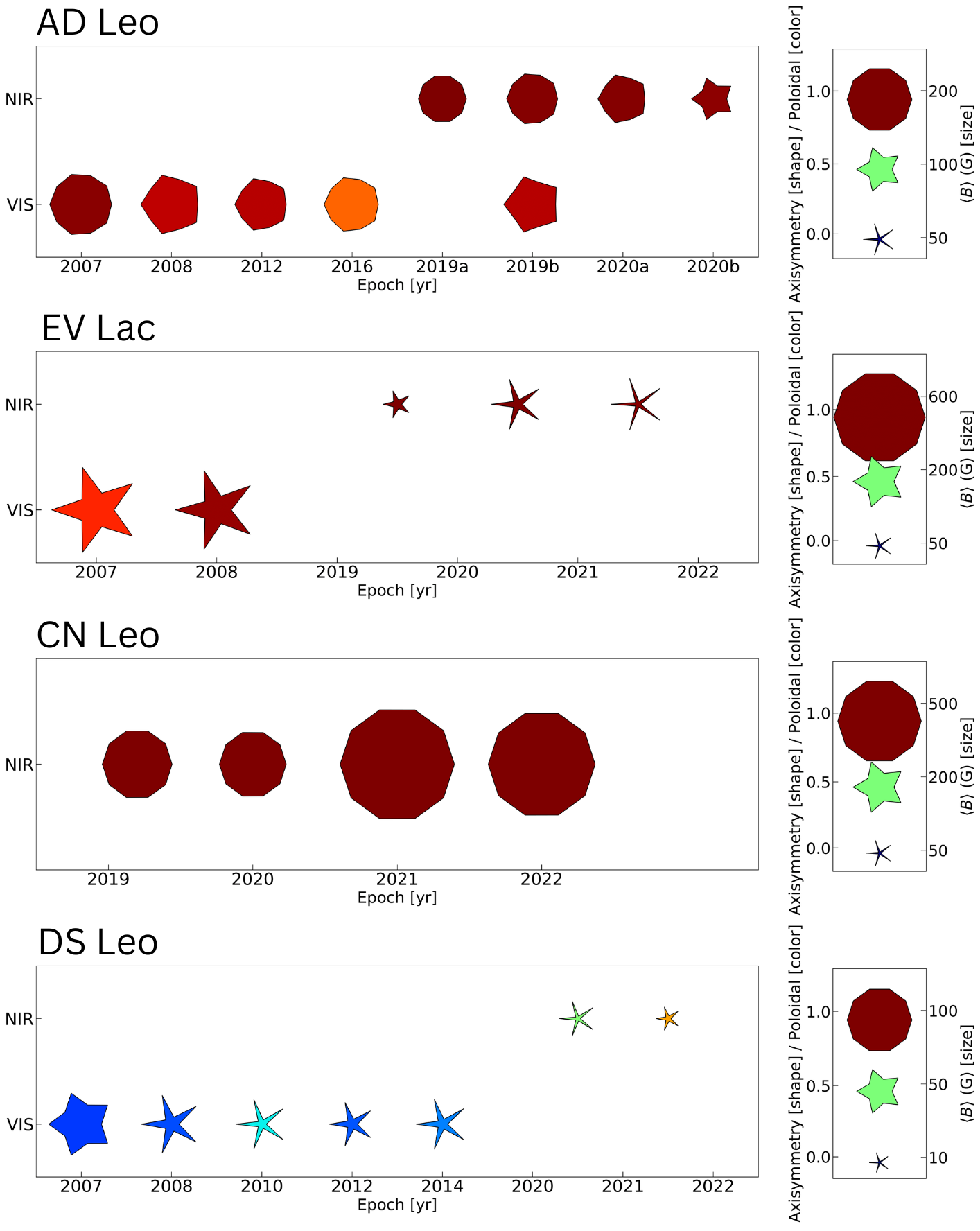}
    \caption{Evolution of the magnetic topology of AD~Leo, EV~Lac, CN~Leo, and DS~Leo. The temporal baseline covers 15~yr, including  the archival optical maps \citep{Donati2008,Morin2008,Hebrard2016} and the near-infrared maps reconstructed with SPIRou data in this work and in \citet{Bellotti2023b} for AD~Leo. The field strength, level of axisymmetry, and dominant field component (poloidal or toroidal) are encoded with the symbol size, shape, and colour, respectively. The red and blue data points represent poloidal- and toroidal-dominated field geometries, whereas round and star-like shapes indicate mostly axisymmetric and non-axisymmetric configurations.}
    \label{fig:zdievol}%
\end{figure*}

\subsection{Summary of the ZDI reconstructions}

The evolution of the large-scale magnetic field for EV~Lac, DS~Leo, and CN~Leo is summarised in Fig.~\ref{fig:zdievol}. For completeness, we also included the evolution of AD~Leo as reconstructed by \citet{Bellotti2023b}, which sees a decrease of field strength accompanied by an increase of the obliquity of the magnetic field axis.

For EV~Lac, the magnetic topology was found to be predominantly non-axisymmetric using optical data \citep{Morin2008}, and we observed an even less axisymmetric field from SPIRou data, with the negative pole of the dipolar component lying approximately at equator. From the SPIRou observations, we also noted a rise in field strength from 57\,G in 2019b to around 200\,G in 2020b2021a and 2021b, following a similar evolution with respect to B$_l$.

For DS~Leo, the poloidal component saw a progressive increase with respect to the toroidal one (from 20 to 70\% of the total magnetic energy), and it remained highly non-axisymmetric throughout \citep{Donati2008,Hebrard2016}. The complexity of the poloidal component of the field has also decreased in the same timescale, since the field is predominantly dipolar. Moreover, the average reconstructed field strength has reduced, from an initial 100\,G in 2007 to 15\,G in the 2021b2022a SPIRou epoch. The panel of DS~Leo in Fig.~\ref{fig:zdievol} includes two field topologies, in 2010 and 2012, which we reconstructed using unpublished Narval data (see Appendix~\ref{app:dsleo_app}).

For CN~Leo, there is no substantial evolution in four years, the topology being mainly poloidal, dipolar and axisymmetric. The only feature is a varying field strength, whose oscillations correlate with the longitudinal field evolution (in absolute value). 

Similarly to what reported by \citet{Bellotti2023b}, we caution that the field strength reported may be underestimated, owing to the limitation of the ZDI model at reproducing the shape of the Stokes~$V$ lobes (see e.g.  Fig.~\ref{fig:zdi_stokesV_evlac}). The near-infrared Stokes~$V$ profiles manifest evident noise that is not rotationally modulated, and which prevents the ZDI model to capture all the information present in the profile. 

{ To describe the magnetic field vector, we employed the formalism of \citet{Lehmann2022}, which sees the substitution of $\beta_{lm}$ with $\alpha_{lm}+\beta_{lm}$ in the spherical harmonics equations. The main effect of using the latter is to change the appearance of the meridional and azimuthal magnetic field maps, making them resemble the radial field map, and avoid unnecessary large values of $\beta_{lm}$. If we use the standard formalism \citep[see e.g.][]{Donati2006} with $\alpha_{lm}$ and $\beta_{lm}$ disjointed, the reconstructed maps of EV~Lac vary less than 3\% in poloidal energy fraction, less than 5\% in dipolar, quadrupolar, and octupolar components, and less than 2\% in axisymmetric fraction. The average field strength decreases by at most 100\,G. For CN~Leo, the variations are similar or lower. This is in agreement with what was previously found also for AD~Leo \citep[see Appendix of][]{BellottiPhD2023}. Overall, the properties of the magnetic topology and its long-term evolution are reconstructed consistently adopting either of the two spherical harmonics formalisms.}

\section{Discussion and conclusions}\label{sec:conclusions}

In this work we presented the long-term spectropolarimetric monitoring of three well-studied, active M dwarfs: EV~Lac, DS~Leo, and CN~Leo. We used archival optical data collected with ESPaDOnS and Narval, as well as near-infrared SPIRou data obtained as part of the Legacy Survey between 2019 and 2022. We carried out distinct analyses to capture the evolution of the magnetic field, employing the longitudinal field as a proxy of the large-scale field component and the FWHM of Stokes~$I$ LSD profiles as a proxy for the small-scale component. We analysed qualitatively the secular evolution of the large-scale magnetic field via principal component analysis and reconstructed its topology by means of Zeeman-Doppler imaging.

We found different trends in the magnetic field behaviour, potentially hinting at a variety of magnetic cycles. Our conclusions are the following:

\begin{enumerate}
    \item The longitudinal magnetic field analysis showed a pulsating trend for EV~Lac with variations between $\pm$400\,G around 2007, then between $\pm$200\,G in 2016, and finally between $\pm$300\,G in 2020. However, the dearth of observations between 2010 and 2019 prevent us from drawing firm conclusions on this behaviour. DS~Leo did not manifest any specific trend as the mean field remained reasonably stable around 0\,G, and the measurements within $\pm$50\,G. CN~Leo exhibited a sine-like trend, characterised by a period of 2.7\,yr, an amplitude of 100\,G, and an average of $-$500\,G.
    \item We observed rotational modulation of the FWHM of Stokes~$I$ for optical observations of EV~Lac and DS~Leo, and near-infrared observations of DS~Leo, likely stemming from distinct contributions of the Zeeman effect in the two wavelength domains.  The FWHM of near-infrared line profiles for stars with intense magnetic fields is expected to be more impacted than at optical wavelengths, resulting in larger scatter in the time series.
    \item For the optical epochs of EV~Lac and DS~Leo where the rotational modulation of the FWHM is evident, we observed both negative and positive correlations with $|$B$_l|$, whereas in some cases this correlation was absent. The interpretation of these results is not straightforward. They indicate an underlying complexity in the link between the small-scale field (traced by FWHM) and the large-scale field (traced by B$_l$), which may not be encapsulated in a simple scaling.
    \item The long-term behaviour of the epoch-averaged FWHM of Stokes~$I$ reflects the secular evolution of the longitudinal field. This is enhanced in particular when high-g$_\mathrm{eff}$ lines are used, whereas low-g$_\mathrm{eff}$ lines tend to yield stable FWHM values throughout the time series.
    \item Using PCA, we obtained the variations in axisymmetry, poloidal-to-toroidal fraction of the axisymmetric component, and complexity for EV~Lac, DS~Leo, and CN~Leo directly from the LSD Stokes~$V$ time series, without prior assumptions on stellar parameters or the need of elaborate field topology models. 
    \item The ZDI reconstructions also revealed distinct secular changes in the magnetic topology. For EV~Lac the field remained mainly poloidal, dipolar, and with a significant contribution of quadrupolar field. The axisymmetry has decreased almost to 0\% in the most recent epoch. For DS~Leo, the poloidal component increased over time with a simultaneous decrease in axisymmetry. The toroidal component has a substantial contribution to the energy budget still, and is mainly axisymmetric, similarly to the optical reconstructions \citep{Donati2008,Hebrard2016}. For CN~Leo, the field maintained a poloidal, dipolar, and axisymmetric configuration, with the field strength following a fluctuating trend similar to B$_l$.
\end{enumerate}

If we group the long-term evolution of the large-scale field geometry according to spectral type, it seems that the cyclic variability decreases towards later spectral types, as also pointed out by \citet{Fuhrmeister2023} from a long-term monitoring of chromospheric activity indicators. From our work we see that the field of DS~Leo (early M type) undergoes rapid variations on short timescales; the field of EV~Lac still manifests an evident topological change, but on longer timescales (similar to AD~Leo, \citealt{Bellotti2023b}); and for CN~Leo (late M type), there is no substantial evolution. This phenomenon could be correlated to the magnetic field intensity, with stronger fields (i.e. later spectral types) being able to quench surface shear and induce intrinsic stability, as already pointed out by \citet{Donati2008} and \citet{Morin2008}. 

With the exception of DS~Leo, activity cycles were reported for EV~Lac and CN~Leo from photometric or chromospheric activity monitoring. For EV~Lac, \citet{Mavridis1986} constrained a 5 yr activity cycle based on flare activity and $B$-band photometric oscillations over a baseline of 9\,yr. Given the gap in our B$_l$ time series between 2010 and 2016, we cannot easily compare the reported timescale with our B$_l$ variations. For CN~Leo, photometric analysis of the All Sky Automated Survey (ASAS) light curves revealed an activity cycle of 8.9$\pm$0.2\,yr \citep{SuarezMascareno2016} or 12.0$\pm$4.4\,yr \citep{Irving2023}. These are 3.3 and 4.4 times longer than the 2.7 yr sine-like variations we recorded for B$_l$. In contrast, \citet{Fuhrmeister2023} and \citet{Mignon2023} did not report any significant long-term periodicity from chromospheric activity monitoring of CN~Leo. These result may resemble what is described for Sun-like stars because there are cases in which magnetic cycles and polarity flips are phased with chromospheric cycles \citep{BoroSaikia2016,Jeffers2017}, whereas in other cases regular chromospheric oscillations are not reflected in polarimetric variations \citep{BoroSaikia2022}. At the same time, different activity indexes may be sensitive to different regions in the stellar atmospheres, and therefore timescales, which may explain why they did not capture long-term variability for CN~Leo \citep[e.g.][]{CortesZuleta2023,Mignon2023}.

In conclusion, the peculiar evolution of the large-scale magnetic field for the stars examined here provides us with a glimpse of a potential variety in cyclic trends for M~dwarfs. This advocates for additional spectropolarimetric monitoring of active M~dwarfs for a better constraint on the timescales involved and the extent of such variety over distinct stellar masses and rotation periods.

\begin{acknowledgements}

We thank the anonymous referee for the thorough review of the manuscript and the suggestions. We acknowledge funding from the French National Research Agency (ANR) under contract number ANR-18-CE31-0019 (SPlaSH). SB acknowledges funding from the European Space Agency (ESA), under the visiting researcher programme. XD acknowledges funding within the framework of the Investissements d'Avenir programme (ANR-15-IDEX-02), through the funding of the `Origin of Life' project of the Univ. Grenoble-Alpes. EM acknowledges funding from FAPEMIG under project number APQ-02493-22and research productivity grant number 309829/2022-4 awarded by the CNPq, Brazil. BK acknowledges funding from the European Research Council (ERC) under the European Union's Horizon 2020 research and innovation programme (Grant agreement No. 865624). Based on observations obtained at the Canada-France-Hawaii Telescope (CFHT) which is operated by the National Research Council (NRC) of Canada, the Institut National des Sciences de l'Univers of the Centre National de la Recherche Scientifique (CNRS) of France, and the University of Hawaii. The observations at the CFHT were performed with care and respect from the summit of Maunakea which is a significant cultural and historic site. We gratefully acknowledge the CFHT QSO observers who made this project possible. This work has made use of the VALD database, operated at Uppsala University, the Institute of Astronomy RAS in Moscow, and the University of Vienna; Astropy, 12 a community-developed core Python package for Astronomy \citep{Astropy2013,Astropy2018}; NumPy \citep{VanderWalt2011}; Matplotlib: Visualization with Python \citep{Hunter2007}; SciPy \citep{Virtanen2020}. 

\end{acknowledgements}

%
%

\bibliographystyle{aa}
\bibliography{biblio}

\begin{appendix}

\section{ZDI analysis of unpublished DS~Leo data}\label{app:dsleo_app}

\begin{figure}
    \centering
    \includegraphics[width=\columnwidth]{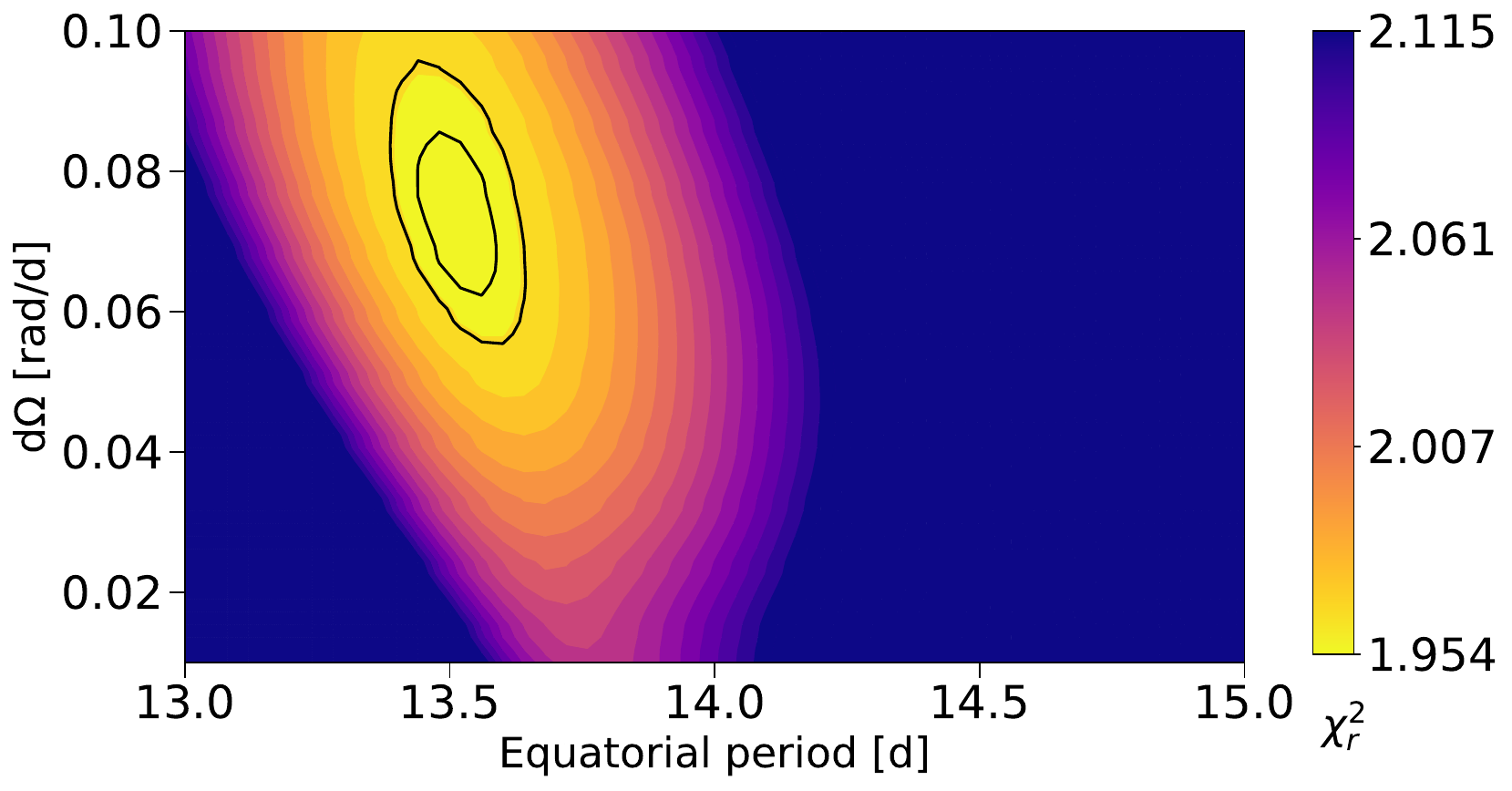}
    \caption{Differential rotation search for DS~Leo for the 2010 Narval data set. The plots show the $\chi^2_r$ landscape over a grid of (P$_\mathrm{rot,eq}$,$d\Omega$) pairs, with the same format as in Fig.~\ref{fig:diff_rot_dsleo}.}
    \label{fig:diff_rot_dsleo_app}%
\end{figure}

In this appendix, we provide the ZDI reconstruction of the large-scale magnetic field of DS~Leo for two Narval epochs with unpublished data, namely 2010 and 2012. The input parameters were $i=$\,60$^{\circ}$ and $v_\mathrm{eq}\sin(i)=$\,2\,km\,s$^{-1}$ for both epochs. For 2010, we successfully constrained differential rotation with a procedure similar to the one described in Sec.~\ref{sec:magnetic_imaging}. As shown in Fig.~\ref{fig:diff_rot_dsleo_app}, the optimised parameters are P$_\mathrm{rot,eq}=13.51\pm0.08$\,d and $d\Omega=0.074\pm0.012$\,rad\,d$^{-1}$. These values are compatible with those estimated from Narval 2008 observations \citep{Donati2008} and 2014 observations \citep{Hebrard2016} within 1$\sigma$, and are 3.9 and 1.7 times larger than the SPIRou 2020b2021a and 2021b2022a values, respectively. This confirms that the differential rotation rate of DS~Leo may be transitioning to slower values over time. For the 2012 Narval epoch, the differential rotation search is inconclusive, hence we fixed the stellar input parameters for ZDI to P$_\mathrm{rot}=$\,13.91\,d and $d\Omega=0.0$\,rad\,d$^{-1}$. The maximum degree of the harmonic expansion was set to $l_\mathrm{max}=$\,8, and the linear limb darkening coefficient to 0.6964 \citep{Claret2011}. 

\begin{figure}[!t]
    \centering
    \includegraphics[width=\columnwidth]{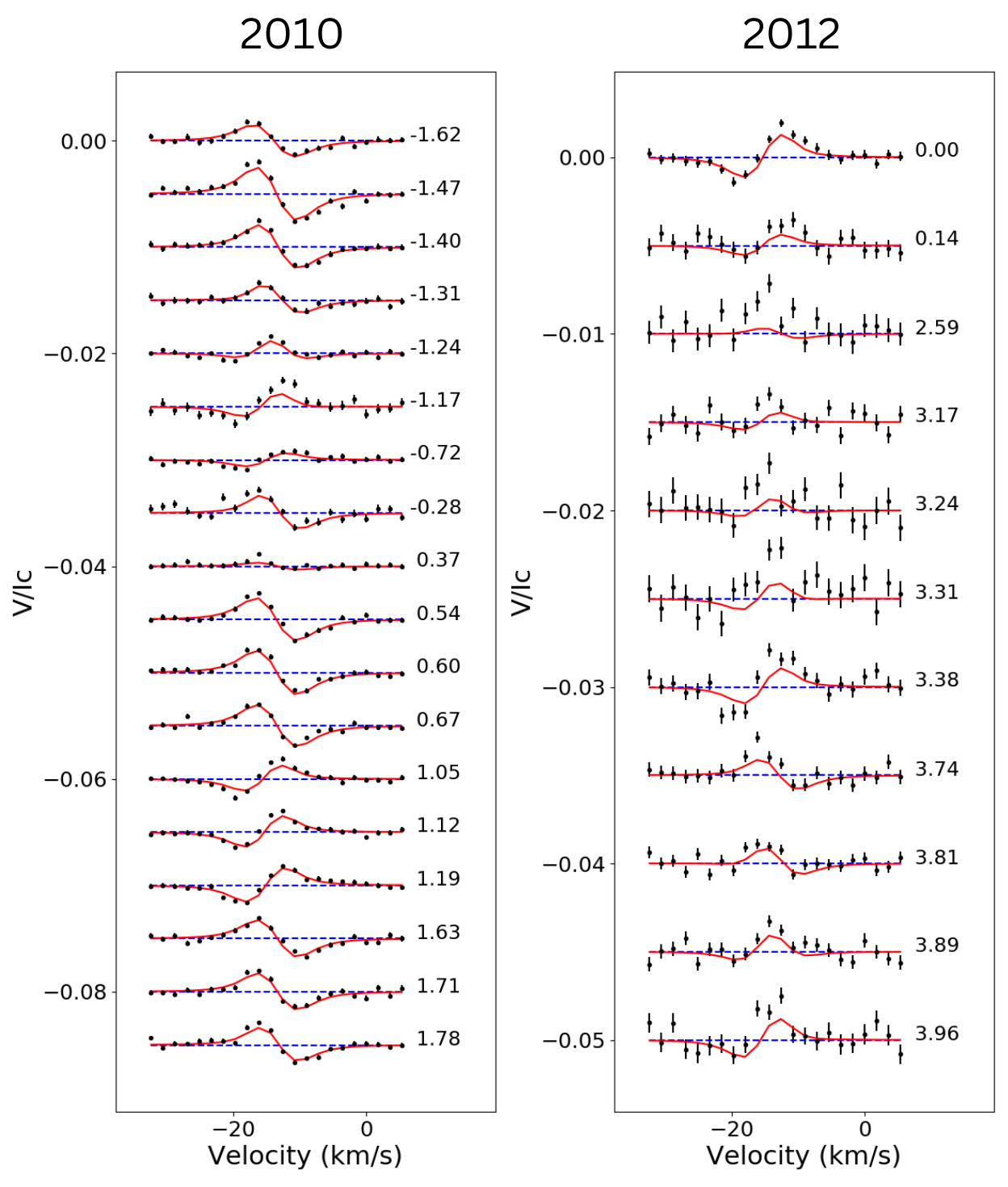}
    \caption{Narval time series of Stokes~$V$ profiles of DS~Leo. Top left: 2010. Bottom: 2012. The cycles in this plot are computed with Eq.~\ref{eq:ephemeris} while using the median HJD for each epoch. The format is the same as in Fig.~\ref{fig:zdi_stokesV_evlac}.}
    \label{fig:zdi_stokesV_dsleo_app}%
\end{figure}

The profiles are shown in Fig.~\ref{fig:zdi_stokesV_dsleo_app}, where we observe clear Zeeman signatures in 2010 and noisier Stokes~$V$ profiles in 2012. While we were not able to optimise the filling factor $f_V$ for the 2010 epoch, we constrained it to be $f_V=14\%$ for the 2012 epoch. However, setting $f_V=100\%$ for the reconstruction of the 2012 time series does not alter the map appreciably, hence we proceeded with $f_V=100\%$ for both 2010 and 2012 Stokes~$V$ time series. The profiles were fitted to a $\chi^2_r$ level of 1.9 from an initial value of 15.8 and 3.7 for 2010 and 2012, respectively. 

\begin{figure}[!t]
    \centering
    \includegraphics[width=\columnwidth]{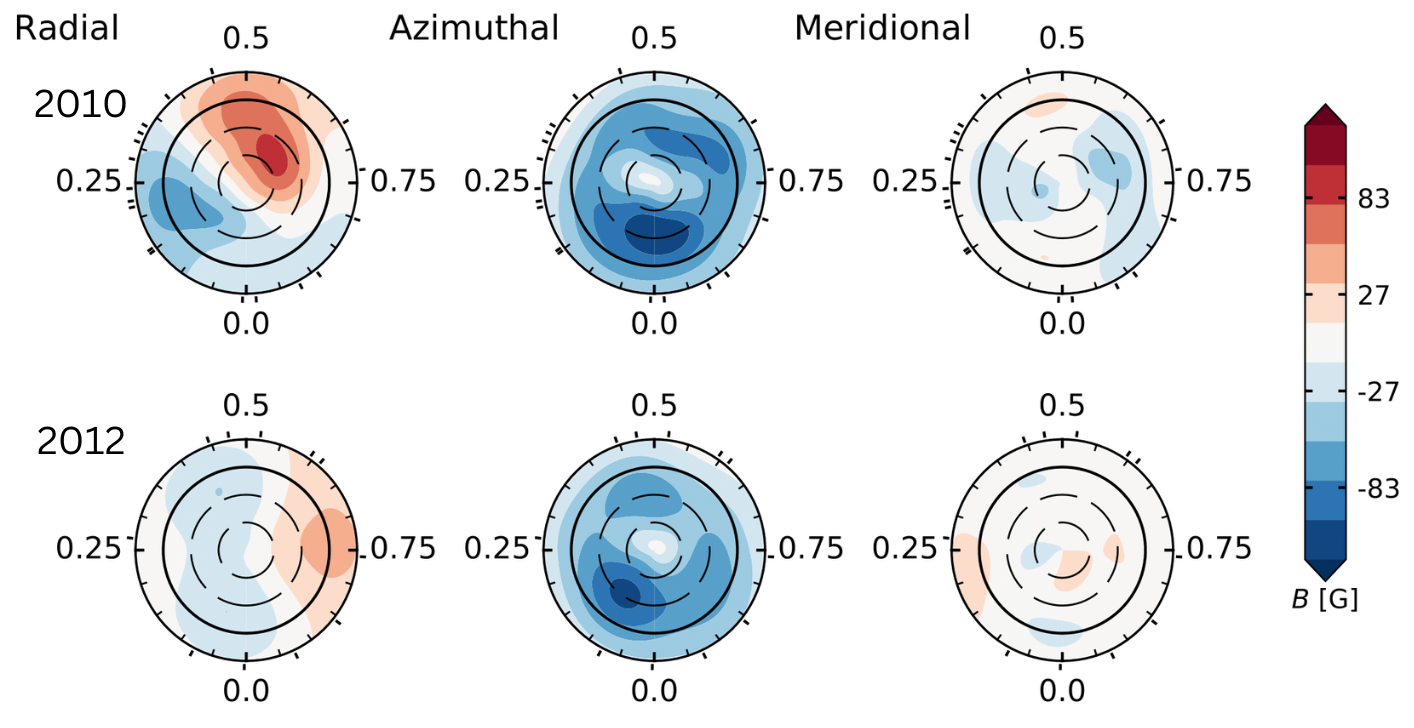}
    \caption{Reconstructed ZDI maps in flattened polar view of DS~Leo. Left: 2010. Right: 2012. The 2010 map accounts for the constrained differential rotation. The format is the same as in Fig.~\ref{fig:zdi_evlac}.}
    \label{fig:zdi_dsleo_app}%
\end{figure}

The maps are given in Fig.~\ref{fig:zdi_dsleo_app} and the corresponding field characteristics are reported in Table~\ref{tab:zdi_output_dsleo_app}. Overall, the characteristics reveal a complex topology, consistently with previous field reconstructions \citep{Donati2008,Hebrard2016}. For both 2010 and 2012, the field is predominantly toroidal ($>60\%$), and the poloidal component is mostly dipolar ($>50\%$), with a significant quadrupolar component (21 and 36\% for 2010 and 2012, respectively). The poloidal component is also mostly non-axisymmetric, accounting for 9\% of the magnetic energy in 2010 and 15\% in 2012. The mean field strength is approximately 64 and 49\,G for 2010 and 2012, compatible with the decrease in field strength reported by \citet{Hebrard2016} for the 2014 Narval epoch.

\begin{table}[b]
\caption{Properties of the magnetic map for the 2010 and 2012 Narval unpublished epochs.} 
\label{tab:zdi_output_dsleo_app}     
\centering                       
\begin{tabular}{l c c}    
\hline
& 2010 & 2012\\
\hline     
B$_\mathrm{mean}$ [G]   & 64.7 & 49.4\\
B$_\mathrm{max}$ [G]    & 129.5 & 113.9\\
B$_\mathrm{pol}$ [\%]   & 36.7 & 21.1\\
B$_\mathrm{tor}$ [\%]   & 63.3 & 78.9\\
B$_\mathrm{dip}$ [\%]   & 71.6 & 47.3\\
B$_\mathrm{quad}$ [\%]  & 21.5 & 36.8\\
B$_\mathrm{oct}$ [\%]   & 4.2 & 14.0\\
B$_\mathrm{axisym}$ [\%]& 64.7 & 81.8\\
B$_\mathrm{axisym,pol}$ [\%] & 8.8 & 15.1\\
\hline                                 
\end{tabular}
\tablefoot{The following quantities are listed: mean magnetic strength; maximum magnetic strength; poloidal and toroidal magnetic energy as a fraction of the total energy; dipolar, quadrupolar, and octupolar magnetic energy as a fraction of the poloidal energy; axisymmetric magnetic energy as a fraction of the total energy.}
\end{table}

\section{Additional FWHM analyses and figures}\label{app:fwhm_app}

In Fig.~\ref{fig:FWHM_geff_highlow}, we show the different Stokes~$I$ profiles for EV~Lac, DS~Leo and CN~Leo when computed with magnetically sensitive (g$_\mathrm{eff}>1.2$) and insensitive (g$_\mathrm{eff}<1.2$) lines. These profiles were used for the FWHM analyses described in Sec.~\ref{sec:FWHM}.

\subsection{Proxy of the unsigned magnetic field}

\begin{figure}[t]
    \centering
    \includegraphics[width=\columnwidth]{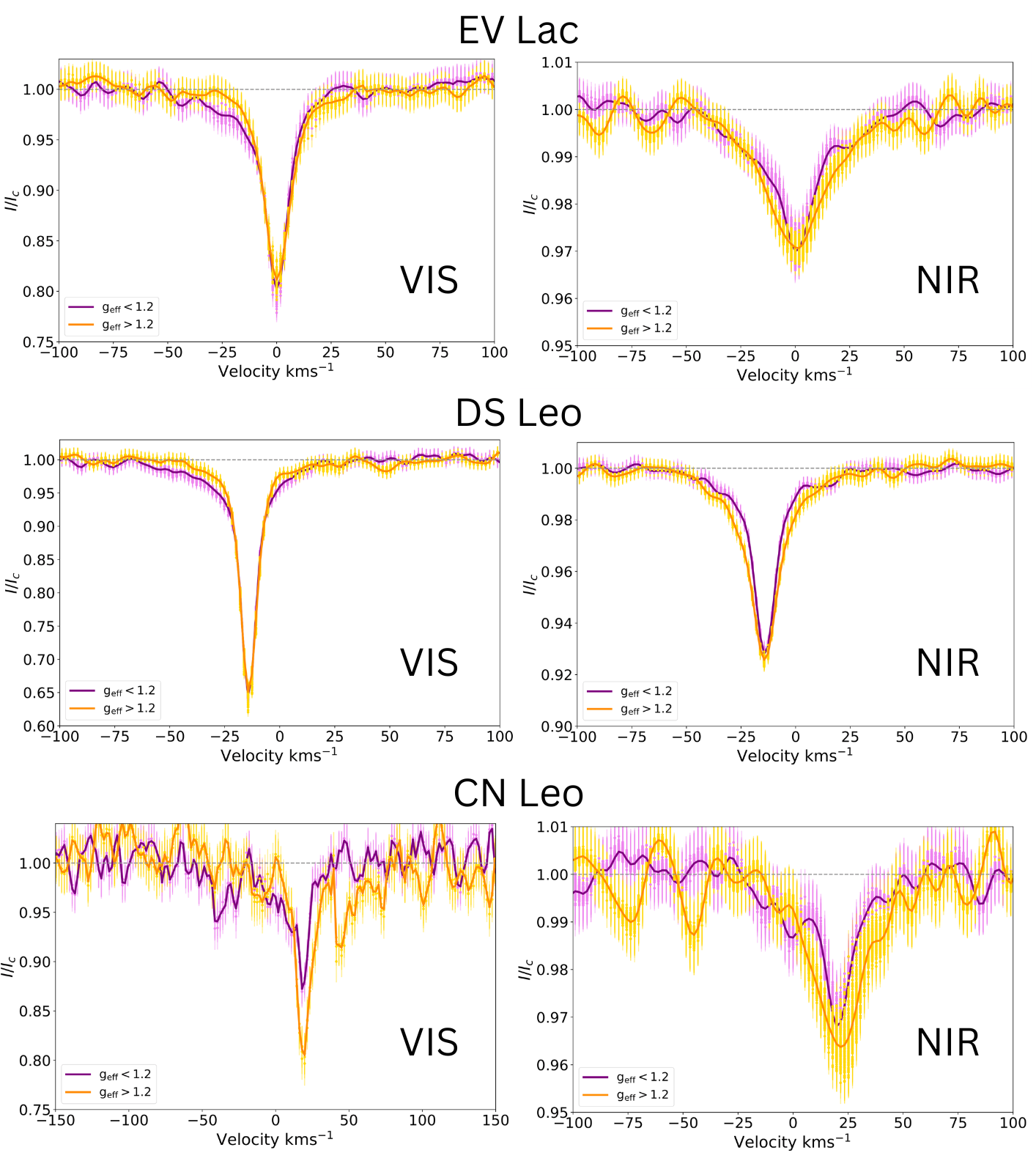}
    \caption{Comparison of Stokes~$I$ profiles for EV~Lac, DS~Leo, and CN~Leo when computed with different line selections based on magnetic sensitivity. Each panel contains all the observations and their median for high-g$_\mathrm{eff}$  (yellow) and low-g$_\mathrm{eff}$ (purple) line selections. }
    \label{fig:FWHM_geff_highlow}%
\end{figure}

For all three stars, the mean FWHM and the dispersion of individual epochs is larger when employing high-g$_{\rm eff}$ lines than the full and the low-g$_{\rm eff}$ mask, owing to a larger influence of the Zeeman effect on the broadening of the lines. The values are reported in Table~\ref{tab:fwhm_quadbro} and an illustration of the different Stokes~$I$ profiles for the two line selections is given in Fig.~\ref{fig:FWHM_geff_highlow}. Taking the epoch-averaged Stokes~$I$ profiles, we computed the quadratic differential broadening between the low- and high-g$_{\rm eff}$ masks. 

For example, the mean FWHM for EV~Lac in the SPIRou 2019b epoch when using low- and high-g$_{\rm eff}$ is 17 and 29\,km\,s$^{-1}$, respectively. The corresponding quadratic differential broadening is $\delta v_B=24$~km\,s$^{-1}$. We then solve the following formula
\begin{equation}\label{eq:zeeman}
    \delta v_B = 1.4\cdot10^{-6}\lambda_0g_\mathrm{eff}B,
\end{equation}
for $B$, with $\delta v_B$ in km\,s$^{-1}$, $\lambda_0$ in nm and $B$ in { G}. The value of $B$ { obtained from Eq.\ref{eq:zeeman}} is { then} divided by two because the Zeeman effect is symmetric, acting in both the positive and negative direction with respect to line centre. The above value of differential quadratic broadening translates into $B=$5.4\,kG. The wavelength $\lambda_0$ is the normalisation wavelength of the low-Land\'e lines (700\,nm for optical and 1700\,nm for near-infrared) and $g_\mathrm{eff}$ the normalisation Land\'e factor. For EV~Lac, the latter is 0.85 in optical and 0.95 in near-infrared, for DS~Leo is 0.82 and 0.94, and for CN~Leo it is 0.98 in near-infrared. The proportionality factor encapsulates fundamental constants such as speed of light, electron mass and electron charge \citep{Zeeman1897,Landi1992}.

For EV~Lac, we obtained $B$ within 4.9-5.4\,kG in near-infrared and 4.2-4.8\,kG in optical, which are similar to the estimates reported in the literature from Zeeman broadening and intensification modelling \citep{Saar1994,Shulyak2017,Shulyak2019}. For DS~Leo, $B$ is 2\,kG and 2.5\,kG for optical and near-infrared. For CN~Leo, $B$ is 4.9\,kG for optical and 3.7-4.7\,kG for near-infrared. The values for DS~Leo and CN~Leo are larger than the literature estimates, that is, twice for DS~Leo \citep{Shulyak2011} and 1.6 times for CN~Leo \citep{Shulyak2019}. These discrepancies are unlikely to be associated with a magnetic field evolution, as we did not capture a significant temporal change of FWHM (see Sec.~\ref{sec:longfwhm}). Instead, they probably stem from the fact that, for instance, the identification of the continuum in the vicinity of the lines is not straightforward, affecting the modelling from which the FWHM is estimated. Identifying the continuum is complicated by the unaccounted molecular lines in the LSD line list and the lower S/N due to the low number of spectral lines used in LSD, as displayed in Fig.~\ref{fig:FWHM_geff_highlow}. For DS~Leo in near-infrared, we also carried out this exercise starting from a 4000\,K mask, but we found values of $B$ around 1.7\,kG, thus still higher than the literature values.

\begin{table*}[ht]
\caption{Unsigned magnetic field estimates from differential broadening measurements for EV~Lac, DS~Leo, and CN~Leo.} 
\label{tab:fwhm_quadbro}     
\centering                       
\begin{tabular}{c c c c c c}       
\hline     
Epoch & Domain & $\langle \textrm{FWHM}_{\textrm{geff}<1.2}\rangle$ & $\langle \textrm{FWHM}_{\textrm{geff}>1.2}\rangle$ & $\delta\textrm{FWHM}$ & $B$\\
& & [km\,s$^{-1}$] & [km\,s$^{-1}$] & [km\,s$^{-1}$] & [kG]\\
\hline
\multicolumn{6}{c}{EV~Lac}\\
\hline
2005 & VIS & 11.5 & 14.1 & 8.3 & 4.8\\
2006 & VIS & 11.7 & 13.9 & 7.4 & 4.4\\
2007 & VIS & 11.6 & 13.7 & 7.1 & 4.2\\
2010 & VIS & 12.0 & 14.6 & 8.3 & 4.8\\
2012 & VIS & 12.0 & 14.1 & 7.5 & 4.4\\
2019b & NIR & 16.6 & 29.3 & 24.1 & 5.4\\
2020b2021a & NIR & 16.2 & 28.4 & 23.3 & 5.2\\
2021b & NIR & 16.2 & 27.5 & 22.2 & 4.9\\
\hline
\multicolumn{6}{c}{DS~Leo}\\
\hline
2006 & VIS & 8.4 & 9.4 & 4.2 & 2.6\\
2007 & VIS & 8.2 & 8.9 & 3.5 & 2.2\\
2008 & VIS & 8.2 & 9.0 & 3.6 & 2.2\\
2009 & VIS & 8.2 & 9.1 & 3.9 & 2.4\\
2010 & VIS & 8.2 & 9.1 & 3.9 & 2.4\\
2011 & VIS & 8.2 & 9.0 & 3.8 & 2.2\\
2012 & VIS & 8.0 & 9.2 & 4.4 & 2.8\\
2014 & VIS & 8.1 & 9.0 & 4.0 & 2.5\\
2020b2021a & NIR & 11.2 & 14.4 & 9.0 & 2.0\\
2021b2022a & NIR & 11.2 & 14.3 & 8.9 & 2.0\\
\hline
\multicolumn{6}{c}{CN~Leo}\\
\hline
2008 & VIS & 7.0 & 12.1 & 9.9 & 4.9\\
2019a & NIR & 11.9 & 20.9 & 17.3 & 3.7\\
2019b2020a & NIR & 11.9 & 24.9 & 21.9 & 4.7\\
2020b2021a & NIR & 12.1 & 21.6 & 17.8 & 3.8\\
2021b2022a & NIR & 12.1 & 21.8 & 18.1 & 3.9\\
\hline  
\end{tabular}
\tablefoot{The columns are: 1) epoch of the observations, 2) wavelength domain, 3) average FWHM of low-g$_\mathrm{eff}$ lines, 4) average FWHM of high-g$_\mathrm{eff}$ lines, 5) quadratic differential broadening, and 6) total unsigned magnetic field.}
\end{table*}

\subsection{Complementary figures to the FWHM analysis} 

We also present the analysis investigating the rotational modulation of the FWHM obtained with the different masks, in optical and near-infrared, for EV~Lac and CN~Leo. For EV~Lac, we observe rotational modulation in the optical epochs, whereas in near-infrared we cannot capture such behaviour due to a larger dispersion of the data. Likewise, CN~Leo's near-infrared time series of FWHM does not reveal rotational modulation (see Sec.~\ref{sec:FWHM} for a detailed description). We finally illustrate the correlation analysis between FWHM and $|$B$_l|$ for DS~Leo, for which we note a variety of correlations depending on the epoch considered.

\begin{figure*}[t]
    \centering
    \includegraphics[width=0.94\textwidth]{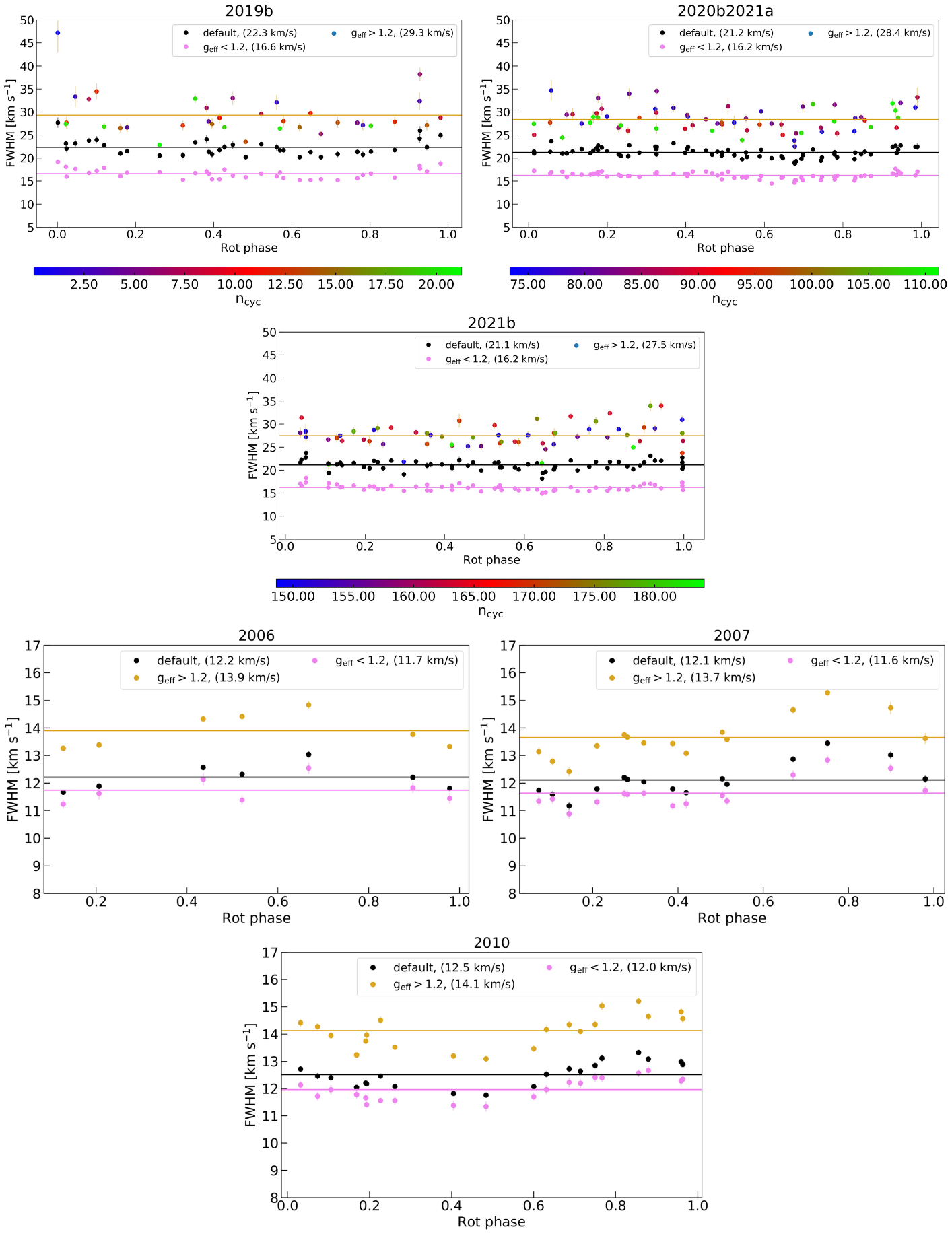}
    \caption{Rotational modulation analysis of FWHM for EV~Lac. The three upper panels correspond to the near-infrared epochs (2019b, 2020b2021a, and 2021b), whereas the three lower panels correspond to the optical epochs containing enough data points (2006, 2007, and 2010). Plotted in  each panel is   the phase-folded time series of FWHM computed with the full, low-g$_\mathrm{eff}$, and high-g$_\mathrm{eff}$ mask with a horizontal line representing the mean FWHM value. For the near-infrared epochs the high-g$_\mathrm{eff}$ time series is colour-coded by rotational cycle to inspect signs of short-term variability. The latter can be seen as data points that are at significantly different values of FWHM at different cycles, but sharing the same rotational phase (e.g. phase 0.05 or 0.30 of the 2020b2021a epoch).}
    \label{fig:FWHM_rotmod_app_evlac}%
\end{figure*}

\begin{figure*}[t]
    \centering
    \includegraphics[width=\textwidth]{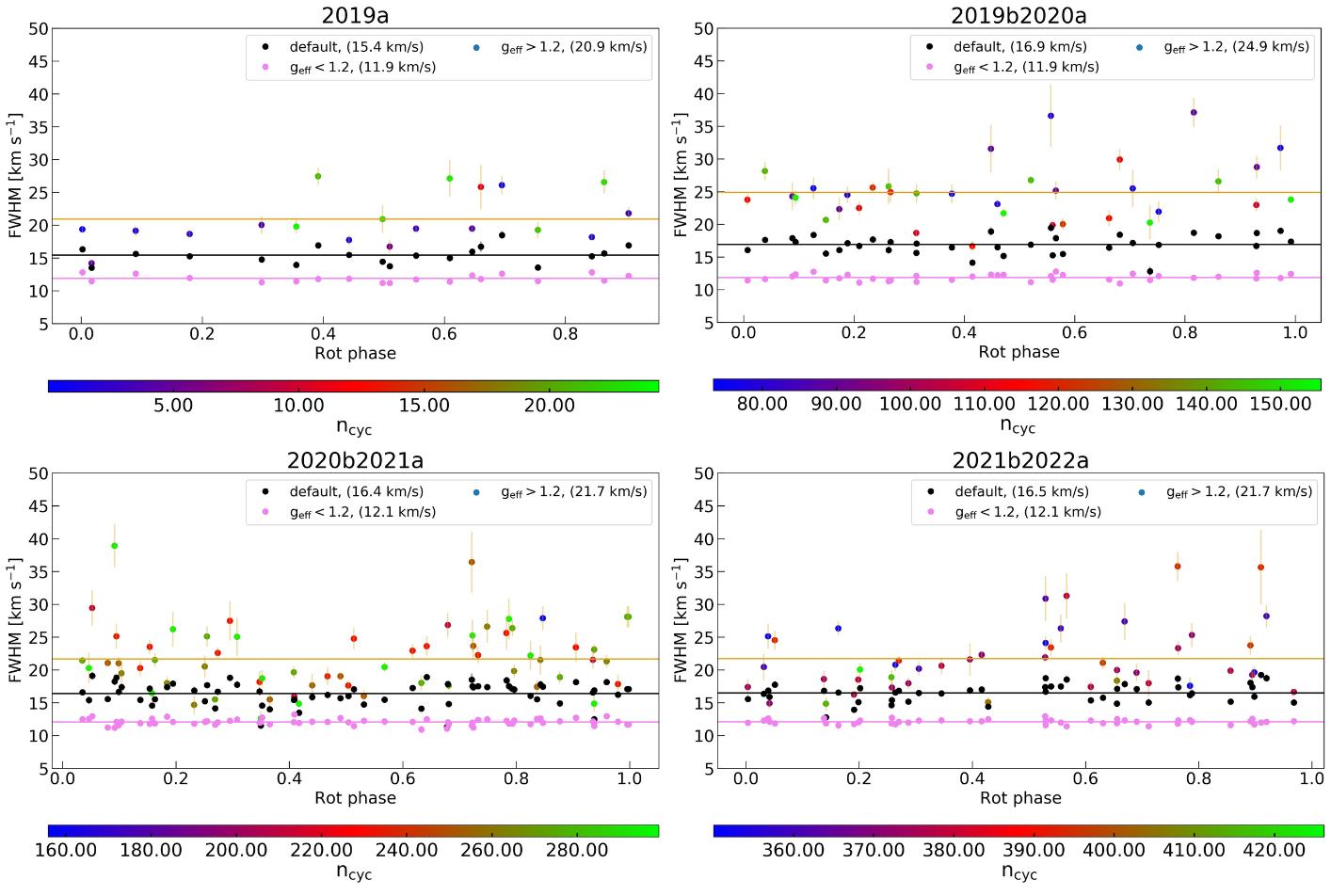}
    \caption{Rotational modulation analysis of FWHM for CN~Leo. The panels illustrate the near-infrared epochs: 2019a, 2019b2020a, 2020b2021a, and 2021b2022a. The  format is the same as in Fig.~\ref{fig:FWHM_rotmod_app_evlac}.}
    \label{fig:FWHM_rotmod_app_cnleo}%
\end{figure*}

\begin{figure*}[t]
    \centering
    \includegraphics[width=\textwidth]{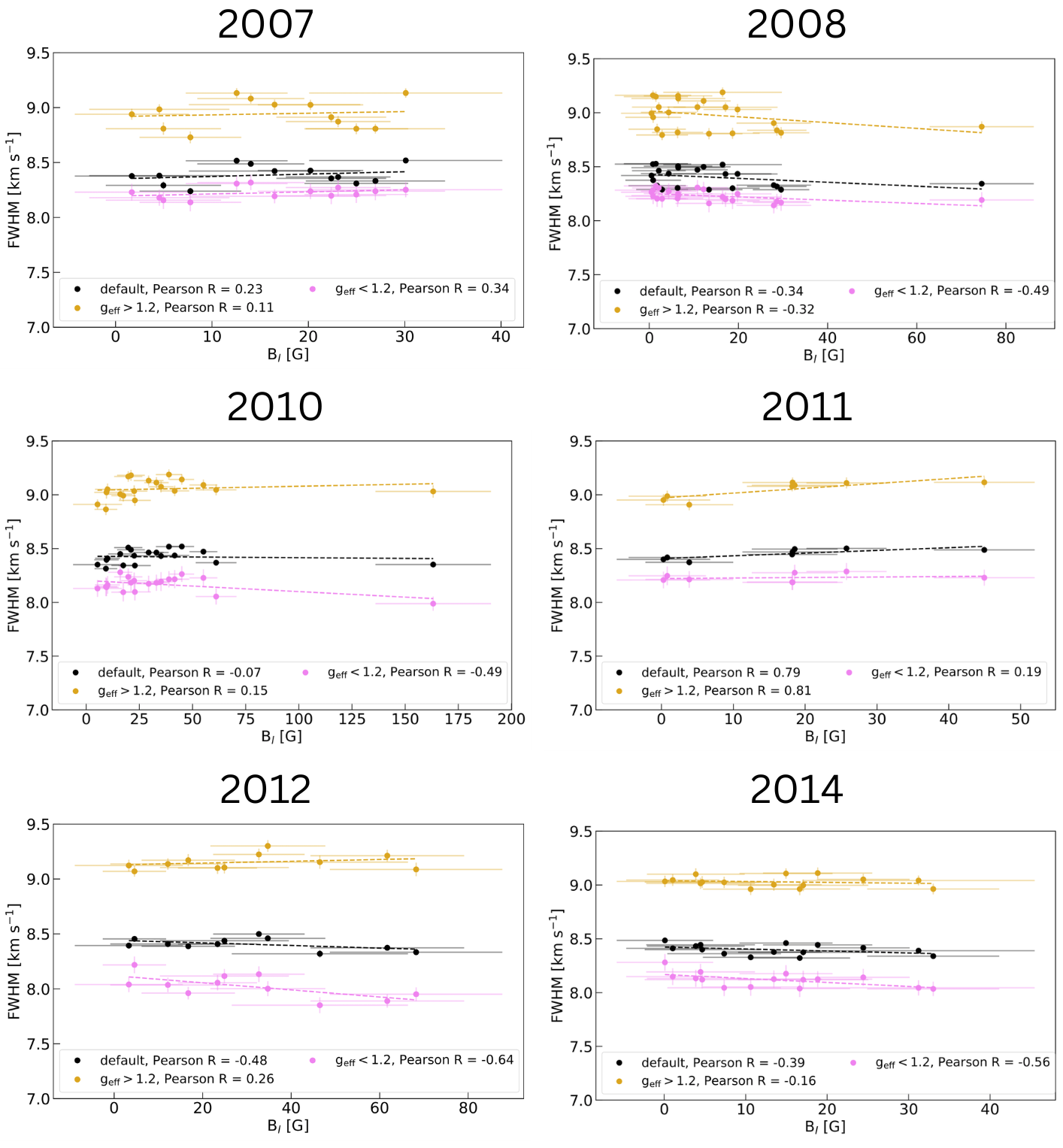}
    \caption{Correlation analysis between FWHM and $|$B$_l|$ for DS~Leo. The six panels correspond to the optical epochs in which rotational modulation is present: 2007, 2008, 2010, 2011, 2012, and 2014. In all panels the data points are colour-coded based on the line mask used for LSD: full (black), low-g$_\mathrm{eff}$ (purple), and high-g$_\mathrm{eff}$ (yellow).}
    \label{fig:FWHM_correl_DSLeo}%
\end{figure*}

\section{Complementary figures to PCA}\label{app:pca_app}

\begin{figure*}
        \raggedright \textbf{a.} \hspace{6.8cm} \textbf{b.} \\
        \centering
        \includegraphics[width=0.55\columnwidth, trim={0 0 0 0}, clip]{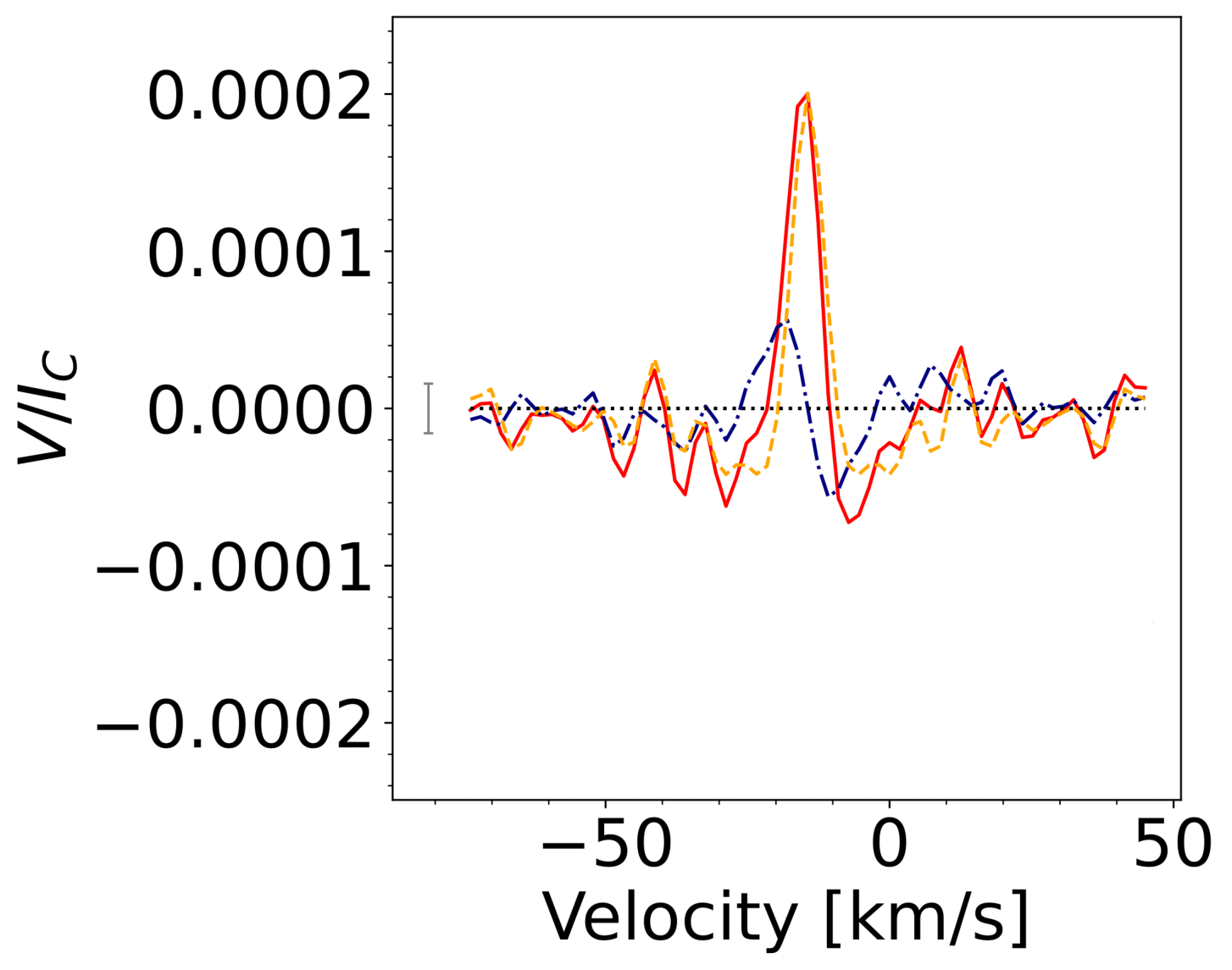}
        \includegraphics[width=\columnwidth, trim={0 400 445 0}, clip]{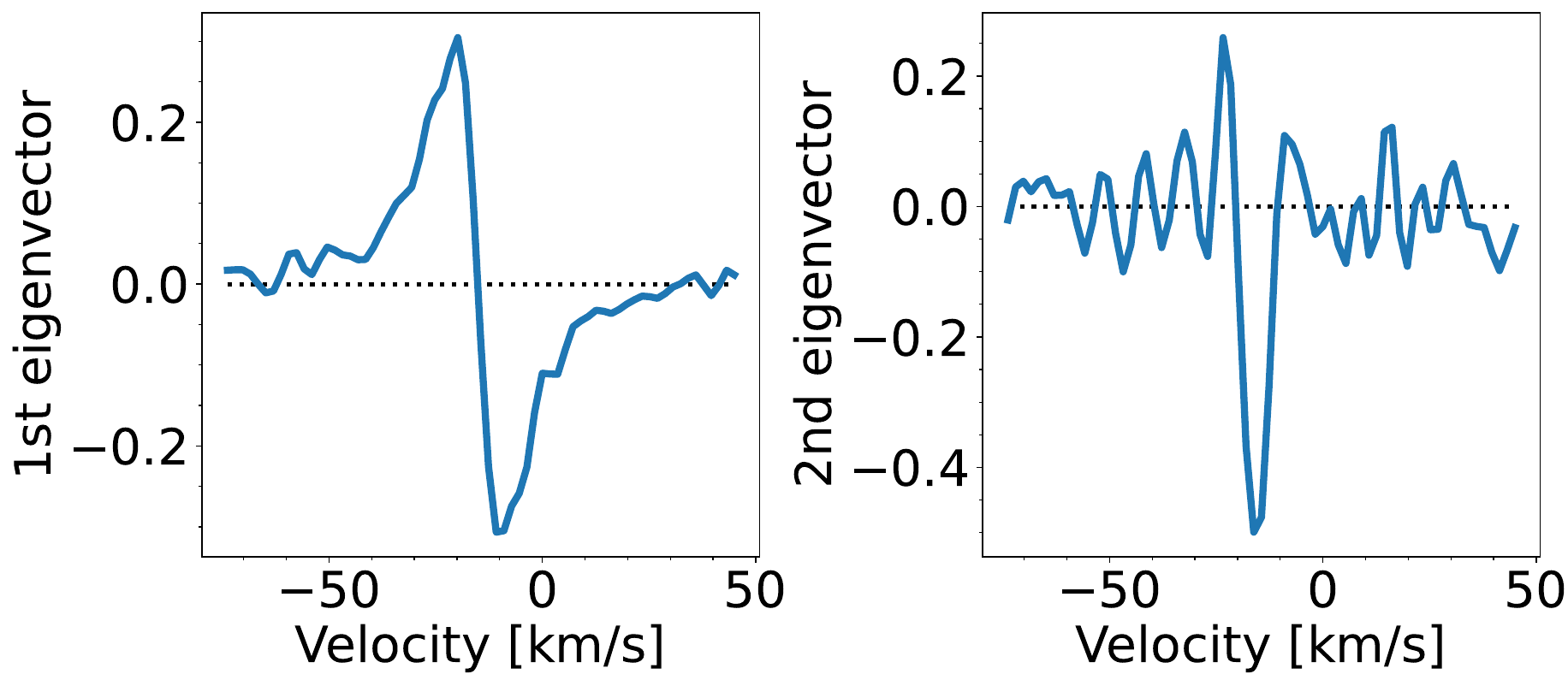}
	 
        \rule{7cm}{0.3mm}\\
        \raggedright \textbf{c.} \\
        \centering
 \textbf{2020b2021a}\\
        \includegraphics[width=0.55\columnwidth, trim={0 0 0 0}, clip]{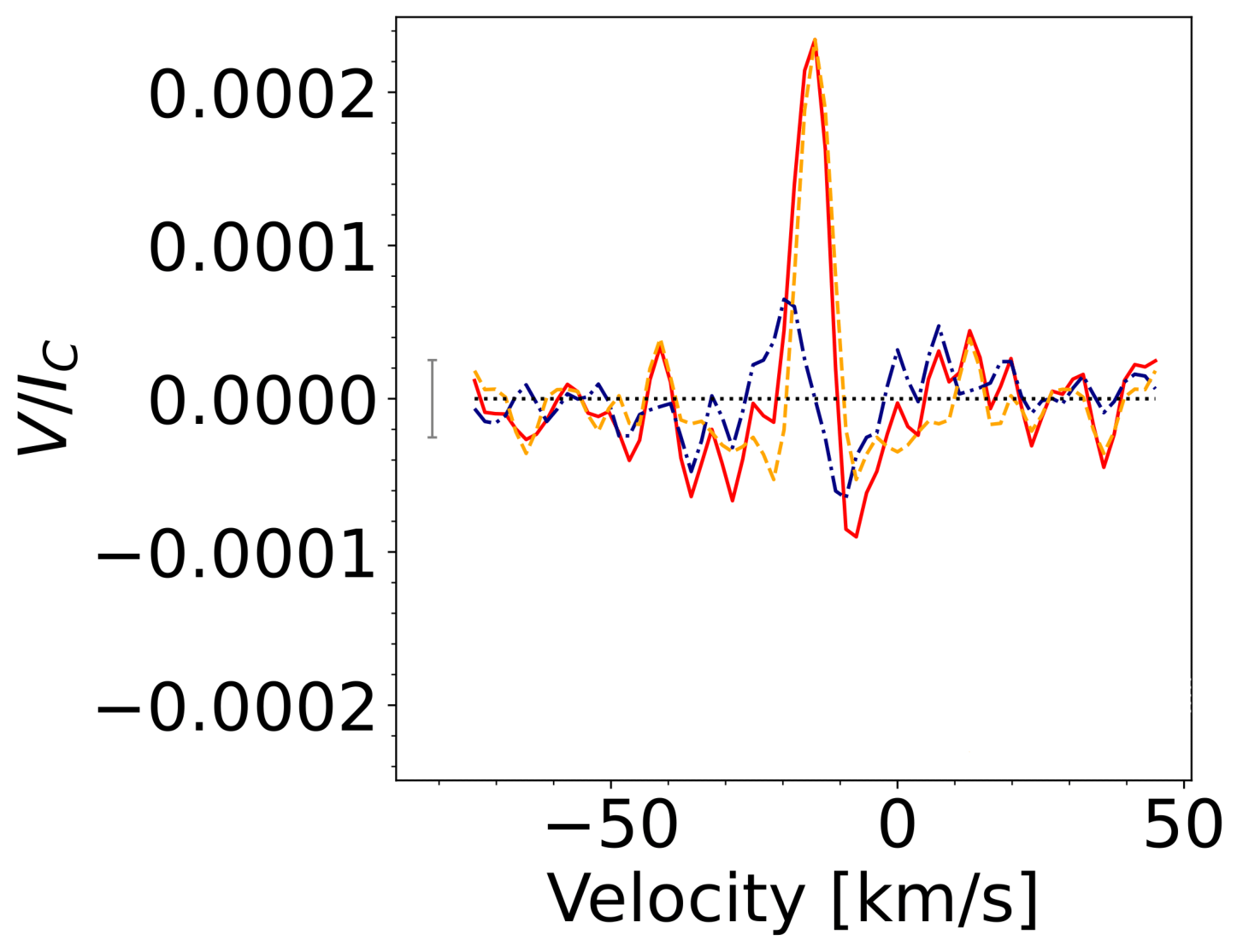}
        \includegraphics[width=\columnwidth, trim={0 400 450 0}, clip]{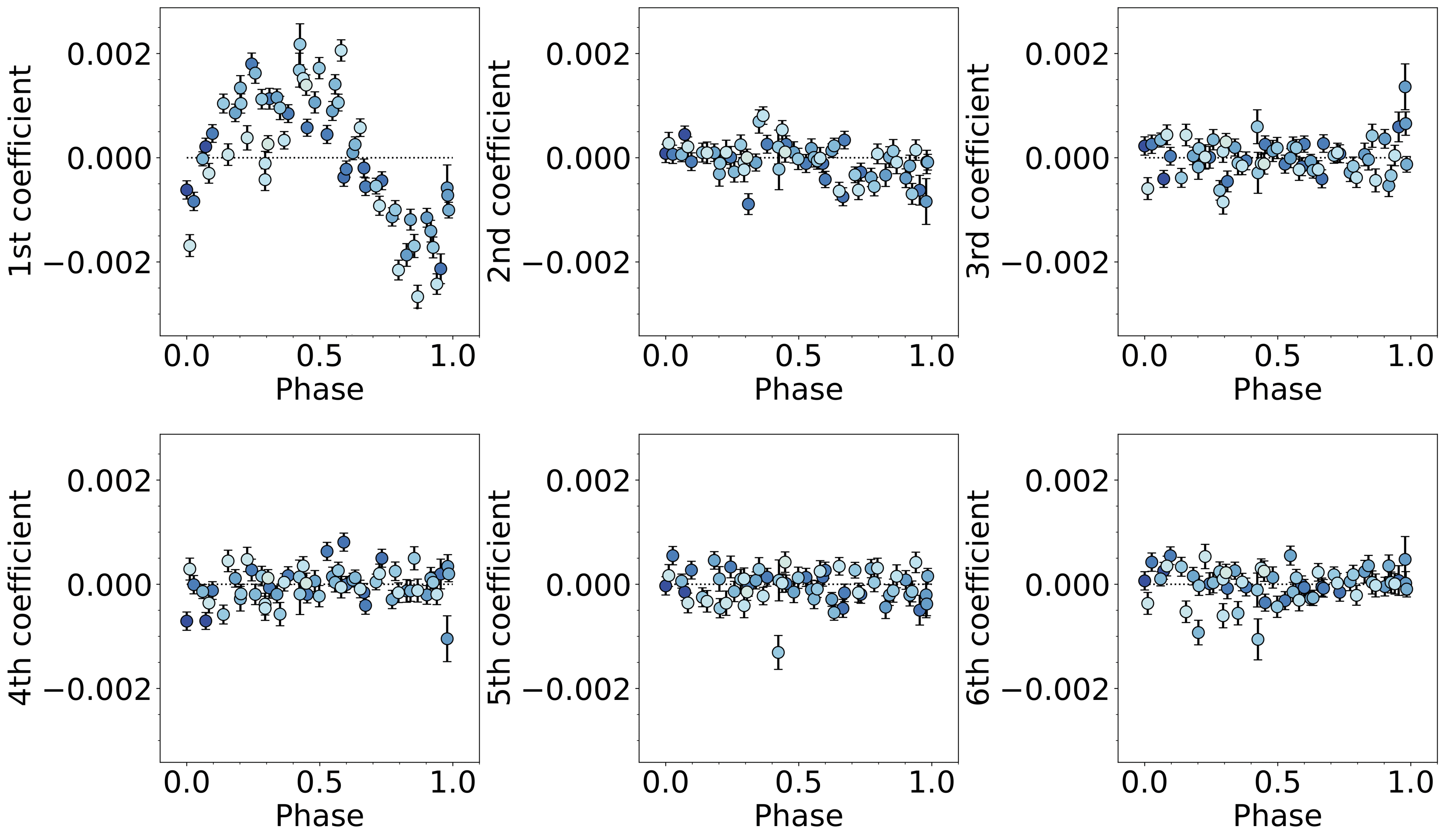}
	
 \textbf{2021b2022a}\\
                \includegraphics[width=0.55\columnwidth, trim={0 0 0 0}, clip]{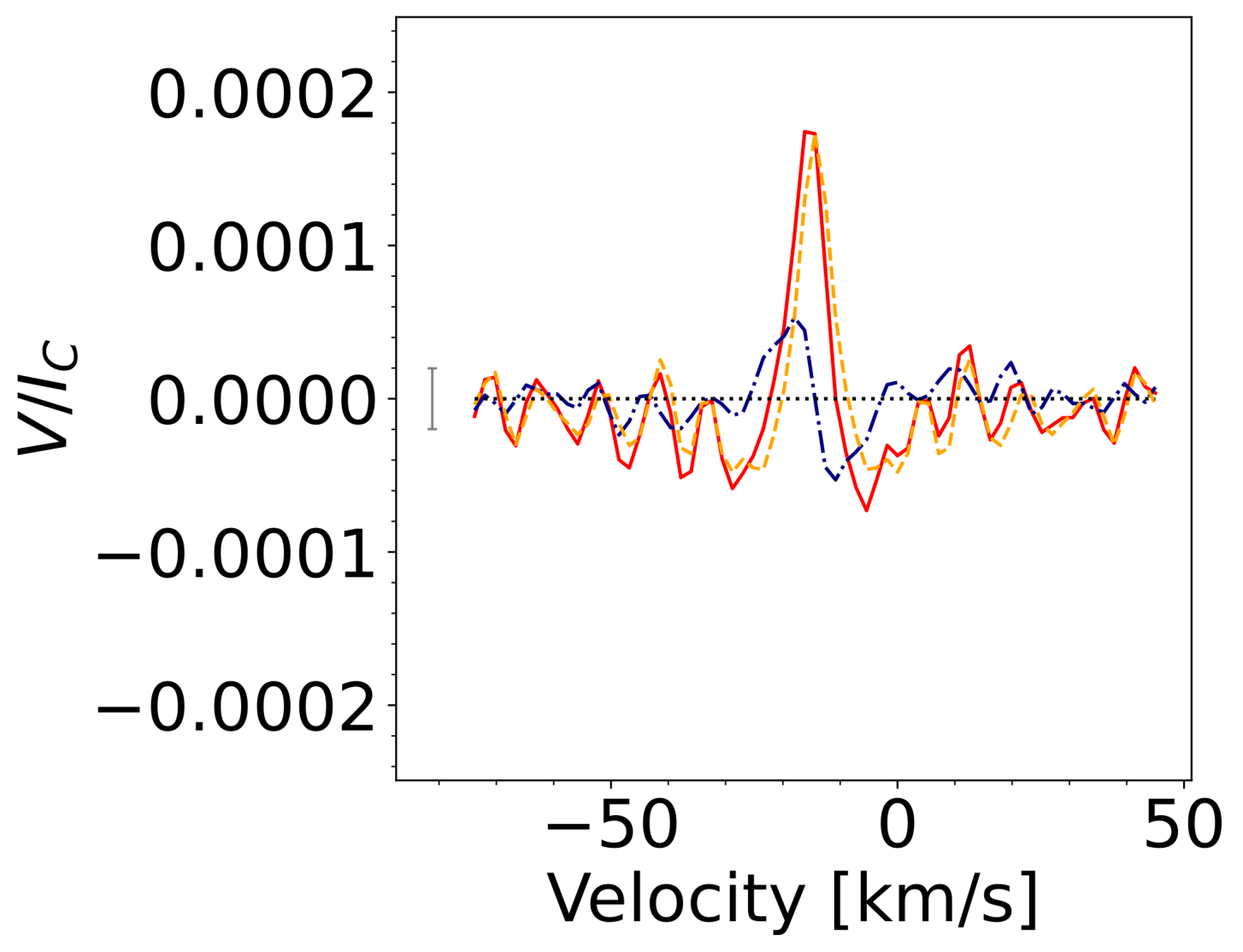}
        \includegraphics[width=\columnwidth, trim={0 400 450 0}, clip]{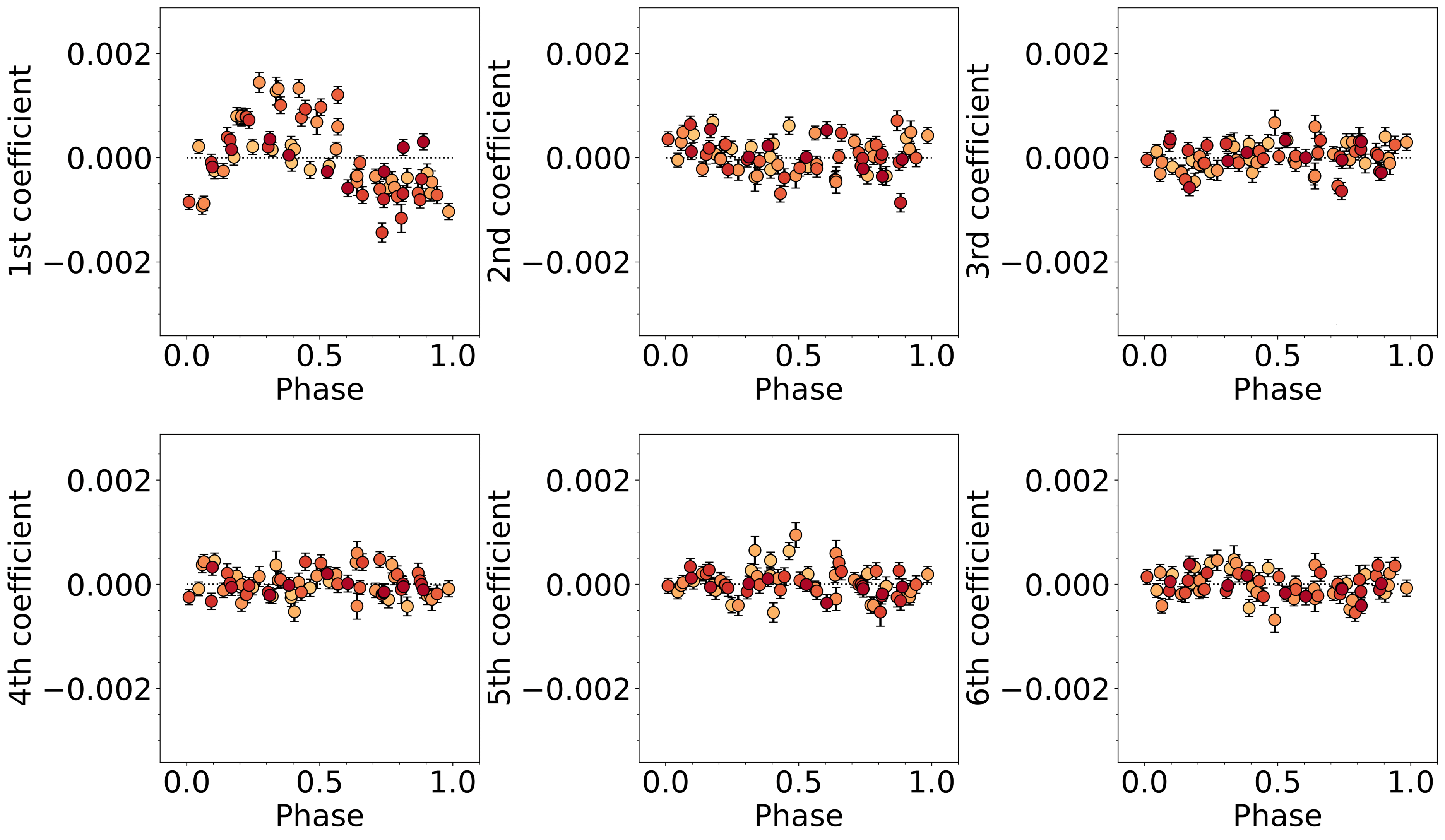}
    \caption{Same as Fig.~\ref{Fig:EVLac_PCA}, but for DS~Leo.}
    \label{Fig:DSLeo_PCA}
\end{figure*}

\begin{figure*}
        \raggedright \textbf{a.} \hspace{6.8cm} \textbf{b.} \\
        \centering
        \includegraphics[width=0.55\columnwidth, trim={0 0 0 0}, clip]{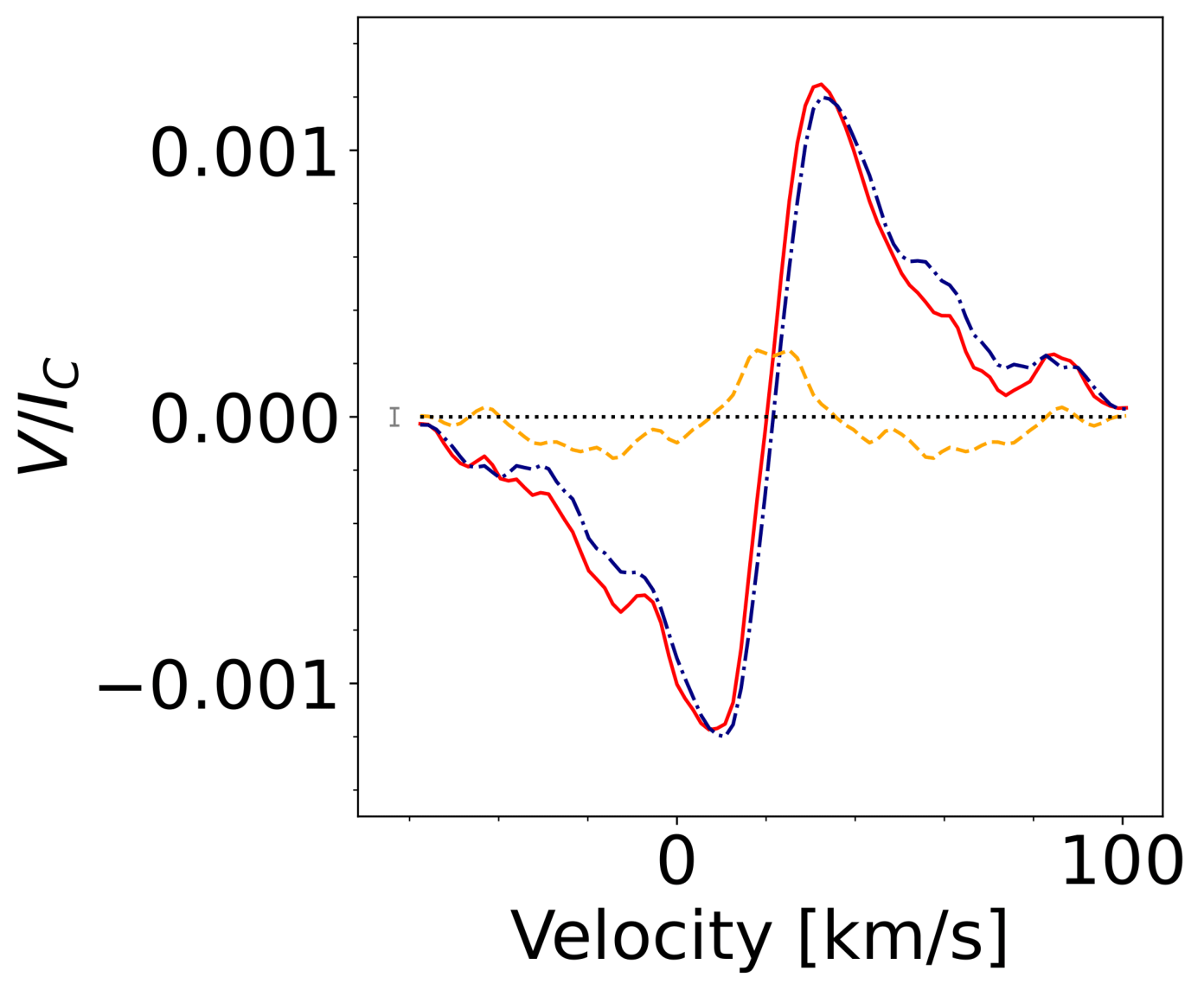}
        \includegraphics[width=\columnwidth, trim={0 400 445 0}, clip]{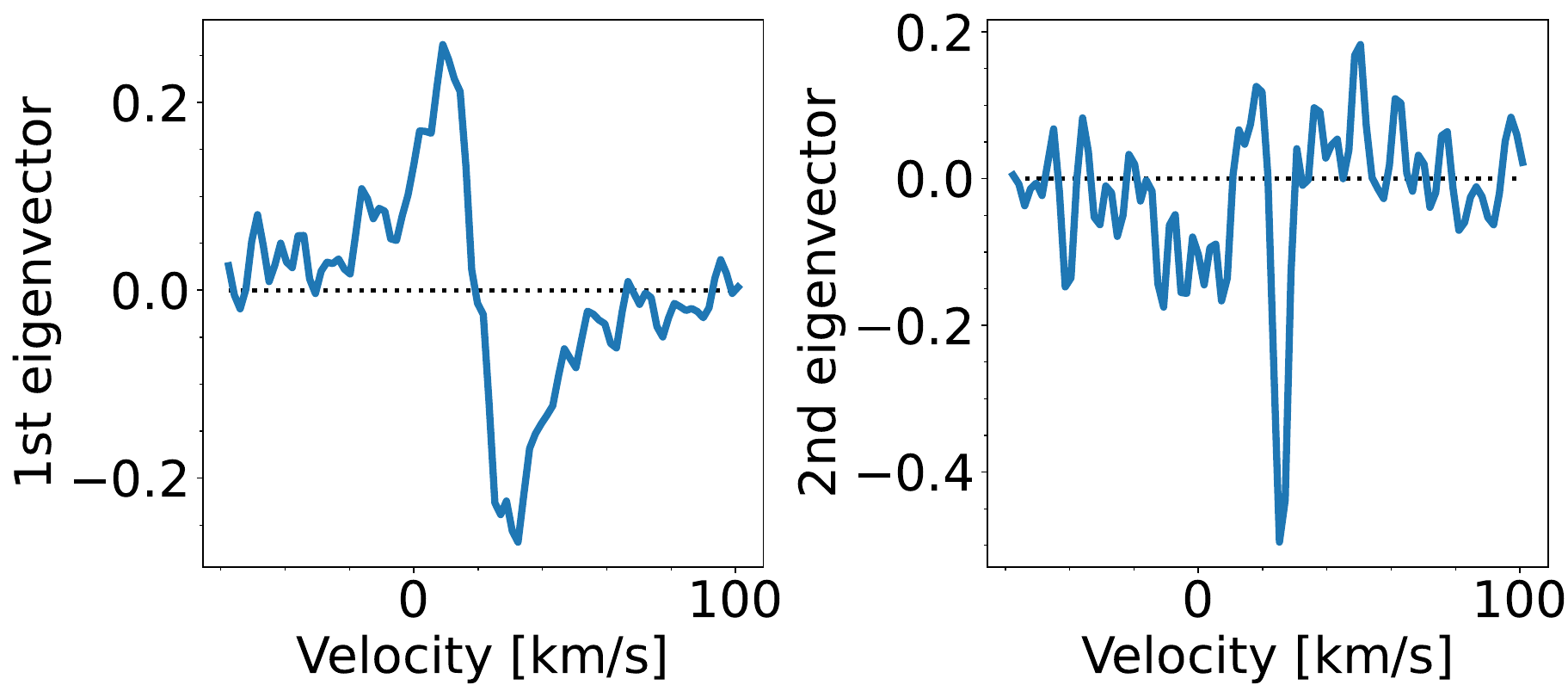}
	 
        \rule{7cm}{0.3mm}\\
        \raggedright \textbf{c.} \\
        \centering
 \textbf{2019a}\\
        \includegraphics[width=0.55\columnwidth, trim={0 0 0 0}, clip]{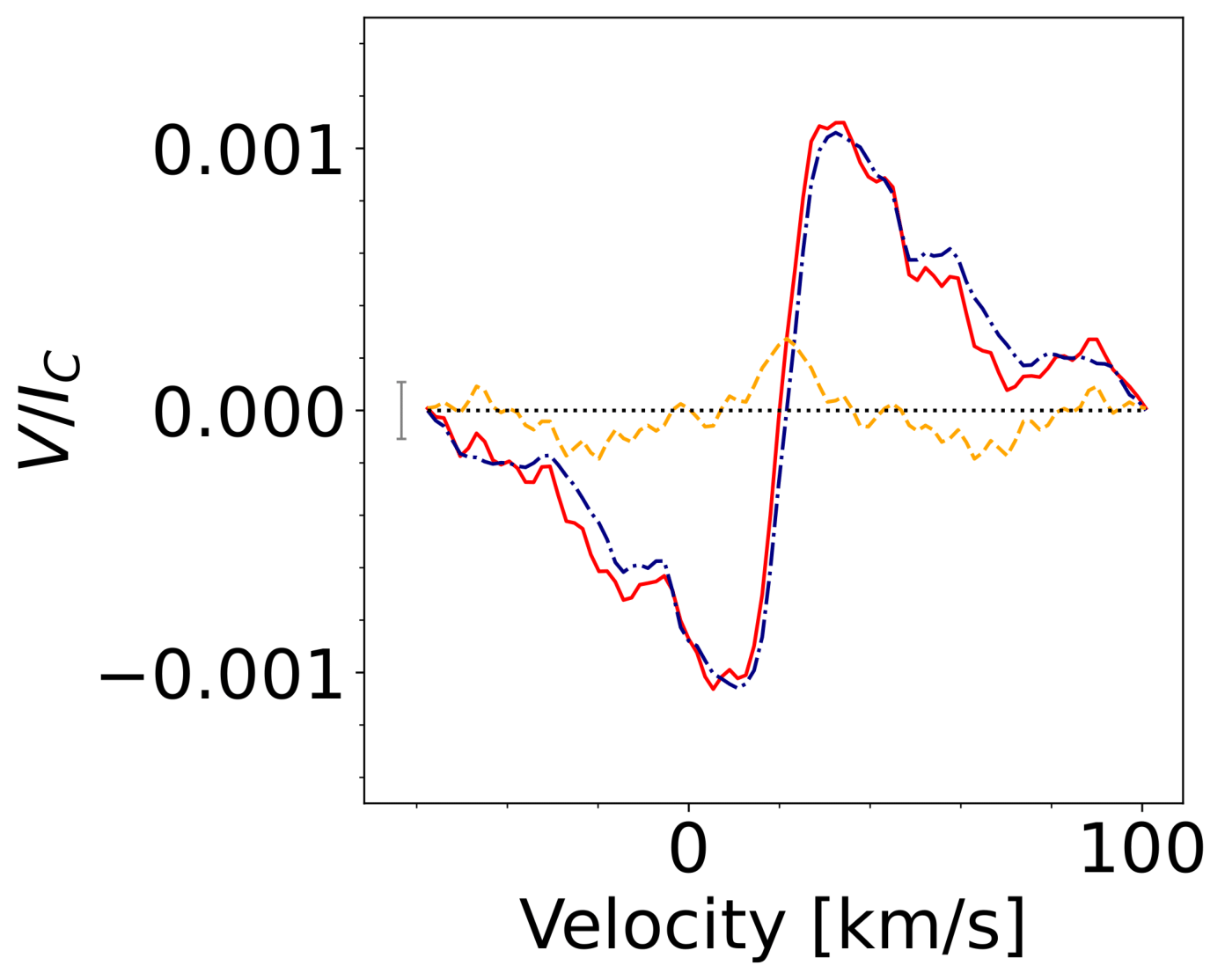}
        \includegraphics[width=\columnwidth, trim={0 400 450 0}, clip]{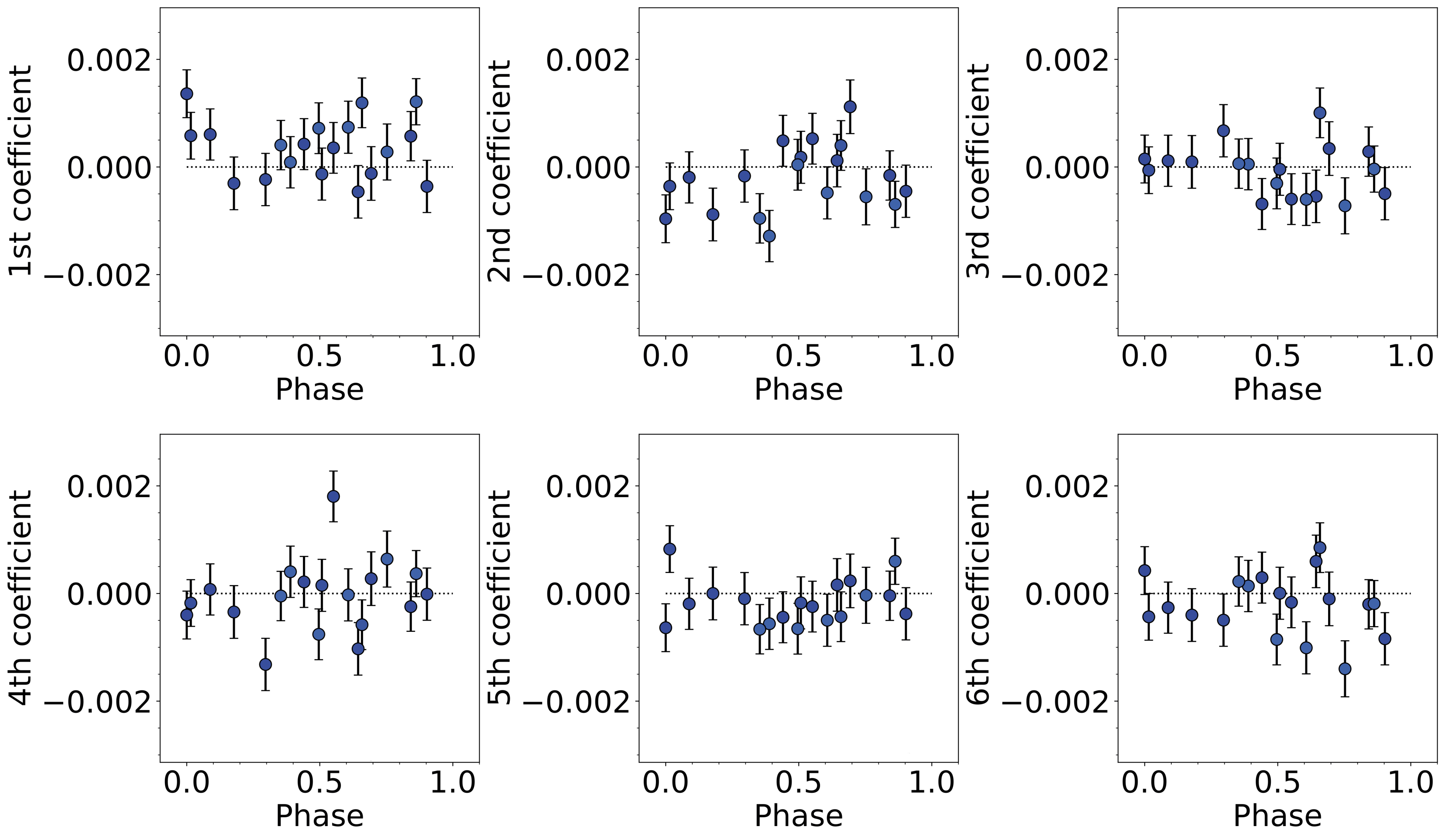}
	
 \textbf{2019b2020a}\\
        \includegraphics[width=0.55\columnwidth, trim={0 0 0 0}, clip]{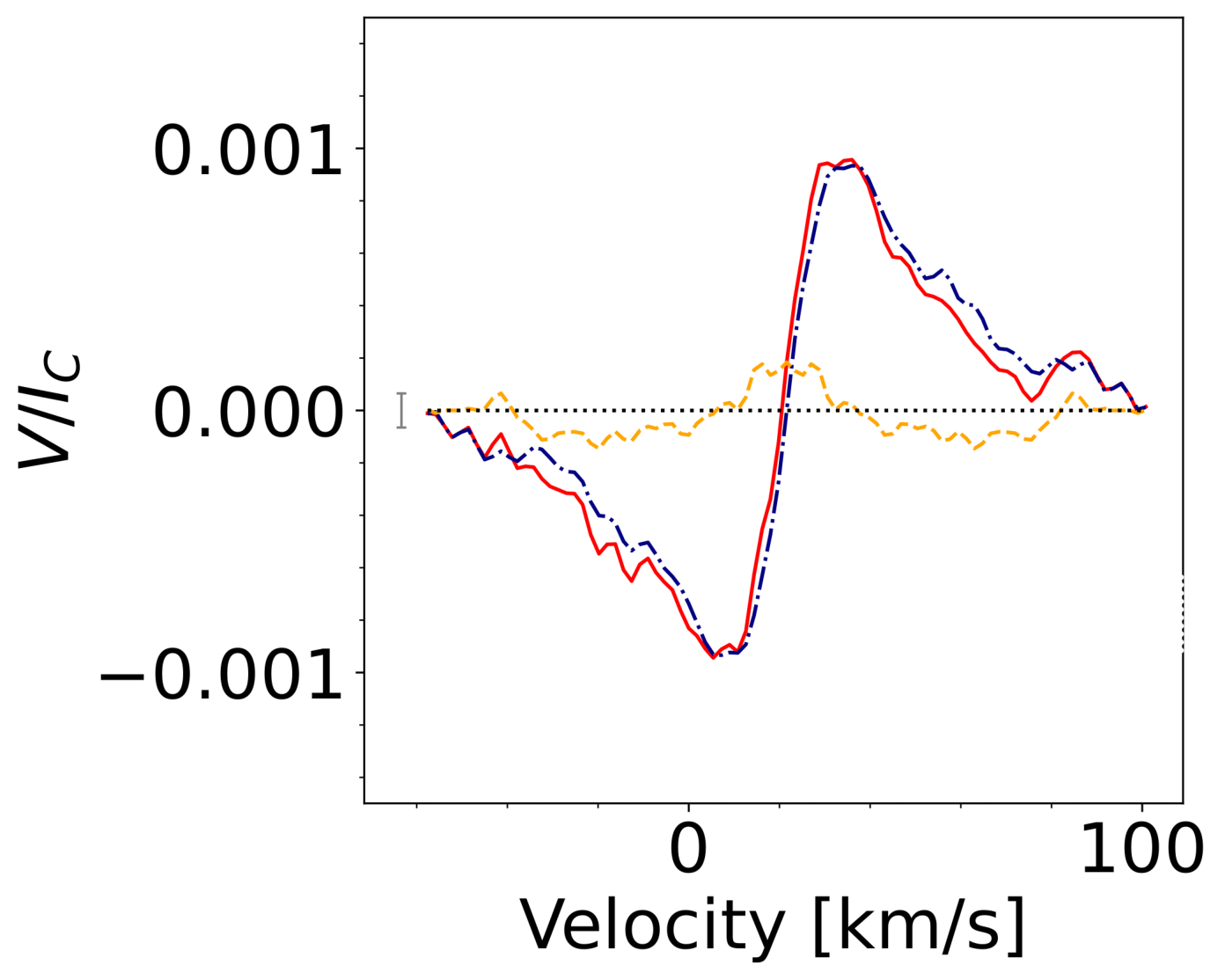}
        \includegraphics[width=\columnwidth, trim={0 400 450 0}, clip]{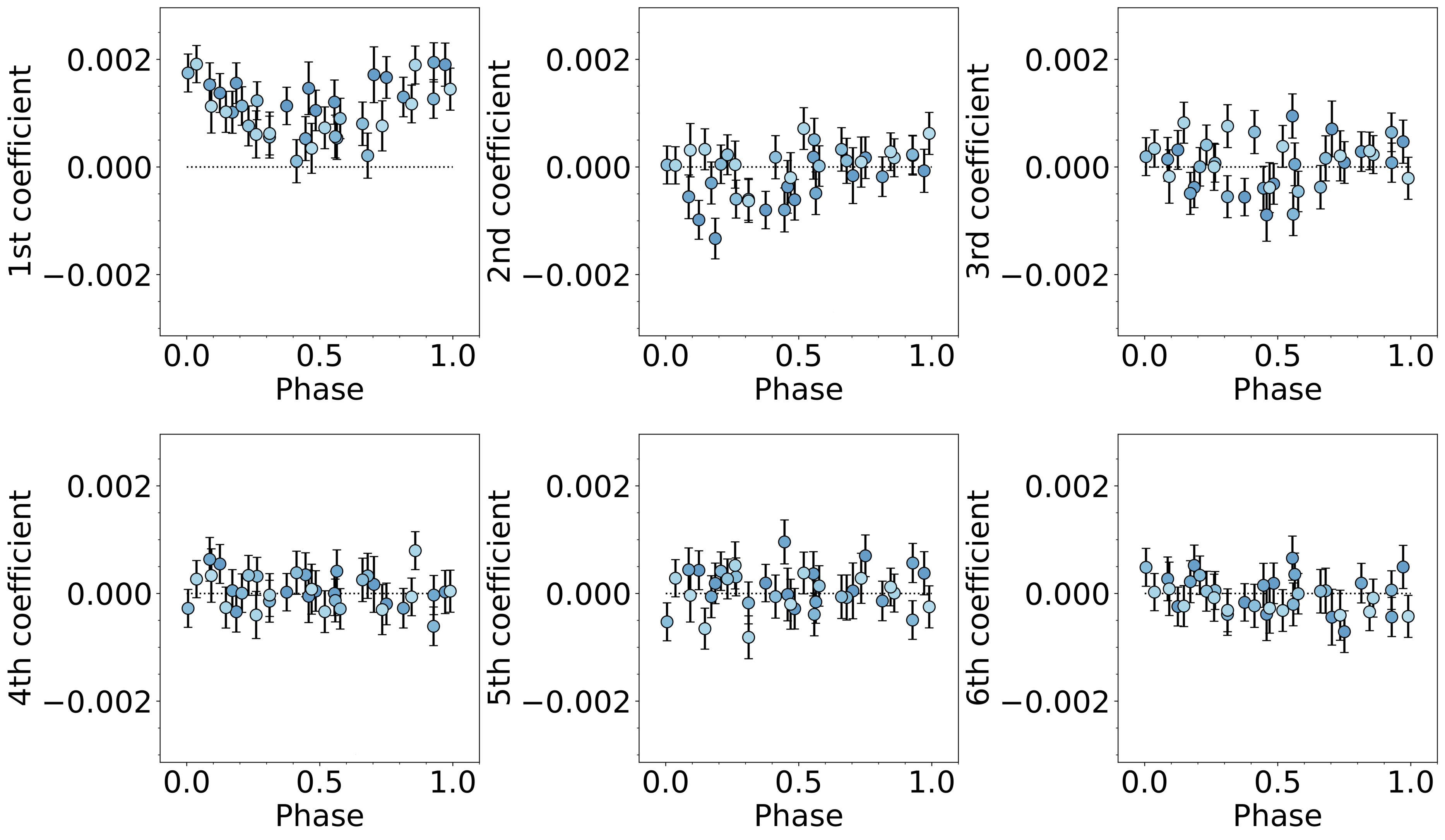}
	
 \textbf{2020b2021a}\\
        \includegraphics[width=0.55\columnwidth, trim={0 0 0 0}, clip]{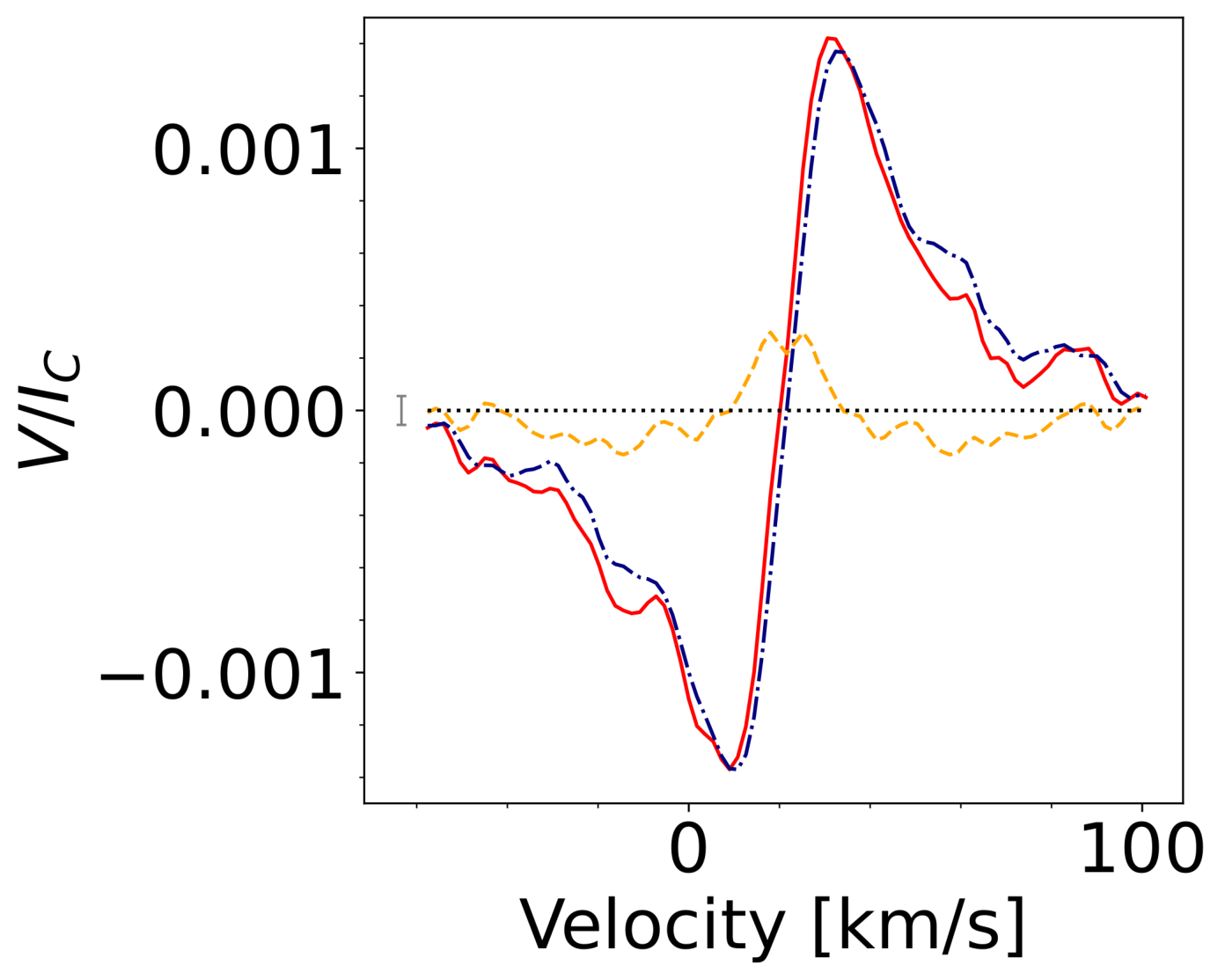}
        \includegraphics[width=\columnwidth, trim={0 400 450 0}, clip]{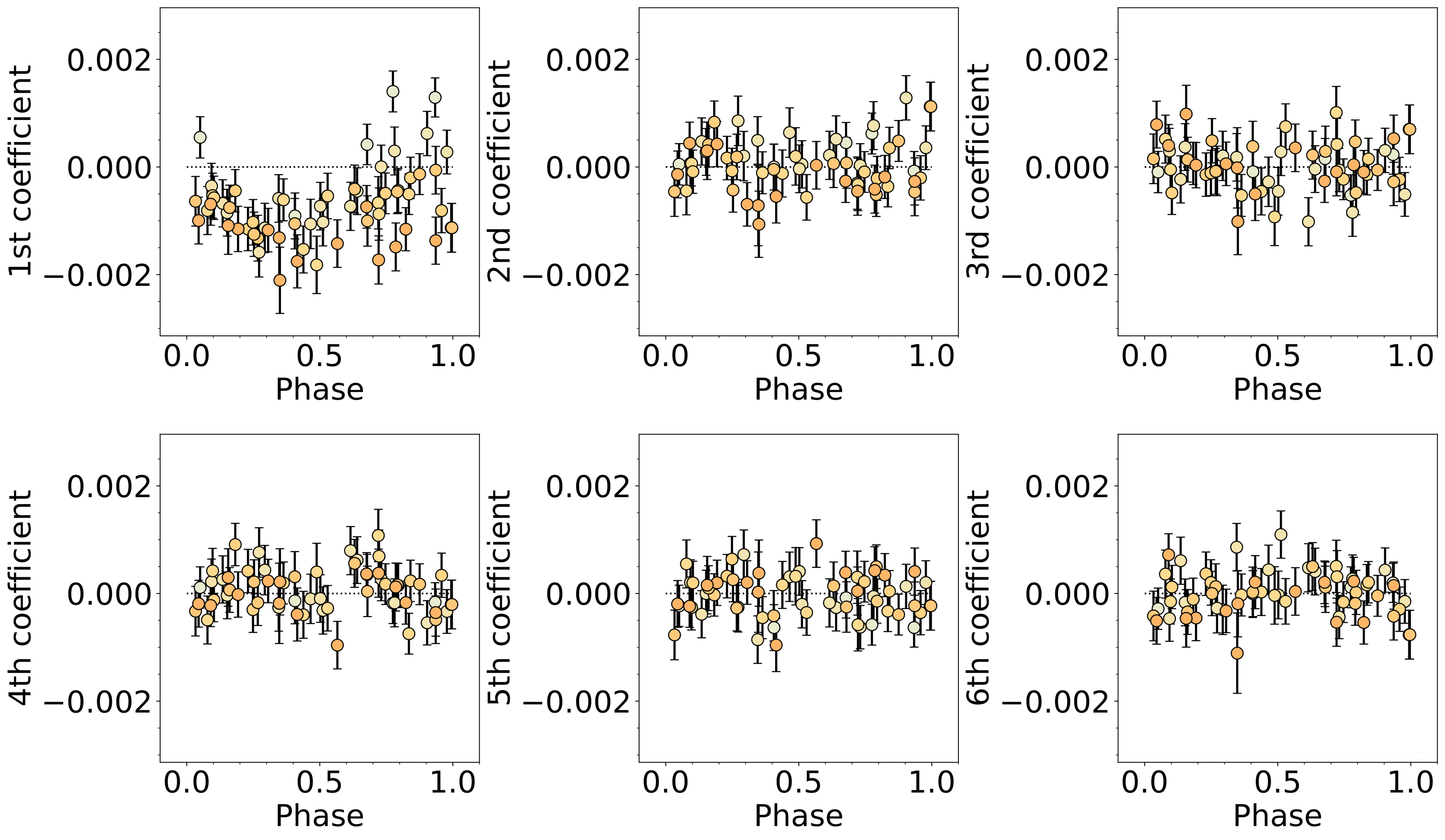}
	
 \textbf{2021b2022a}\\
        \includegraphics[width=0.55\columnwidth, trim={0 0 0 0}, clip]{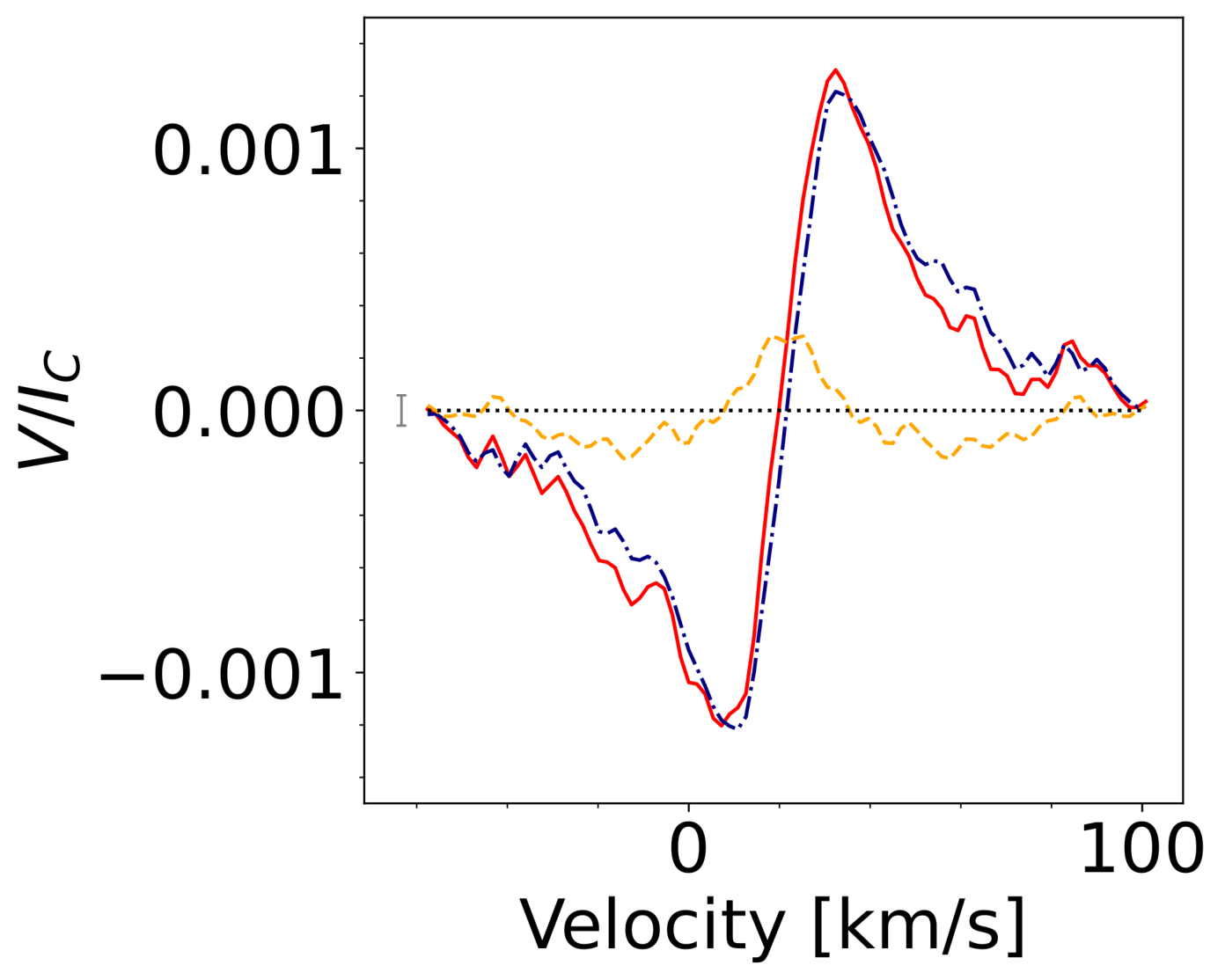}
        \includegraphics[width=\columnwidth, trim={0 400 450 0}, clip]{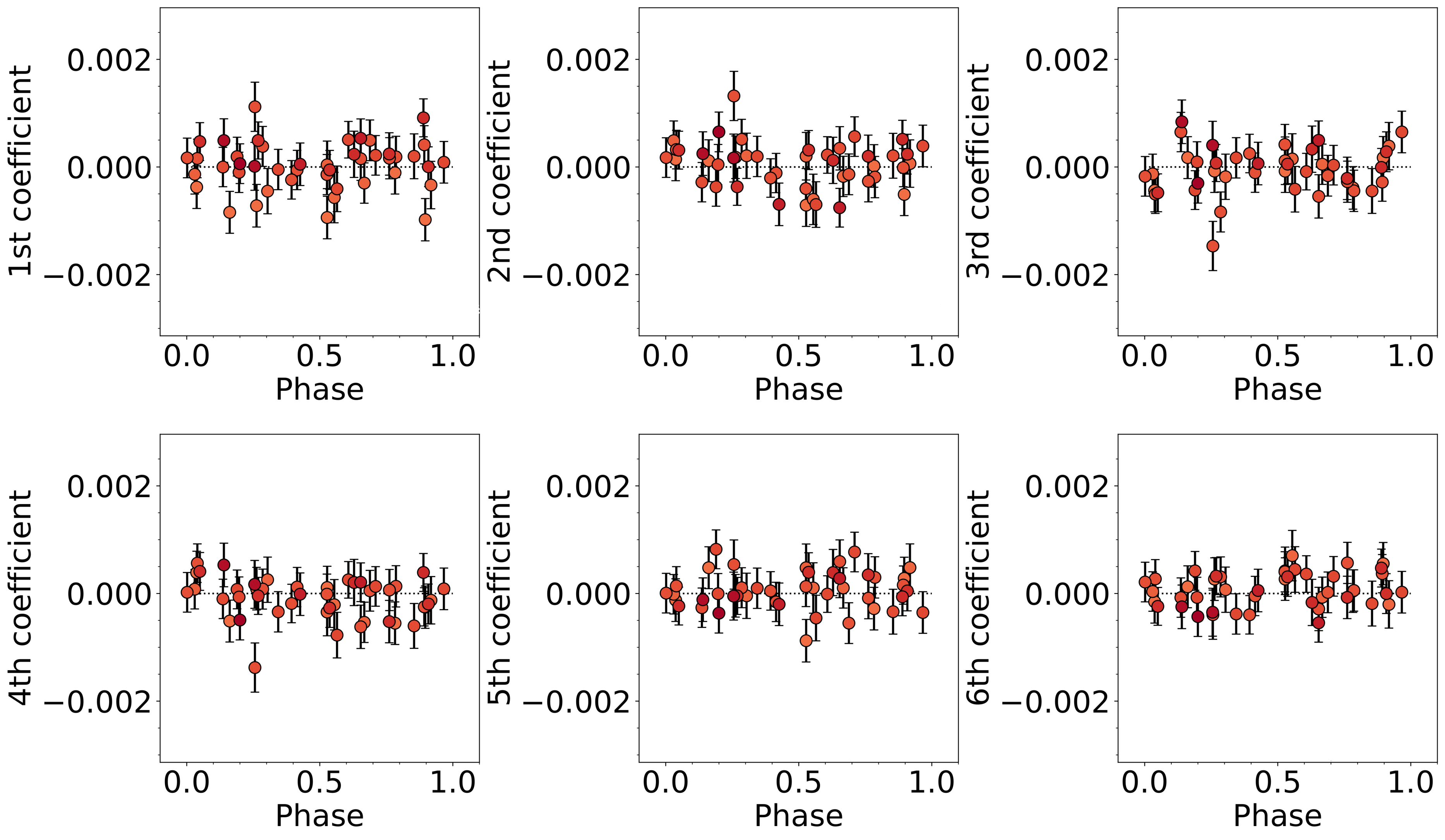}
    \caption{Same as Fig.~\ref{Fig:EVLac_PCA}, but for CN~Leo.}
    \label{Fig:CNLeo_PCA}
\end{figure*}

In this appendix, we illustrate the PCA analysis applied to DS Leo and CN Leo outlined in Sec.~\ref{sec:pca}. The mean Stokes~$V$ profile analysis and the per-epoch analysis of the coefficients are shown in Fig.~\ref{Fig:DSLeo_PCA} and Fig.~\ref{Fig:CNLeo_PCA}.

\section{Complementary figures to the ZDI analysis}\label{app:zdi_app}

\begin{figure*}[t]
    \centering
    \includegraphics[width=\textwidth]{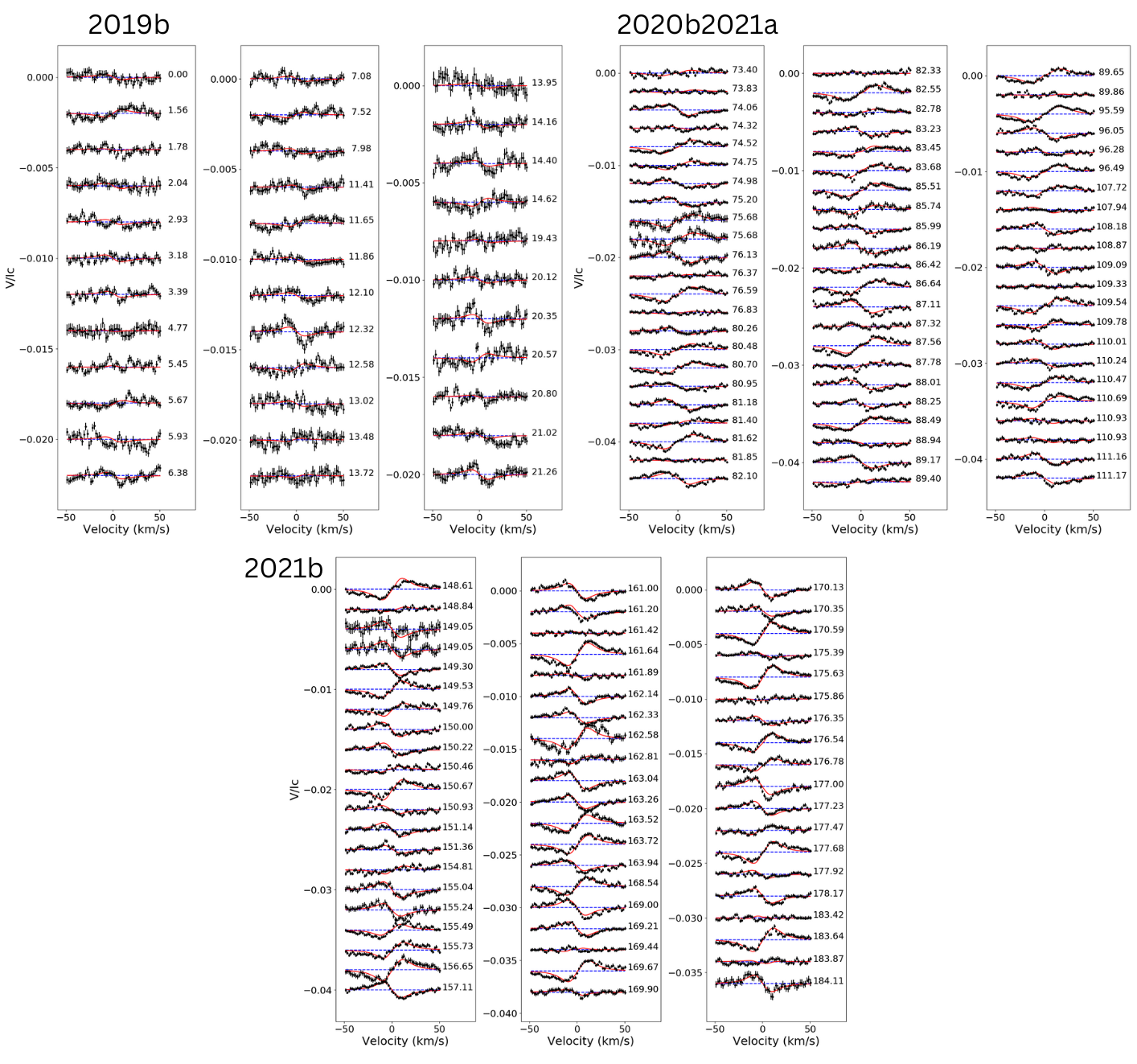}
    \caption{SPIRou time series of Stokes~$V$ profiles of EV~Lac. Top left: 2019b. Top right: 2020b2021a. Bottom: 2021b. The observations are shown in black and the models in red, and the rotational cycle is printed on the right (see Eq.~\ref{eq:ephemeris}). The profiles are offset vertically for visualisation purposes.}
    \label{fig:zdi_stokesV_evlac}%
\end{figure*}

\begin{figure*}[!t]
    \centering
    \includegraphics[width=0.74\textwidth]{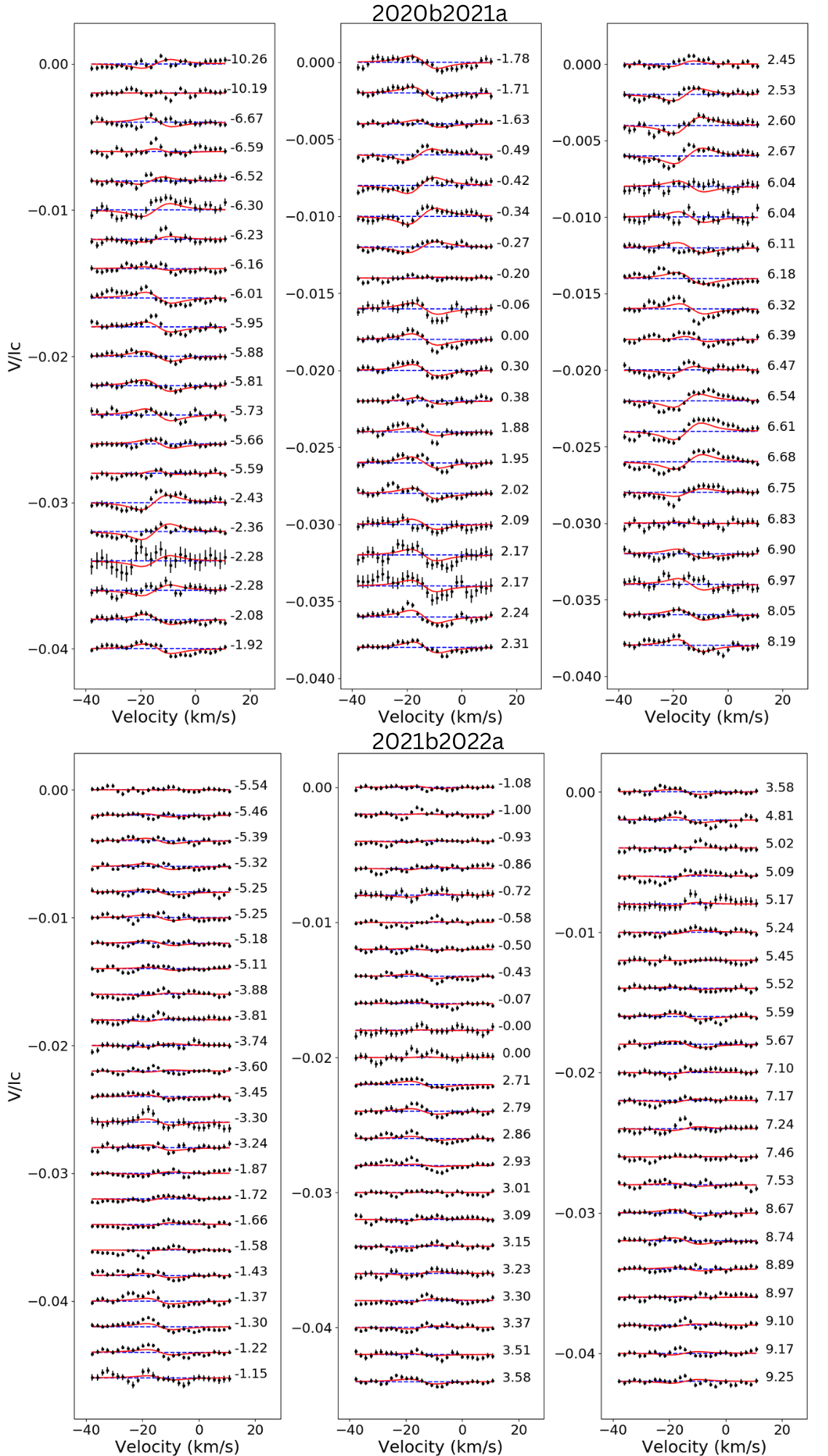}
    \caption{SPIRou time series of Stokes~$V$ profiles of DS~Leo. Top: 2020b2021a. Bottom: 2021b2022a. The cycles in this plot are computed with Eq.~\ref{eq:ephemeris} while using the median HJD for each epoch. The format is the same as in Fig.~\ref{fig:zdi_stokesV_evlac}.}
    \label{fig:zdi_stokesV_dsleo}%
\end{figure*}

\begin{figure*}[!t]
    \centering
    \includegraphics[width=\textwidth]{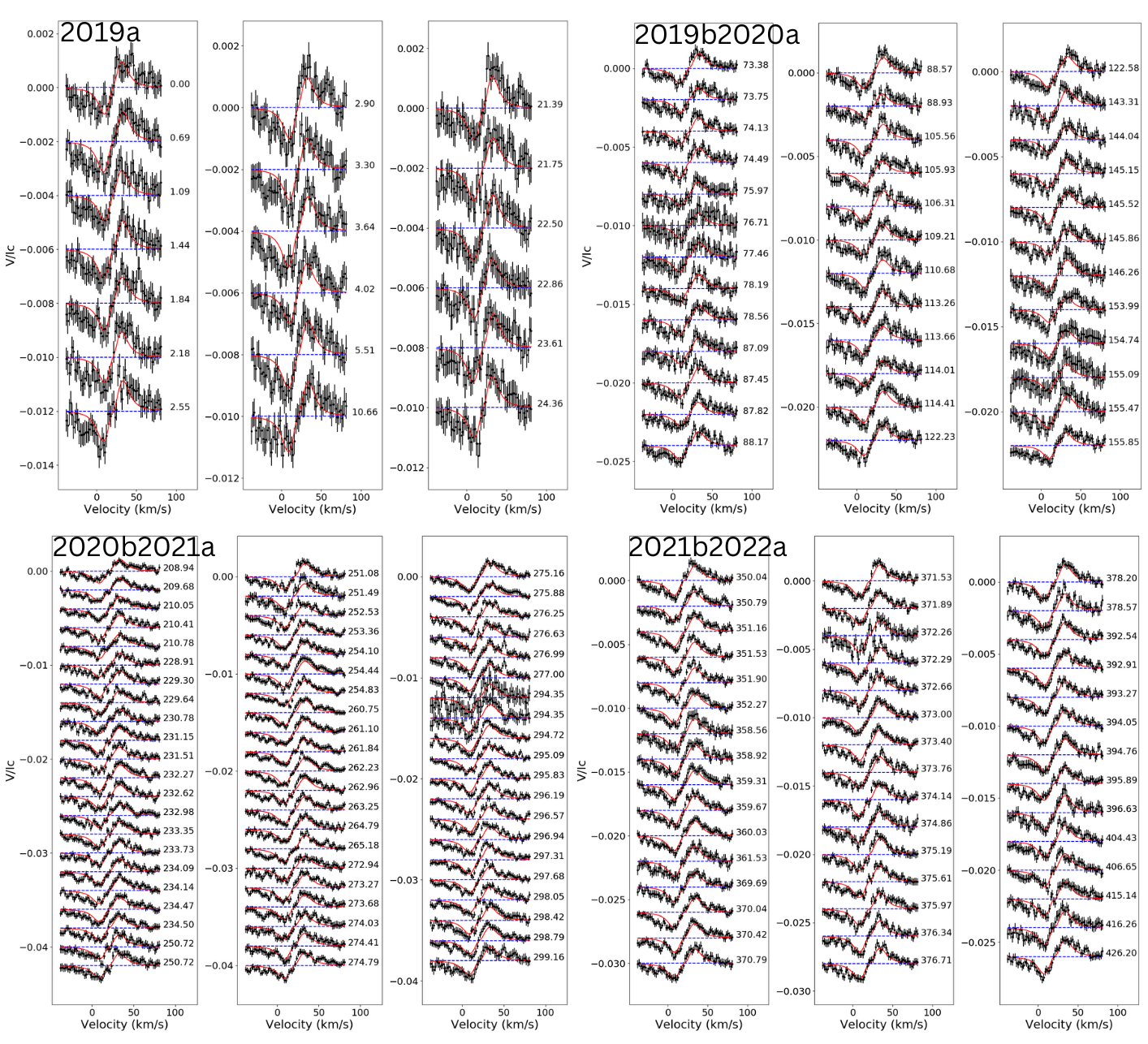}
    \caption{SPIRou time series of Stokes~$V$ profiles of CN~Leo. Top left: 2019a. Top right: 2019b2020a. Bottom left: 2020b2021a. Bottom right: 2021b2022a. The format is the same as in Fig.~\ref{fig:zdi_stokesV_evlac}.}
    \label{fig:zdi_stokesV_cnleo}%
\end{figure*}

In this appendix, we provide complementary figures to the Zeeman-Doppler imaging analysis (see Sec.~\ref{sec:magnetic_imaging}). Fig.~\ref{fig:zdi_stokesV_evlac}, Fig.~\ref{fig:zdi_stokesV_dsleo} and Fig.~\ref{fig:zdi_stokesV_cnleo} show the near-infrared time series of circularly polarised Stokes~$V$ profiles for EV~Lac, DS~Leo, and CN~Leo. 

\section{Observing log}\label{app:logs}

This appendix lists the spectropolarimetric observations of EV~Lac, DS~Leo and CN~Leo collected with ESPaDOnS, Narval and SPIRou. Tables with longitudinal magnetic field values analysed in Sec.~\ref{sec:Blon} are also provided.

\onecolumn

\begin{longtable}{lccrccc}
\caption{\label{tab:log_evlac} EV~Lac observations collected with SPIRou. The columns are: (1 and 2) date and universal time of the observations, (3) heliocentric Julian date, (4) rotational cycle of the observations found using Eq.~\ref{eq:ephemeris}, (5) exposure time of a polarimetric sequence, (6) signal-to-noise ratio at 1650 nm per polarimetric sequence, (7) RMS noise level of Stokes $V$ signal in units of unpolarised continuum.} \\
\hline\hline
Date & UT & HJD & $n_\mathrm{cyc}$ & $t_{exp}$ & S/N & $\sigma_\mathrm{LSD}$\\
 & [hh:mm:ss] & $[-2450000]$ & & [s] & & [$10^{-4}I_c$]\\
\hline
\endfirsthead
\caption{Continued.}\\
\hline\hline
Date & UT & HJD & $n_\mathrm{cyc}$ & $t_{exp}$ & S/N & $\sigma_\mathrm{LSD}$\\
 & [hh:mm:ss] & $[-2450000]$ & & [s] & & [$10^{-4}I_c$]\\
\hline
\endhead
\hline
\endfoot
\multicolumn{6}{c}{2019}\\
\hline
September 11 & 13:55:51.10 & 8738.0805 & 0.00 & 4x61 & 177 & 1.5\\
September 18 & 9:14:04.52 & 8744.8848 & 1.56 & 4x61 & 258 & 1.5\\
September 19 & 8:18:23.93 & 8745.8461 & 1.78 & 4x61 & 273 & 1.6\\
September 20 & 11:55:01.36 & 8746.9965 & 2.04 & 4x61 & 219 & 1.8\\
September 24 & 8:10:20.29 & 8750.8405 & 2.93 & 4x61 & 277 & 1.5\\
September 25 & 10:26:04.22 & 8751.9348 & 3.18 & 4x61 & 274 & 1.5\\
September 26 & 8:19:18.57 & 8752.8467 & 3.39 & 4x61 & 279 & 1.8\\
\end{longtable}

\begin{longtable}{lccrccc}
\caption{\label{tab:log_dsleo} DS~Leo observations collected with SPIRou. The columns are the same as in Table~\ref{tab:log_evlac}.} \\
\hline\hline
Date & UT & HJD & $n_\mathrm{cyc}$ & $t_{exp}$ & S/N & $\sigma_\mathrm{LSD}$\\
 & [hh:mm:ss] & $[-2450000]$ & & [s] & & [$10^{-4}I_c$]\\
\hline
\endfirsthead
\caption{Continued.}\\
\hline\hline
Date & UT & HJD & $n_\mathrm{cyc}$ & $t_{exp}$ & S/N & $\sigma_\mathrm{LSD}$\\
 & [hh:mm:ss] & $[-2450000]$ & & [s] & & [$10^{-4}I_c$]\\
\hline
\endhead
\hline
\endfoot
\multicolumn{6}{c}{2020}\\
\hline
November 04 & 14:41:01.65 & 9158.1132 & 0.00 & 4x66 & 242 & 1.3\\
November 05 & 14:32:49.50 & 9159.1076 & 0.07 & 4x66 & 248 & 1.3\\
December 24 & 13:11:16.12 & 9208.0545 & 3.59 & 4x55 & 250 & 1.5\\
December 25 & 14:26:36.37 & 9209.1069 & 3.67 & 4x66 & 255 & 1.4\\
December 26 & 12:54:42.75 & 9210.0431 & 3.73 & 4x83 & 221 & 1.4\\
December 29 & 14:46:25.61 & 9213.1208 & 3.95 & 4x83 & 172 & 2.0\\
December 30 & 14:48:29.36 & 9214.1222 & 4.03 & 4x78 & 241 & 1.5\\
\end{longtable}

\begin{longtable}{lccrccc}
\caption{\label{tab:log_cnleo} CN~Leo observations collected with SPIRou. The columns are the same as in Table~\ref{tab:log_evlac}.} \\
\hline\hline
Date & UT & HJD & $n_\mathrm{cyc}$ & $t_{exp}$ & S/N & $\sigma_\mathrm{LSD}$\\
 & [hh:mm:ss] & $[-2450000]$ & & [s] & & [$10^{-4}I_c$]\\
\hline
\endfirsthead
\caption{Continued.}\\
\hline\hline
Date & UT & HJD & $n_\mathrm{cyc}$ & $t_{exp}$ & S/N & $\sigma_\mathrm{LSD}$\\
 & [hh:mm:ss] & $[-2450000]$ & & [s] & & [$10^{-4}I_c$]\\
\hline
\endhead
\hline
\endfoot
\multicolumn{6}{c}{2019}\\
\hline
April 16 & 12:13:42.22 & 8590.0095 & 22.56 & 4x128 & 176 & 2.5\\
April 18 & 9:09:19.88 & 8591.8815 & 23.25 & 4x117 & 171 & 2.8\\
April 19 & 10:44:28.81 & 8592.9476 & 23.65 & 4x128 & 163 & 2.8\\
April 20 & 9:35:28.57 & 8593.8996 & 24.00 & 4x128 & 171 & 2.7\\
April 21 & 11:35:37.16 & 8594.9831 & 24.40 & 4x128 & 170 & 2.8\\
April 22 & 9:19:34.48 & 8595.8886 & 24.74 & 4x111 & 172 & 2.6\\
April 23 & 9:33:44.59 & 8596.8984 & 25.11 & 4x111 & 172 & 2.7\\
\end{longtable}

\begin{longtable}{ccc}
\caption{\label{tab:Bl_log_evlac} Optical and near-infrared measurements of longitudinal magnetic field for EV~Lac. The columns are: (1) heliocentric Julian date of the observation, (2) B$_l$ with formal error bar (see Eq.~\ref{eq:Bl}), and (3) instrument used.} \\
\hline\hline
HJD & B$_l$ & Instrument\\
 $[-2450000]$ & [G] &\\
\hline
\endfirsthead
\caption{Continued.}\\
\hline\hline
HJD & B$_l$ & Instrument\\
 $[-2450000]$ & [G] &\\
\hline
\endhead
\hline
\endfoot
3569.1119 &    144.4$\pm$8.3 & ESPaDOnS \\
3606.1478 & $-$305.9$\pm$20.4 & ESPaDOnS \\
3631.9065 &    23.9$\pm$9.4 & ESPaDOnS \\
3631.9447 & $-$106.2$\pm$11.0 & ESPaDOnS \\
3631.9828 & $-$82.9$\pm$9.6 & ESPaDOnS \\
3953.0731 & $-$529.5$\pm$20.6 & ESPaDOnS \\
3955.0663 &    319.8$\pm$15.1 & ESPaDOnS \\
3956.0600 & $-$354.3$\pm$13.8 & ESPaDOnS \\
\end{longtable}

\begin{longtable}{ccc}
\caption{\label{tab:Bl_log_dsleo} Optical and near-infrared measurements of longitudinal magnetic field for DS~Leo. The columns are the same as in Table~\ref{tab:Bl_log_evlac}.} \\
\hline\hline
HJD & B$_l$ & Instrument\\
 $[-2450000]$ & [G] &\\
\hline
\endfirsthead
\caption{Continued.}\\
\hline\hline
HJD & B$_l$ & Instrument\\
 $[-2450000]$ & [G] &\\
\hline
\endhead
\hline
\endfoot
3748.1023 &    4.7$\pm$5.9 & ESPaDOnS \\
3748.9053 &    23.6$\pm$13.1 & ESPaDOnS \\
3774.0359 &    9.8$\pm$6.0 & ESPaDOnS \\
3781.1526 & $-$7.9$\pm$9.1 & ESPaDOnS \\
4126.6502 & $-$16.5$\pm$8.9 & NARVAL \\
4127.6349 & $-$4.5$\pm$7.3 & NARVAL \\
4128.6469 &    1.7$\pm$5.9 & NARVAL \\
4129.6107 &    22.4$\pm$5.7 & NARVAL \\
\end{longtable}

\begin{longtable}{ccc}
\caption{\label{tab:Bl_log_cnleo} Optical and near-infrared measurements of longitudinal magnetic field for CN~Leo. The columns are the same as in Table~\ref{tab:Bl_log_evlac}.} \\
\hline\hline
HJD & B$_l$ & Instrument\\
 $[-2450000]$ & [G] &\\
\hline
\endfirsthead
\caption{Continued.}\\
\hline\hline
HJD & B$_l$ & Instrument\\
 $[-2450000]$ & [G] &\\
\hline
\endhead
\hline
\endfoot
4546.9935 & $-$633.8$\pm$52.7 & ESPaDOnS\\
4547.9162 & $-$669.7$\pm$61.8 & ESPaDOnS\\
4547.9584 & $-$779.1$\pm$64.6 & ESPaDOnS\\
4548.8749 & $-$825.1$\pm$64.4 & ESPaDOnS\\
8590.0095 & $-$576.1$\pm$73.6 & SPIRou \\
8591.8815 & $-$632.6$\pm$70.9 & SPIRou \\
8592.9476 & $-$592.8$\pm$77.2 & SPIRou \\
8593.8996 & $-$577.7$\pm$71.6 & SPIRou \\
\end{longtable}

\end{appendix}

\end{document}